\documentclass[preprint,12pt,authoryear]{elsarticle}

\usepackage{amssymb}
\usepackage[T1]{fontenc} 
\usepackage{subcaption}
\usepackage{float}
\usepackage{booktabs, caption, longtable, colortbl, array}
\usepackage{lscape}
\usepackage{hyperref}
\usepackage{rotating}
\usepackage{threeparttable}
\hypersetup{
  colorlinks   = true, 
  urlcolor     = red, 
  linkcolor    = blue, 
  citecolor   = blue 
}

\usepackage{xcolor}

\usepackage{multirow}
\usepackage{titletoc}

\usepackage{amsmath}

\usepackage[resetlabels]{multibib}
\newcites{appendix}{References}

\usepackage{chngcntr}

\journal{Social Networks}

\usepackage{todonotes}
\newcommand{\zack}[1]{
  {\color{black}#1}}

\makeatletter
\def\ps@pprintTitle{%
  \let\@oddhead\@empty
  \let\@evenhead\@empty
  \let\@oddfoot\@empty
  \let\@evenfoot\@oddfoot}
\makeatother

\begin{document}

\begin{frontmatter}



\title{Understanding the Personal Networks of People Experiencing Homelessness in King County, WA with Aggregate Relational Data\tnoteref{published}}

\tnotetext[published]{This is the authors' version of a manuscript published in \emph{Social Networks}. The final version of record is available at \url{https://doi.org/10.1016/j.socnet.2026.02.003}. Please cite as: Z. W. Almquist, I. Kahveci, O. Kajfasz, J. Rothfolk, and A. Hagopian (2026). Understanding the Personal Networks of People Experiencing Homelessness in King County, WA with Aggregate Relational Data. \emph{Social Networks} 86, 150--172. \url{https://doi.org/10.1016/j.socnet.2026.02.003}.}


\author[inst1]{Zack Almquist\footnote{Corresponding Author: \href{mailto:zalmquist@uw.edu}{zalmquist@uw.edu}; ORCID: \href{https://orcid.org/0000-0002-1967-123X}{0000-0002-1967-123X}}}
\author[inst1]{Ihsan Kahveci} 
\author[inst2]{Owen Kajfasz}
\author[inst2]{Janelle Rothfolk}
\author[inst1]{Amy Hagopian}

\affiliation[inst1]{organization={University of Washington},
            city={Seattle},
            state={WA},
            country={USA}
}

\affiliation[inst2]{organization={King County Regional Homelessness Authority},
            city={Seattle},
           state={WA},
            country={USA}}

%

\begin{abstract}
\noindent
Personal networks among people experiencing homelessness are vital but often overlooked. Drawing on a unique three-year dataset (2022–2024) from over 3,000 unhoused individuals in King County, Washington, this study analyzes the structure and change of four different personal networks: acquaintances, close friends, kinship (household), and peer referrals. Using this respondent-driven sample data set, we find that the mean size of the acquaintance networks ranged from 22 to 27 people, while the mean number of close friendships declined from 4.9 to 4.19. This drop, despite population growth, suggests a decrease in network density and potentially an increase in social isolation or a shift in composition (e.g., more people experiencing homelessness for the first time). However, we do see growth in the kinship networks, which grew slightly, indicating a greater prevalence of family co-homelessness or reliance on kin. These shifts highlight the need for policies that foster social connection and community stability.
\end{abstract}



\end{frontmatter}


\section{Introduction}

\noindent
The social networks of people experiencing homelessness are an understudied and important aspect of the lives of unhoused people in the United States (US). The personal networks of social and economic support \citep{small2013weak,small2017someone} play an important role in policy, support, and interventions (e.g., improved sleep locations or housing) for this population \citep{joly2014supporting,kennedy2022restructuring}. Personal networks for the unhoused population provide access to information (e.g., shelters, new programs, resources), amenities (e.g., access to phones, food, or shelter), emotional support (e.g., someone to talk to), and safety (e.g., physical safety). These networks nudge decisions about where people spend their time, where they work, who they go to for help, and ultimately whether they remain in their housing once they obtain it \citep{vsimon2019activity}. Social support is known to help people move out of substance use situations and help them not use in the first place \citep{rapier2019inverse}. There are, however, only few studies on the networks between people experiencing homelessness \cite[e.g.,][]{anderson2021ecology,anderson2023norms,almquist2020large,green2013social}, and qualitative and quantitative studies on the importance of social support networks for the unhoused population broadly \cite[e.g.,][]{cummings2022social,solarz1990social,green2013social,groton2019social}. The social relations of people experiencing homelessness are complex and often evolve over their experience without housing. There is some evidence from qualitative studies that we would expect different personal network sizes within the community of people experiencing homelessness, for example, it has been shown that people who are new to homelessness often focus their ties on their housed community, eschewing ties to those also living unhoused \citep{snow1987identity}. This multi-year, large-scale study will enable us to develop a deeper understanding of the personal networks of individuals experiencing homelessness.
 
Specifically, this article aims to build on the existing qualitative and quantitative literature on the networks of people experiencing homelessness by utilizing a population-level sample of individuals experiencing homelessness in King County, Washington (WA). This sample exists, because King County, WA has adopted a network sampling-based approach to counting the number of people experiencing unsheltered homelessness \citep{almquist2024innovating}, which we can leverage to understand the social networks between people experiencing homelessness over a large metro area (i.e., Seattle, WA).\footnote{King County which contains Seattle, WA is the twelfth largest county by population in the US with 2.3 Million people and the fourth largest population of unhoused people, with more than 16,000 people living outdoors or in temporary shelters on any given night.}

People living unhoused are typically considered a hard-to-reach population \citep{almquist2024innovating, killworth1998estimation}, but they are a growing and socially important community across the US \citep{almquist2025does}. The US Department of Housing and Urban Development's (HUD) Annual Homeless Assessment Report (AHAR) -- presented to the US Congress in December 2024 -- reported 771,480 people experiencing homelessness in the United States on any given night. This number is estimated from two key measurements: (1) the sheltered counts from the Homeless Management Information System (HMIS) databases and allied service provider records contained in the HUD-designated administrative/geographic unit known as continuum-of-care \cite[see for details][]{almquist2020connecting}, and (2) the unsheltered count \citep{richards2023unsheltered}, which comes from the Point-in-Time (PIT) count conducted biennially every January within a given federally defined local jurisdiction\footnote{These jurisdictions are known as Continuum of Care (COC) by HUD.}. Traditionally, this PIT count is conducted on a single night in January by volunteers and is performed as a visual census of people sleeping in the streets \citep{almquist2020connecting}. For example, in King County, WA, 2020, this was conducted from 2:00 AM to 6:00 AM with around 1,000 volunteers, followed in the next few weeks by a demographic survey in day centers and emergency shelters \citep{home2020seattle}. This process has been scrutinized recently \cite[see,][]{tsai2022annual}, and new methods have been proposed in recent years. \cite{almquist2024innovating} introduced a method to count the unsheltered people experiencing homelessness using network-based methods. This method was implemented using network sampling — specifically respondent-driven sampling (RDS) \citep{heckathorn1997respondent,gile2010respondent} — in 2022 and again in 2024, with a large-scale, unofficial county-wide RDS conducted in 2023.

Over the last three years, in King County, WA, this network-based approach has yielded the most extensive set of network data on the US unhoused population that we are aware of. In this paper, we will analyze the resulting network data to understand the social networks of the unhoused population in King County, WA, with an eye on how this can be used to improve social policy and outcomes of the unhoused community. The data collected in this study included several aggregate personal network measures of the interpersonal network people experiencing homelessnes (i.e., the node set is limited to people also experiencing homelessness): (1) an \emph{Acquaintanceship} network (the total number of people experiencing homelessness who ego\footnote{The focal node in a network, in this case, the survey respondent \citet{almquist2012random,wasserman1994social}.} knows by sleep location); (2) a \emph{Close Friendship} network (measure as a list of close alters\footnote{A person connected to ego \citet{almquist2012random,wasserman1994social}.} experiencing homelessness by their sleep location); (3) a kinship network (measured by ego's household relationships based on sleep location), and (4) a peer referral network (which is measured through a coupon recruitment process leveraged for the data collection process). These network measures are often known as Aggregate Relational Data (ARD) \citep{breza2020using} and can be employed to learn about population-level characteristics of the social network of interest \citep{breza2023consistently,breza2020using}, in this case, people experiencing homelessness.

Our goal in this paper is threefold: (i) to provide a comprehensive description of the social network of people experiencing homelessness in King County, WA, from 2022 to 2024 and any resulting temporal differences; and (ii) to model the characteristics of the population which correlate with the size of people's personal networks, and (iii) to simulate a complete network based on (i)'s characteristics. Finally, we will focus on the differences and similarities of the three core personal networks across demographics and other characteristics with an eye to how these networks can inform policy (e.g., improve information flow during a natural hazard).   

The paper is organized as follows. We begin with a concise background on homelessness, social networks, demographics, and health, concluding with our core research questions. This is followed by a brief overview of the data collection process and a presentation of key variables along with sample descriptive statistics. Next, we describe the methods employed in the study. Finally, we present the results and discuss their implications.

\section{Homelessness, social networks, demographics, and health: Literature review and study goals}

\subsection{Personal networks, methods for measurement and issues around people experiencing homelessness}

\zack{There is a rich and long literature of methods for surveying personal networks \citep{mccallister1978procedure,bernard1990comparing,marsden1990network}, and it is known to be one of the largest and most impactful areas of study within the field of social network analysis is that of personal networks \citep{small2017someone,fischer1982dwell,moore1990structural}. Individuals are known to leverage their personal networks for resources, including social capital \citep{bhandari2009social}, emotional intimacy, and mental health support \citep{berkman2000social}. Furthermore, a large body of work demonstrates the relationship between the size of friends and family and access to social support \cite[e.g.,][]{roberts2011communication}. Classic work in the space of personal networks has historically focused on access to jobs \cite[e.g.,][]{granovetter1973strength} and access to information or resources \cite[e.g.,][]{burt2018structural}, or its impact \citet{fischer1982dwell} on cultural influence. For a complete review, see \citet{small2021personal}.

Recent work by \citet{small2017someone} and others has shifted the focus of personal networks from close ties to ``weak'' ties and other social relations to which people have access. \citet{small2021personal} calls explicitly for more quantitative, temporal, and varied personal network data and theory to advance the field. 

People experiencing homelessness's personal networks are broader than simply those with those who are also experiencing homelessness, which this paper limits its analysis to. The social support system of people experiencing homelessness often includes social relations to housed individuals (e.g., friends, family, adopted kin), however int his work we only consider the social relations between those who are experiencing homelessness -- though this experience can range from unsheltered homelessness (e.g., sleeping outside or in a tent or RV), emergency shelters (e.g., church basements) to transitional housing provided by federal funds \citep{snow1987identity}. For some, other homeless people are the dominant social ties, but for many others, their homeless friends and acquaintances are marginal, and their housed friendships are more important \citet[e.g.,][]{corinth2018social}. Other key frameworks include \citet{desmond2012disposable} 's work on ``disposable ties,'' which develop quickly and dissolve just as quickly and can be a key relationship for those in poverty, thus making household and family ties more stable and important \cite[e.g.,][]{stack1997all}.} 

\subsection{Key demographic and health outcomes in the population of people experiencing homelessness}

\zack{There is a large body of work on the experience of the unhoused population. Most recently, \cite{richards2023unsheltered} has provided a rather complete review of the literature. Key characteristics considered include demographics (age, gender, race, ethnicity, veteran status), health (chronically homeless, physical and mental disability, self-rated health, pain, and substance use), sleep location (typically as defined by HUD, where a person is considered sleeping sheltered if they use an emergency shelter and unsheltered if they are living outside in locations not commonly considered for human habitation like a park bench or a car). \cite{richards2023unsheltered} further points out that household composition, i.e., who one travels with and sleeps with on any given night, is often a key social variable. \citep{kuhn2020homelessness} has examined sleep and sleep location, focusing on where individuals sleep (e.g., shelter, unsheltered, doubled up) and their associated sleep quality. Further work by \citet{rice2025sleep} has demonstrated with watches that sleep quality varies greatly by sleep location (e.g., tents, tiny homes, or emergency shelters). }

\subsection{Understanding homelessness through networks}

\zack{While this is an extensive ethnographic literature on the importance of social networks for people experiencing homelessness \cite[e.g.][]{cummings2022social,stablein2011helping,meanwell2012experiencing}, there is a much more limited set of studies that look at large-scale empirical networks of people experiencing homelessness. In the ethnographic literature, we see importance around social support \cite[e.g., safety, resources, companionship][]{cummings2022social}, access to and maintaining housing -- for example, in the housing first literature we see that people who maintain their ties with people experiencing homelessness after being housed often fall back into homelessness \citep{padgett2016housing}; \cite{lachaud2024social} finds that social isolation and mental health go hand-in-hand; and there is ethnographic work showing the indigenous and native populations often require distinct social network based interventions for care centered around local practices and trust \citep{browne2016enhancing}. 

The more limited empirical network studies have focused on issues of network composition (e.g., age, race, or gender), structure, and risk \citep{baker1994gender}. In the case of gender differences, homeless men who are long-term or chronically homeless tend to have more fragmented networks that include higher-risk contacts compared to men with intermittent homelessness \citep{evans2004risk}. For women, the experience of homelessness substantially alters social networks. Before entering a shelter, women typically maintain informal support systems composed of family and friends, yet they often perceive these ties as unable to provide housing assistance. After being rehoused, the differences between their social networks and those of other low-income mothers generally diminish. However, this recovery is strongly influenced by geographic displacement, as being rehoused far from one’s original community can disrupt these essential connections \citep{mcgrath2023social}. Youth do make use of smartphones and the internet; these technologies can help maintain or form positive ties (e.g., to home‐based peers) and for health/information‐seeking \citep{lal2021scoping}.

Longitudinal and temporal studies of network change and composition are limited \citep{richards2023unsheltered}, with most work focusing on those who transition from homelessness to no longer experiencing homelessness \citep{richards2023unsheltered}. Recent work by \citet{kuhn2023encampment} has developed a cell-based panel, \emph{The Periodic Assessment of Trajectories of Housing, Homelessness, and Health Study (PATHS)}, which allows for regular reporting on key experiences, including responses to wildfires \citep{kuhn2023encampment} and forced displacement \citep{kuhn2023encampment}, however this study is not representative of the population, is limited to those who can use cell phones, and does not actively collect network data. A key piece of this work is to address such challenges in population-level inference about changes in composition and network structure among people experiencing homelessness.}

\subsection{Homelessness and social connectivity: Evidence, gaps, and study aims}

\zack{In this work, we build on key measures of personal networks within the population of people experiencing homelessness, such as the mean degree (the average size of one's personal network) -- which is often considered one of the most important graph statistics in network analysis \citep{wasserman1994social}. The mean degree is the core metric used for capturing information about people's ``personal networks.'' It allows for understanding well-being \citep{wellman1990different}, social capital \citep{lin2017building}, and discussion of important matters \cite{burt1984network}. Another reason the mean degree is essential is that, if we assume the mean degree is stable \cite{anderson1999interaction}, it fixes the density of the social network, which can be viewed as a proxy for the likelihood that any given individual will interact. If the mean degree is fixed, then the probability of interaction necessarily declines as the population grows.  

Overall, our multiple measures of personal networks between people experiencing homelessness capture various aspects of the extensive literature on personal and egocentric networks \citep{brea2023personal}. We generally associate the acquaintance network with a network of weak-tie relations \citep{granovetter1973strength} and a general sense of how embedded the respondent is in the larger homelessness community. }

\subsection{Research questions and study aims}

The core research questions in this article are threefold. First, we are interested in providing a large-scale, three-year profile of social networks among people experiencing homelessness in King County, WA \cite[expanding the limited research on temporal data on people experiencing unsheltered homelessness][]{richards2023unsheltered}. To our knowledge, this is the most extensive set of network studies on the unhoused population in King County, WA, and likely the United States. Second, we want to investigate the relationship between individual characteristics and personal network size in this population \citep{richards2023unsheltered}. Lastly, we aim to utilize the available network statistics to simulate a comprehensive baseline network for this population. Here, we specify each research question (RQ) and subquestions we will analyze below:

\begin{itemize}
    \item[RQ \#1] Describe the networks between people experiencing homelessness in King County, WA, from 2022 to 2024.
    \begin{itemize}
        \item[RQ \#1.1] Estimate the basic sample characteristics (sample size, max number of waves), and mean number of recruits by those who recruit other people experiencing homelessness. 
        \item[RQ \#1.2] Estimate the sample visibility and degree distribution, and the population-level degree distribution.
        \item[RQ \#1.3] Estimate the sample and population mean degree. 
        \item[RQ \#1.4] Estimate gender and race mixing. 
        \item[RQ \#1.5] Explore the Kinship network (household family composition). 
         \item[RQ \#1.6] Explore the diffusion of information (coupon passing) in 2023 and 2024 RDS samples. 
    \end{itemize}
    \item[RQ \#2] Model the degree distribution across the four core network measures for 2024 RDS data. 
    \item[RQ \#3] Simulate a complete network based on the 2024 data, leveraging the mean degree (density), degree distribution, and mixing-matrix estimated in RQ \#1.
\end{itemize}

\zack{These research questions build on prior work in large-scale social network research and homelessness to address and integrate several open questions into the literature. RQ~\#1 helps develop key tools and metrics for anyone seeking to reproduce the work in \citet{almquist2024innovating} or to conduct a large-scale network study, such as that in \citet{wesson2025novel}. Each subquestion tackles important issues in the field, including understanding the basic sample characteristics of a large-scale sample of social networks among people experiencing homelessness \citep{almquist2020large}. \citet{almquist2020large} also demonstrates how this information can be used by care providers for resource management and sharing. Other key areas of interest in the social network and homelessness literature include gender, which is known to be an important attribute influencing both service access and risk in experiencing homelessness \citep{richards2023unsheltered}, and kinship, where much work has explored the formation of ``street kin'' and family ties as critical to how individuals manage and survive homelessness \citep{smith2008searching}. RQ~\#2 is motivated by the personal network literature \cite[e.g.,][]{small2013weak}, egocentric network literature \cite[e.g.,][]{wasserman1994social}, and research on people experiencing homelessness \cite[e.g.,][]{anderson2021ecology,anderson2023norms,richards2023unsheltered}, all of which require extensive knowledge on the degree distribution and its relation to individual (or node) characteristics. Finally, RQ~\#3 extends work on power analysis and simulation of large-scale social networks among people experiencing homelessness to better understand challenges in deploying large-scale RDS surveys in new contexts and to improve resource estimates for information diffusion and disaster response \citep{almquist2020large,almquist2024innovating}.}





\section{Study design and data collection process}
\label{sec:survey_background}

This data set was collected in three rounds of a network-based peer referral method known as respondent-driven sampling. The first survey was fielded from March 9 to April 6, 2022 (24 days); the second was fielded from April 24 to June 1, 2023 (38 days); and the third was fielded from January 22 to February 2 (11 days) \citep{almquist2024innovating}. In this section, we go over the background of the survey (how it came about), the study design, the hub location, the study participants, the network questions employed, and an overview of the demographic and community needs assessment questions asked over the three years.

The US Department of Housing and Urban Development, starting in the 1980s, through the McKinney-Vento Act (1987), then starting in 1994, HUD began to establish the Continuum of Care (CoC) program and a formal application for federal funds based on CoC jurisdictions (formal catchments for federal funding and responsibility for establishing care for people experiencing homelessness). A detailed discussion of this can be found in \citet{almquist2025does}, and a formal definition is available in \citet{HUD_Homelessness_Manual}. One of the requirements for funding that emerged in the early 2000s was a complete count of people experiencing homelessness in each local jurisdiction that wants federal funds. This count has become known as the Point-in-Time count or ``PIT'', and is measured on a single day in the last week of January. 

HUD defines homelessness as individuals and families who lack a fixed, regular, and adequate nighttime residence. This definition of homelessness includes those staying in emergency shelters, transitional housing, and those sleeping in ``places not meant for human habitation'' (e.g., streets, cars, parks, etc). To obtain a count of people experiencing the HUD definition of homelessness, it has developed mainly into two parts: (1) a count of all people sleeping in an emergency shelter or transitional housing (this is known as the ``sheltered count''); and (2) a visual census of people sleeping outside (known as the ``unsheltered count''). This process has several critiques, primarily focused on the count of people living outside \cite[see, for example][]{tsai2022annual}. 

In 2022, King County CoC in Washington state began employing a respondent-driven sampling (RDS) approach to count the unsheltered population of people experiencing homelessness in the county. Full details of this process can be found in \citet{almquist2024innovating} and \citet{authorityking}. The data discussed in this paper are derived from the RDS surveys conducted in 2022, 2023, and 2024.

\subsection{Study design}

The method employed for sampling the unhoused population in King County, WA is based on the \textbf{respondent-driven sampling} method, which was designed for studying hard-to-reach or hidden populations, such as people who inject drugs, sex workers, or even people experiencing homelessness \citep{heckathorn1997respondent}. The King County Regional Homelessness Authority, in collaboration with the University of Washington, conducted an annual RDS for King County, WA, from 2022 to 2023. 

RDS sampling frame is focused on \emph{seed selection}, \emph{hub location}, \emph{peer recruitment}, \emph{incentives}, and \emph{waves}. Below, we sketch out the choices made by the King County Regional Homelessness Authority from 2022 to 2024. For full details, please read \citet{almquist2024innovating}.

\begin{figure}[htp!]
    \caption{Illustration of the longest chain in 2024 with the degree distribution and geospatial location of the interview (labeled as RDS hub location).}
    \label{fig:wave}
    \centering
    \includegraphics[width=1\linewidth]{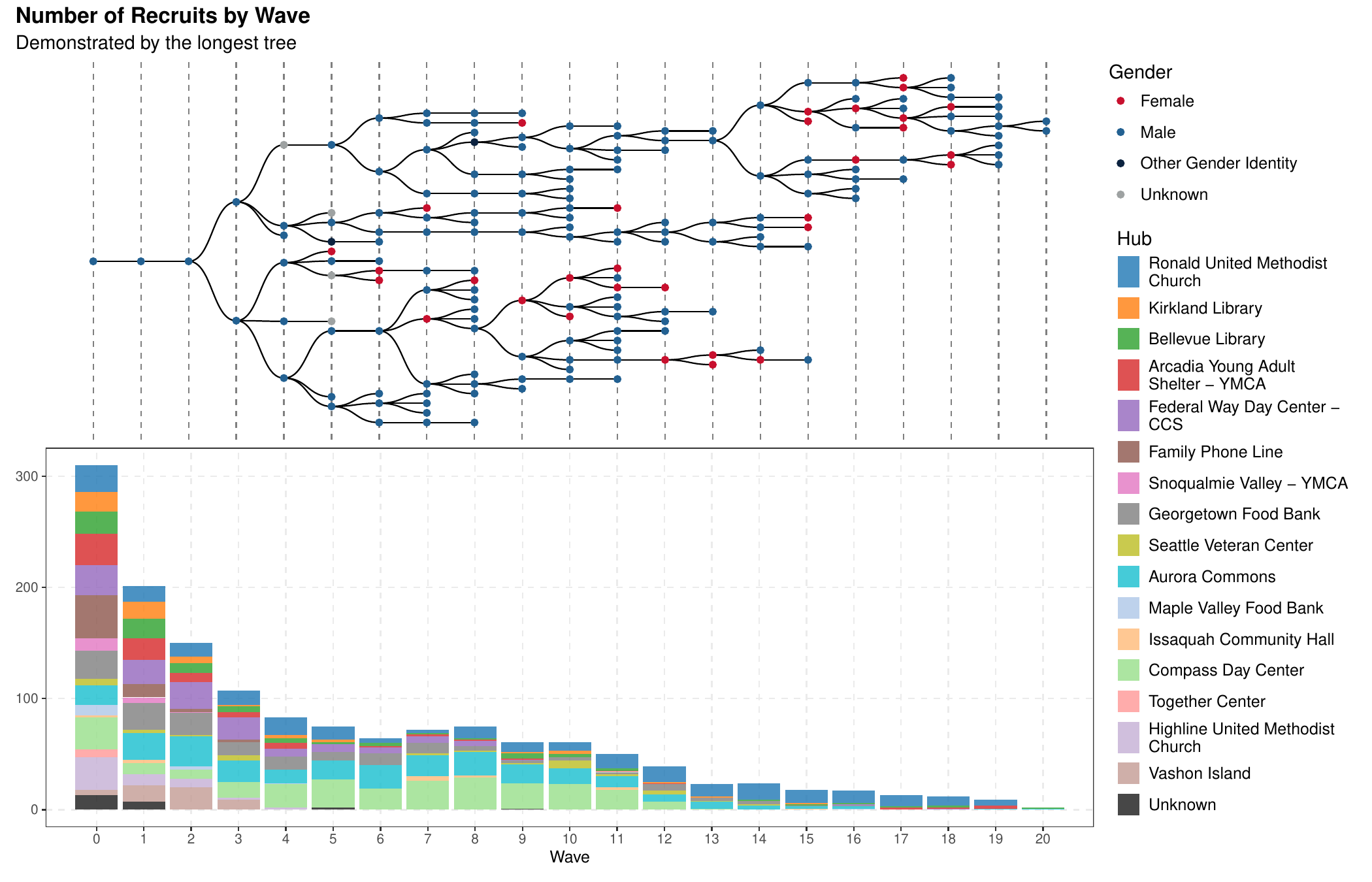}
\end{figure}

\begin{enumerate}[i)]
    \item \textbf{Seed Selection}: Seeds were selected by local outreach organizations in King County. The method largely employed the outreach workers handing out ``seed coupons'' to people sleeping outside. 

    \item \textbf{Hub location}: Hub locations were chosen based on historic PIT counts, outreach worker recommendations, accessibility (e.g., distance from bus or train lines), and community input. See Section~\label{sec:hub} for a discussion on hub locations.

    \item \textbf{Peer Recruitment}: Each seed receives a set number of \emph{coupons} (3) to recruit peers from their social network of people experiencing homelessness.

    \item \textbf{Incentives}: In 2022, we only provided incentives for survey completion (\$25). In 2023 and 2024, we provided dual incentives: one for survey completion (\$20) and another for each eligible person recruited by the participant (\$5).

    \item \textbf{Waves}: Each recruited participant received coupons to recruit others, resulting in a series of recruitment \emph{``waves''}. The final sample size was based on simulation analysis \cite[see]{almquist2024innovating}, and an example of the longest chain can be found in Figure~\ref{fig:wave}. 
\end{enumerate}

To illustrate the RDS procedure, we have mapped out the longest wave labeled by gender of the respondent in Figure~\ref{fig:wave} along with the histogram of recruitment chain size for each hub employed in 2024. 

 Following the work of \cite{heckathorn1997respondent} and follow-up work by \citet{salganik20045} and further work by \cite{gile2015diagnostics,gile2011improved,gile2010respondent} and others, we performed convergence diagnostics and followed the RDS estimators in \cite{handcock2024package,gile2010respondent}. Another critical aspect of RDS, from an IRB perspective, is that data collection occurs at a fixed location, known as a ``hub.'' Hubs were selected based on historically significant locations, community input, and personal experience. The initial chosen respondents (known in the RDS literature as ``seeds'') were recruited by service care workers who worked in the locations surrounding the selected hubs. We explain further how hub locations were chosen in the next section.

\subsection{Hub locations}
\label{sec:hub}

\begin{figure}[htp!]
    \caption{King County, WA plotted with the RDS HUBs for all three years. Seattle is in Red, urban areas in grey. King County, WA, which has around 2.3 million people, contains Seattle, WA, the largest city at around 750,000 people, and is the region's economic hub.}\label{fig:hubloc2024}
    \centering
    \includegraphics[width=.7\linewidth]{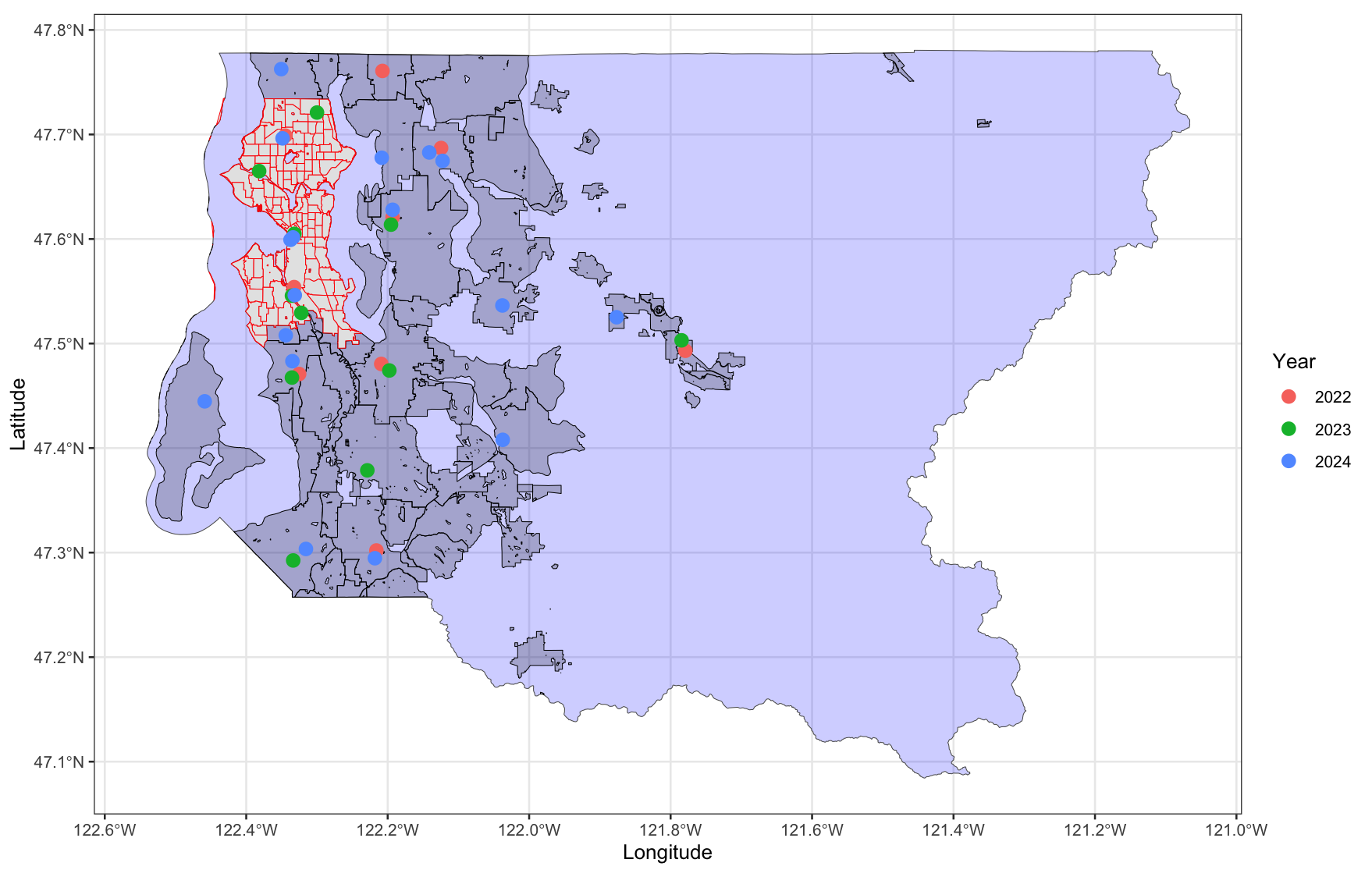}
\end{figure}

\begin{figure}[htp!]
    \caption{UW/King County Regional Homelessness Authority RDS PIT Count 2024 median distance traveled by hub location.}
    \label{fig:hubloc2024dist}
    \centering
    \includegraphics[width=1\linewidth]{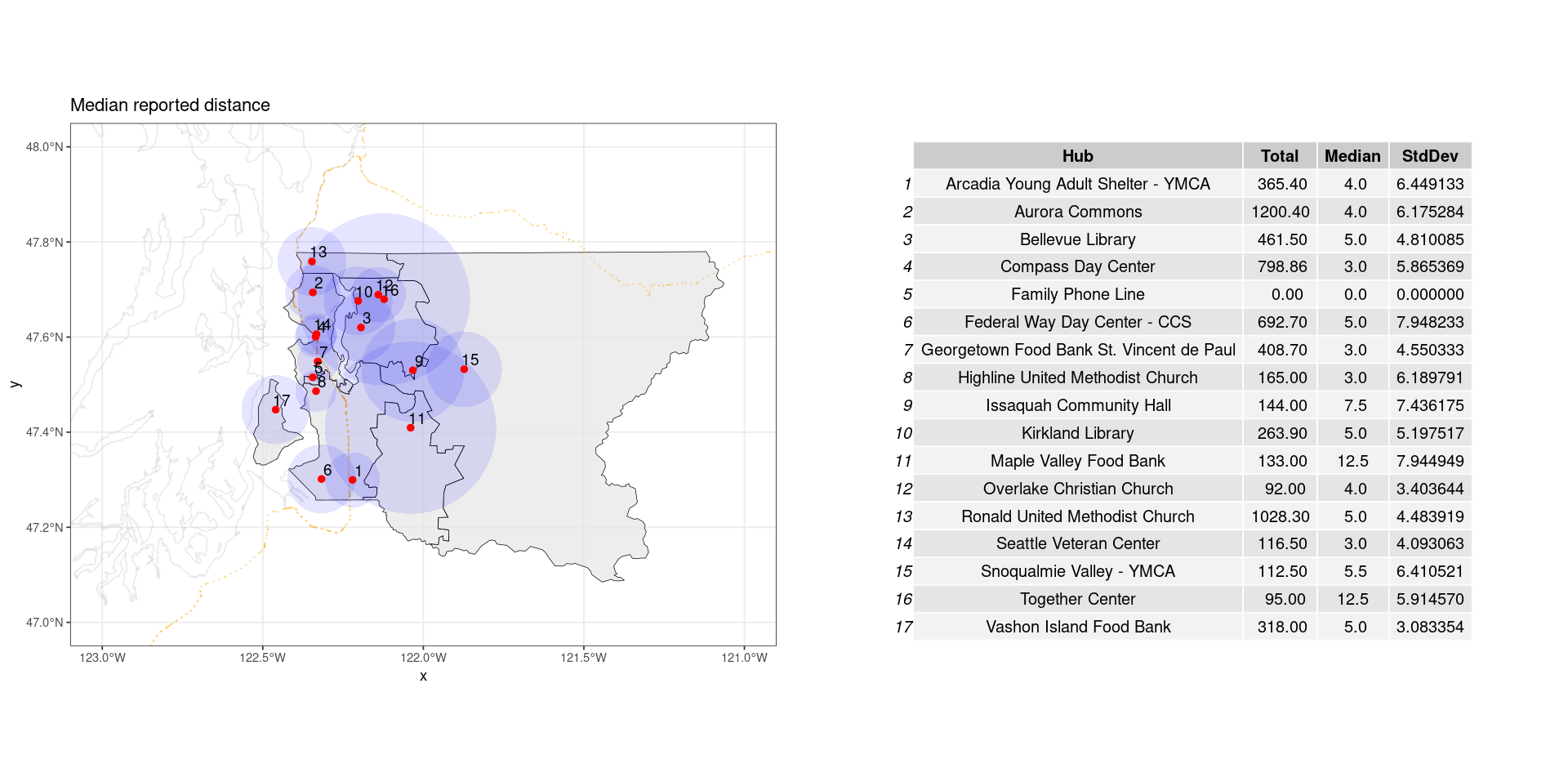}
\end{figure}

An essential aspect of any RDS design is the spatial location of hub or hubs in data collection \citep{toledo2011putting,jenness2014spatial}. The data used in this paper were selected based on historical PIT data, outreach workers, individuals with lived experience, community input, and travel logistics. The hub locations for all three years can be seen in Figure~\ref{fig:hubloc2024}. In 2024, we will add questions about distance, time, and mode of travel. The general takeaway from 2024 is that individuals tended to travel about 20 minutes plus or minus 10 minutes primarily by foot, bike, or bus. The median distance traveled by hub location in 2024 can be visualized in Figure~\ref{fig:hubloc2024dist}.

\subsection{Study participants}

Respondents to the survey were individuals experiencing homelessness recruited through peer referral methods (see Study Design) during two years of HUD mandate point-in-time counts of unsheltered people experiencing homelessness (2022 and 2024) and one unofficial point-in-time count conducted by the University of Washington in 2023. For details, see \citet{almquist2024innovating}. In total, 671 individuals experiencing homelessness were interviewed in 2022, yielding an analytic sample of 616 individuals. In 2023, 1,106 individuals experiencing homelessness were interviewed, and in 2024, 1,466 individuals experiencing homelessness were interviewed. All individuals in the sample were experiencing homelessness following the HUD definition of homelessness as discussed in Section~\ref{sec:survey_background}. 

\subsection{Network instruments}

First, we review the network questions used in these surveys, then proceed to the core set of network statistics generated from our sample, starting with the mean degree and the degree distribution. For 2023 and 2024 data, we collected four network measures (we have included screenshots of the network questions in \ref{sec:network_questions}): 

\paragraph{Acquaintance network} - this is measured through the reported aggregate counts of the number of people experiencing homelessness, and was asked by the following question: 

\begin{itemize}
    \item[2022:] How many people do you personally know who are unhoused or experiencing homelessness with whom you interact at least occasionally? \textbf{Response}: Numeric [top coded to 100].
    \item[2023:] Outside your family living with you, how many people do you personally know who are unhoused or experiencing homelessness?  \textbf{Response}: Numeric [top coded to 100].
    \item[2024:]  Outside your family living with you, how many people do you personally know who are unhoused or experiencing homelessness? \textbf{Response}: Numeric [top coded to 100].
\end{itemize}

\paragraph{Close friendship network} -- is measured as a network list generator of people the respondent (also known as ego in the network literature) knows who are experiencing homelessness. This question was not asked in 2022, and the survey response list was expanded to allow for up to 20 in 2024 (we decided to extend to 20, based on the work of \citet{hill2003social} and others showing that many human groups tend not to scale much beyond 20 in many empirical settings). The question asked each year, and the response type is provided below:

\begin{itemize}
    \item[2022:] --
    \item[2023:] Outside your family living with you, please list the first name, pseudonym, or initials of the person and their relation (e.g., Friend, Family, etc.) of people you personally know who are unhoused or experiencing homelessness? (Please answer for as many people as you know.) \textbf{Response}: List up to 15 names (respondent was not told of this limit).
    \item[2024:] Outside your family living with you, please list the first name, pseudonym, or initials of the person and their relation (e.g., Friend, Family, etc.) of people you personally know who are unhoused or experiencing homelessness? (Please answer for as many people as you know.) \textbf{Response}: List up to 20 names (respondent was not told of this limit).
\end{itemize}

\paragraph{Kinship network} - or family networks was measured by asking people to list all members of their current household. This measure of kinship is limited to the immediate household only, as required by HUD, but does include adopted kin and other non-blood-related kin that live in the same household. In 2022, we gathered only aggregate data and did not collect list-based data on the number of people experiencing homelessness or their families. All three years' questions and responses are provided below.

\begin{itemize}
    \item[2022:] How many people are in your household are living with you now? \textbf{Response}: Numeric.
    \item[2023:] Please list the initials of all the people in your household (if prompted, say this includes family members, partners, children, etc.). \textbf{Response}: List up to 10 names (respondent was not told of this limit).
    \item[2024:] Please list the initials of all the people in your household. (If prompted, say this includes family members, partners, children, etc.). \textbf{Response}: List up to 10 names (respondent was not told of this limit). 
\end{itemize}

\paragraph{Peer referral network via coupons} - this is measured through the linked data provided by the coupons and our internal software. For details on the software used for tracking the peer referral network, see \citet{almquist2024innovating}.




\subsection{Demographics and needs assessment questions}
\label{study:dem}

All demographic questions and basic measurements, as well as the time spent experiencing homelessness, are guided by HUD requirements \citep{HUD_Homelessness_Manual}. Below are the definitions of gender, race, ethnicity, and age, as asked of all respondents and family members across all years. We also add housing status here, again defined by HUD requirements.

\begin{itemize}
   \item[]\emph{Gender} was asked a six-option question in 2022 and 2023 (male, female, questions, A gender other than singularly female or male (e.g., nonbinary, genderfluid, agender, culturally specific gender), transgender and do not know) and in 2024, a seventh option was added by the request of the Chief Seattle Club\footnote{A a Native-led housing and human services agency.} (Two-Spirit). The six-option question is the standard for HUD.
    \item[]\emph{Race}  in 2022 and 2023 followed the standard US Census definitions of race, where one can select all that apply and include (American Indian, Alaskan Native or Indigenous; Asian or Asian American; Black, African American or African; Native Hawaiian or Pacific Islander; White or prefer not state). In 2024, the US Census and HUD added categories for Latina/o/x or Hispanic and Middle Eastern or North African categories. 
    \item[]\emph{Ethnicity} in 2022, 2023, and 2024, the survey followed the HUD and US Census definitions of Latino/a/x or Hispanic origin, with yes/no responses.
    \item[]\emph{Age}  was asked as the year and month of birth. For privacy reasons, the survey did not ask the day. 
    \item[] \emph{Mental disability} was asked in all three years with the HUD-mandated question, ``Do you identify as having a severe mental illness?'' Response options include `yes', 'no', `choose not to answer', and `do not know'.
    \item[] \emph{Physical disability} was asked in all three years with the HUD-mandated question, ``Do you identify as having a disability?'' Response options include `yes', `no', `choose not to answer', and `do not know'.
    \item[] \emph{Substance use disorder} was asked in all three years with the HUD-mandated question, ``Do you identify as having a substance use disorder?'' Response options include 'yes', 'no', 'choose not to answer', and 'do not know'.
    \item[]\emph{Housing status} was asked in all three years with the question, ``Where did you sleep last night?'' with the following response options: Outside in a tent (or tent-like structure); Outside, not in a tent; In a car, truck, or van (smaller vehicle); In an RV, trailer, or boat (larger vehicle); In a park (uncovered, like on a bench); In an overnight shelter (e.g. mission, church, resource shelter, etc.); In a hotel or motel; In an abandoned building/backyard or storage structure; In a public facility or transit (bus/train station, transit center, airport, hospital waiting room); In a tiny home; On public transit (e.g. slept on bus, train, etc.); In jail or prison; In a hospital (stayed as patient overnight); In a drug or alcohol treatment/detox center; In a friend or family member's house/apartment; In another place not listed (write below); Choose not to answer; Do not know.
\item[] \emph{Chronic homelessness} as defined by HUD, consists of three parts: (1) disabling condition (physical, mental of substance use), (2) length of homelessness (continuous for a year or more, or four episodes in the last 12 months), and (3) the individual must be sleeping in a place not meant for human habitation (e.g., emergency shelter or outside). This is a composite measure built from the survey based on HUD guidelines.
\end{itemize}

Full details on the demographic and needs assessment questions fielded in 2022 and 2023 can be found \citet{DSSG2024_Understanding_Homelessness}, and for 2024, they can be found in \citet{authorityking}.

\section{Data and Methods}

In this section, we review the methods and data used to analyze the descriptive statistics derived from our RDS sample and our statistical models for personal network size (degree) and complete network analysis (exponential-family random graph models). 

\subsection{RDS Estimators for descriptive analysis}


To perform our descriptive analysis, we will leverage the methods explicitly developed for RDS \citep{gile2010respondent} and use the R software package \texttt{RDS} \citep{RDS_R}. \citet{fellows2022robustness} and others advocate for the RDS-II estimator, also known as the Voltz-Hackathorn estimator \citep{volz2008probability}, and it is a good default method for estimating the central tendencies and standard errors for population estimates. 

Let $S = \{1, 2, \ldots, n\}$  denote the set of sampled individuals in the RDS study, where $n$ is the total sample size. For each respondent $i \in S$, we observe the outcome of interest $y_i$, which may be a binary indicator (e.g., uses emergency shelter or not) or a continuous measurement. Additionally, each respondent reports their personal network degree $d_i$, defined as the number of eligible contacts they have within the target population. The RDS-II estimator, also known as the Volz-Heckathorn (VH) estimator, uses these quantities to estimate the population mean of $y$ by weighting each observation inversely proportional to its degree to account for differential sampling probabilities. Specifically, the estimator accounts for the probability of inclusion proportional to network size, under the assumption of random recruitment within the network. Then the RDS-II estimator (VH) is as follows,

\begin{align}
\hat{\mu}_{\text{VH}} &= \frac{\sum_{i \in S} \frac{y_i}{d_i}}{\sum_{i \in S} \frac{1}{d_i}} \label{eq:rds2-estimator}
\end{align}


The standard error can be derived from the delta-method and is as follows,
\begin{align}
\text{SE}(\hat{\mu}_{\text{VH}}) = \sqrt{ \frac{1}{n} \cdot \frac{\sum_{i \in S} \left( \frac{y_i}{d_i} - \hat{\mu}_{\text{VH}} \cdot \frac{1}{d_i} \right)^2}{\left( \sum_{i \in S} \frac{1}{d_i} \right)^2} } \label{eq:vh_se}
\end{align}

We use a bootstrap estimator of the RDS-II estimator \cite[for details, see][]{gile2011improved,gile2010respondent} for the standard errors reported in this paper for the descriptive statistics; this estimator can be expressed as follows:

\begin{align}
\text{SE}_{\text{bootstrap}}(\hat{\mu}_{\text{VH}}) &= \sqrt{\frac{1}{B - 1} \sum_{b=1}^B \left( \hat{\mu}_{\text{VH}}^{(b)} - \bar{\mu}_{\text{VH}} \right)^2} \label{eq:bootstrap_se}
\end{align}

Last, a note about the degree weights we use for our descriptive analysis, we follow \citet{mclaughlin2015inference, mclaughlin2024modeling} and use the MLE estimate of the visibility statistics to impute $d_i^{v}$, which we then use as our weights for all three years. These weights are readily computed in the \texttt{RDS} package in the \texttt{R} programming language \citet{RDS_R}.

\subsection{Visualizing RDS Data using Graph Isomorphisms}
\label{sec:iso}
\zack{A common challenge in data visualization is that graphs (networks) with many nodes are difficult to visualize without clutter \citep{herman_graph_2000}. While RDS data consists of rooted trees that individually may not contain many nodes, the total number of trees is often less compelling visually and is often easier to categorize with some form of aggregation \citep{holten_hierarchical_2006}. As an alternative approach to the pure census of RDS trees, we, also, visualize only the unique isomorphic tree structures---that is, the distinct structural patterns that appear in the data---and annotate each with its frequency of occurrence. This representation reduces visual clutter while preserving essential structural information about the recruitment patterns. We do this for both the peer-referral and kinship (household) networks.

To formalize this approach, we first define what it means for two graphs to be structurally identical. Two graphs are \textbf{isomorphic} if they have the same structure---that is, they have identical adjacency relations, even if their vertices are labeled differently or positioned differently in a visual representation. In other words, two graphs are isomorphic if one can be transformed into the other simply by relabeling its vertices \citep{wasserman1994social}.

Formally, let $G_1 = (V_1, E_1)$ and $G_2 = (V_2, E_2)$ be two graphs, where $V_i$ denotes the vertex set and $E_i$ denotes the edge set. The graphs $G_1$ and $G_2$ are \textbf{isomorphic} (written $G_1 \cong G_2$) if there exists a one-to-one mapping ($\phi$)from the nodes of $G_1$ to the nodes $G_2$.
\begin{equation}
\phi: V_1 \rightarrow V_2
\end{equation}
such that for any two vertices $u, v \in V_1$:
\begin{equation}
(u, v) \in E_1 \text{ if and only if } (\phi(u), \phi(v)) \in E_2
\end{equation}

This definition has several important components. (1) Every vertex in $G_1$ maps to exactly one vertex in $G_2$, and every vertex in $G_2$ is mapped to by exactly one vertex in $G_1$. (2) No vertices are left out, and no vertex is mapped to more than one. (3) Two vertices are connected by an edge in $G_1$ if and only if their corresponding vertices (under the mapping $\phi$) are connected by an edge in $G_2$. The mapping preserves both the presence and absence of edges. The mapping $\phi$ itself is called an \textbf{isomorphism} between $G_1$ and $G_2$.

To illustrate this concept in the context of Respondent-Driven Sampling (RDS), consider two recruitment trees generated from different seeds. Let $T_1$ be a recruitment tree where seed node $s_1$ recruits three participants $p_1$, $q_1$, and $r_1$. Let $T_2$ be another recruitment tree where seed node $s_2$ recruits three participants $a_2$, $b_2$, and $c_2$. Despite having different participant identifiers, these two RDS trees are isomorphic because we can define a mapping $\phi(s_1) = s_2$, $\phi(p_1) = a_2$, $\phi(q_1) = b_2$, $\phi(r_1) = c_2$ that preserves all recruitment relationships. Both trees have the same structural pattern: each seed recruits exactly three participants, resulting in identical recruitment networks. We compute the isomorphisms in the statistical programming language \textit{R} \citep{rprog}. }

\subsection{Statistical modeling approach for personal networks} 

To understand the effect of node-level covariates \citep{butts2008social} (e.g., basic demographics, length of time experienced homelessness, etc.) on network structure (primarily degree distribution and mean degree), we employ count data models with a generalized linear model (GLM) structure \citep{nelder1972generalized}. We focus on a core class of count data models with a long history of use in the social sciences and social network literature, which is that of the classic Poisson and Negative Binomial models with and without a zero-inflation component \cite[e.g.,][]{zheng2006many}. We consider model classes with a zero-inflation component in the kinship network context (i.e., families of size one are expected to differ systematically from families of two or more people). As \cite{butts2003network} notes, recall and bias issues can systematically reduce the number of personal contacts reported. 

\subsubsection{Models for the personal network size (degree distribution)}

\paragraph{Zero-Inflated Poisson (ZIP) Model} The ZIP model combines a Poisson model with a logistic model for zero inflation (note that this inherently contains the pure Poisson model by zeroing out the logistic model). Mathematically, we can write it as follows,

\begin{equation}
Y_i = \begin{cases} 
      0 & \text{with probability } \pi_i, \\
      \text{Poisson}(\lambda_i) & \text{with probability } 1 - \pi_i,
   \end{cases}
\end{equation}
where:
\begin{itemize}
\item  \(\pi_i\) is the probability of an excess zero, modeled by a logistic function: \(\pi_i = \frac{\exp(\mathbf{z}_i^\top \boldsymbol{\gamma})}{1 + \exp(\mathbf{z}_i^\top \boldsymbol{\gamma})}\),
\item \(\lambda_i\) is the Poisson mean, modeled by a log-link: \(\log(\lambda_i) = \mathbf{x}_i^\top \boldsymbol{\beta}\),
\item \(\mathbf{x}_i\) and \(\mathbf{z}_i\) are vectors of covariates for the count and zero-inflation parts, respectively.

\end{itemize}

\paragraph{Zero-Inflated Negative Binomial (ZINB) Model} The ZINB model generalizes the ZIP by allowing overdispersion through a Negative Binomial distribution for the counts:

\begin{equation}
Y_i = \begin{cases} 
      0 & \text{with probability } \pi_i, \\
      \text{Negative Binomial}(\mu_i, \alpha) & \text{with probability } 1 - \pi_i,
   \end{cases}
\end{equation}
where:
\begin{itemize}
\item \(\pi_i = \frac{\exp(\mathbf{z}_i^\top \boldsymbol{\gamma})}{1 + \exp(\mathbf{z}_i^\top \boldsymbol{\gamma})}\) is the zero-inflation probability,
\item \(\mu_i\) is the Negative Binomial mean, modeled by \(\log(\mu_i) = \mathbf{x}_i^\top \boldsymbol{\beta}\),
\item  \(\alpha\) is the dispersion parameter for the Negative Binomial.

\end{itemize}

\subsubsection{Model selection} 

We employ the classic model selection criterion \cite[e.g.,]{kuha2004aic} to select the best-fitting model, where lower scores indicate a better fit. For our analysis, we focus on the AICc metric \citep{anderson1994aic} because it corrects for finite-population-size issues and is known to handle overdispersion better when used as a decision criterion.

To find the best-fitting model, we created a set of candidate models based on their theoretical relevance (see the demographics and characteristics of homelessness we consider below). Then, we assessed model diagnostics and AICc to determine the most suitable model family (class) on a given degree distribution \cite{AICcmodavg-package}. Then, we employ a (backward) stepwise selection algorithm \citep{olusegun2015identifying} using AICc as the decision criterion to find the best fitting model, including the final set of node-level characteristics. We use the \cite{jackman2015package} in the R statistical programming environment \citep{hornik2012comprehensive}. We consider the following covariates, which are available across surveys and generally available on all respondents in all three-degree distribution models: \emph{age, gender, race, ethnicity, shelter use, veteran status, chronically homeless}\footnote{Using HUD definition of chronically homeless \citep{HUD_Homelessness_Manual}.}, \emph{self identified mental health, self identified substance use,} and \emph{self identified disability}.

\subsection{ERGMs for complete network analysis and simulation}
\label{sec:ergm}

Exponential-family Random Graph Models (ERGMs) are a class of statistical models used to represent and analyze social relations within networks. They provide a flexible framework for modeling the complex dependencies in network data \citep{robins2007introduction}, and have become increasingly common through statistical packages such as \texttt{STATNET} \citep{handcock2008statnet}. ERGMs have a long history in social network analysis, from the p* models \citep{pattison1999logit} to more recent developments \citep{hunter2012computational}.  Unlike traditional models that assume independent observations, ERGMs accommodate the interdependence of network ties by specifying the probability distribution over the set of all possible graphs \citep{robins2007introduction}. ERG models can be written as the probability distribution over a set of graphs $G$ as follows:

\begin{equation}
    P(G = g \mid \theta) = \frac{\exp\left(\theta^{T}s(g)\right)}{Z(\theta)},
\end{equation}
where $g$ is the observed graph, $\theta$ is a vector of parameters, $s(g)$ is a vector of sufficient statistics that capture graph features such as edges, triangles, and degree distributions, and $Z(\theta)$ is a normalizing constant \citep{hunter2012computational}. Given this formulation, the key becomes estimating sufficient statistics from the data. The main challenge in estimating ERGMs arises from the normalizing constant $Z(\theta)$, which is computationally expensive to calculate because it involves summing over all possible graphs. Markov Chain Monte Carlo (MCMC) methods are typically employed to approximate the likelihood function and estimate model parameters \citep{hunter2012computational}. Alternative methods, such as Maximum Pseudo-Likelihood Estimation (MPLE), are also used to speed up computation. However, they may be less accurate than full MCMC-based inference \citep{hunter2012computational} and have generally fallen out of favor as MCMC methods have become more readily available.

\subsubsection{Estimation of network models from sampled data}

In cases where we do not have complete network data, one advantage of ERGMs is that they are exponential-family models and can therefore be estimated directly from population statistics derived from the sampling design. This procedure can be readily employed in \citep{krivitsky2017inference} by specifying the target statistics. Here, we will compute the target statistics from the network descriptive statistics in this paper. See \citet{jenness2017epimodel} for a clear description of how to compute the target statistics to fit ERGMs from population-level statistics.

\subsubsection{Simulation of complete networks from ERGMS}

Using the target statistics estimated from the descriptive statistics (e.g., gender mixing, density, and degree distribution), we simulate complete networks using the MCMC algorithms developed in \texttt{R} package \texttt{STATNET} \citep{handcock2008statnet} -- results can be seen in Figure~\ref{fig:power_ergm}.

\subsection{Data: Key variables and sample descriptive statistics} \label{sec:variables}

\zack{In this subsection, we first review the data and basic RDS sample descriptive statistics. Then we follow up with a review of all the variables and networks explored in the results and analysis section of this article.}

\subsection{Sample characteristics}

\begin{figure}[htp!]
    \caption{RDS Tree plots for 2022, 2023, and 2024 RDS surveys conducted on the population of people experiencing homelessness in King County, WA.}
    \label{fig:rdsplots}
    \centering
    \includegraphics[width=1\linewidth]{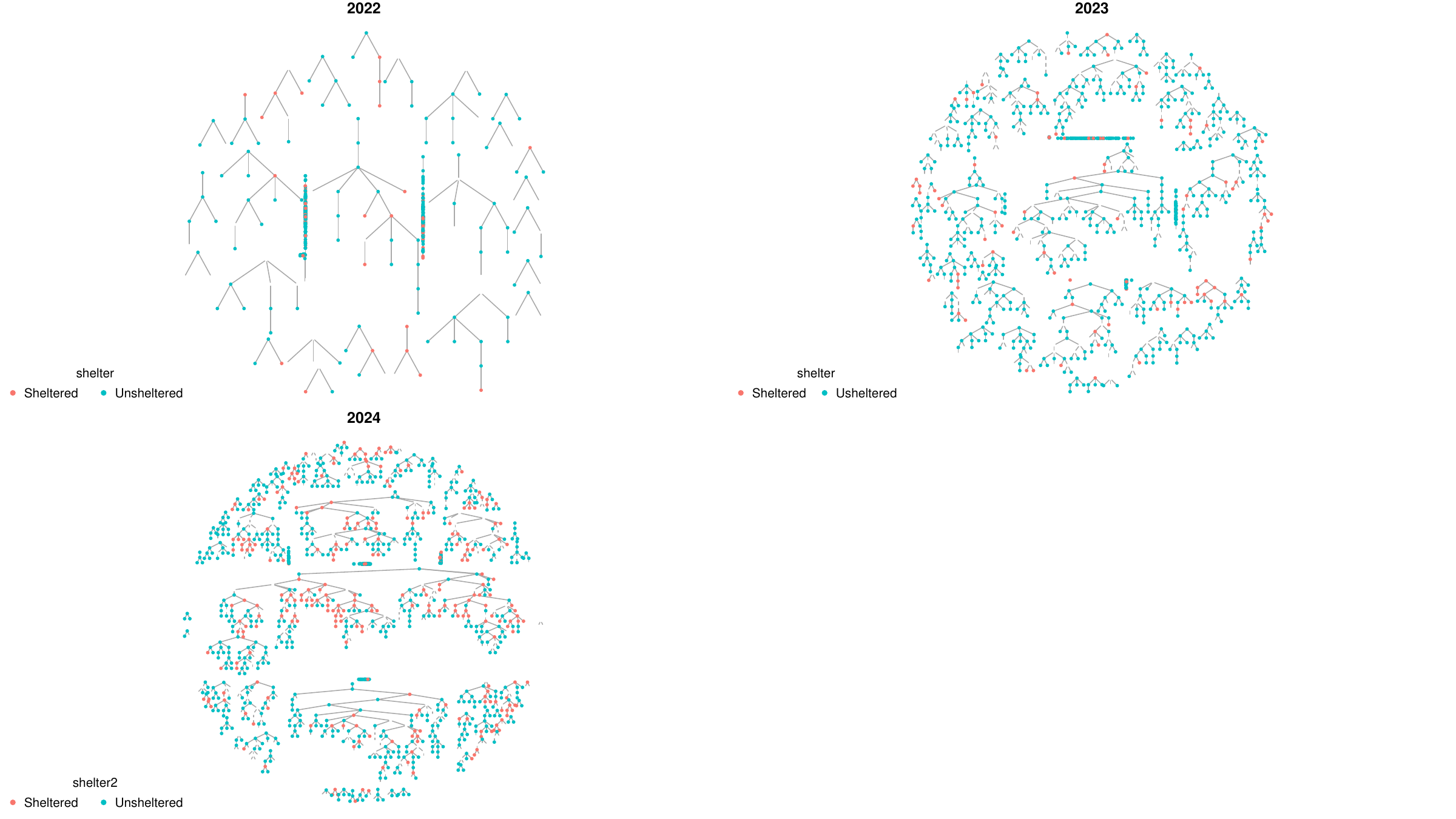}
\end{figure}

\begin{figure}[H]
    \caption{Recruitment trees for the 2023 and 2024 RDS surveys counted by their graph isomorphism.}
    \label{fig:rdstree}
   \centering
    \begin{subfigure}[t]{1\textwidth}
        \centering
        \includegraphics[width=1\linewidth]{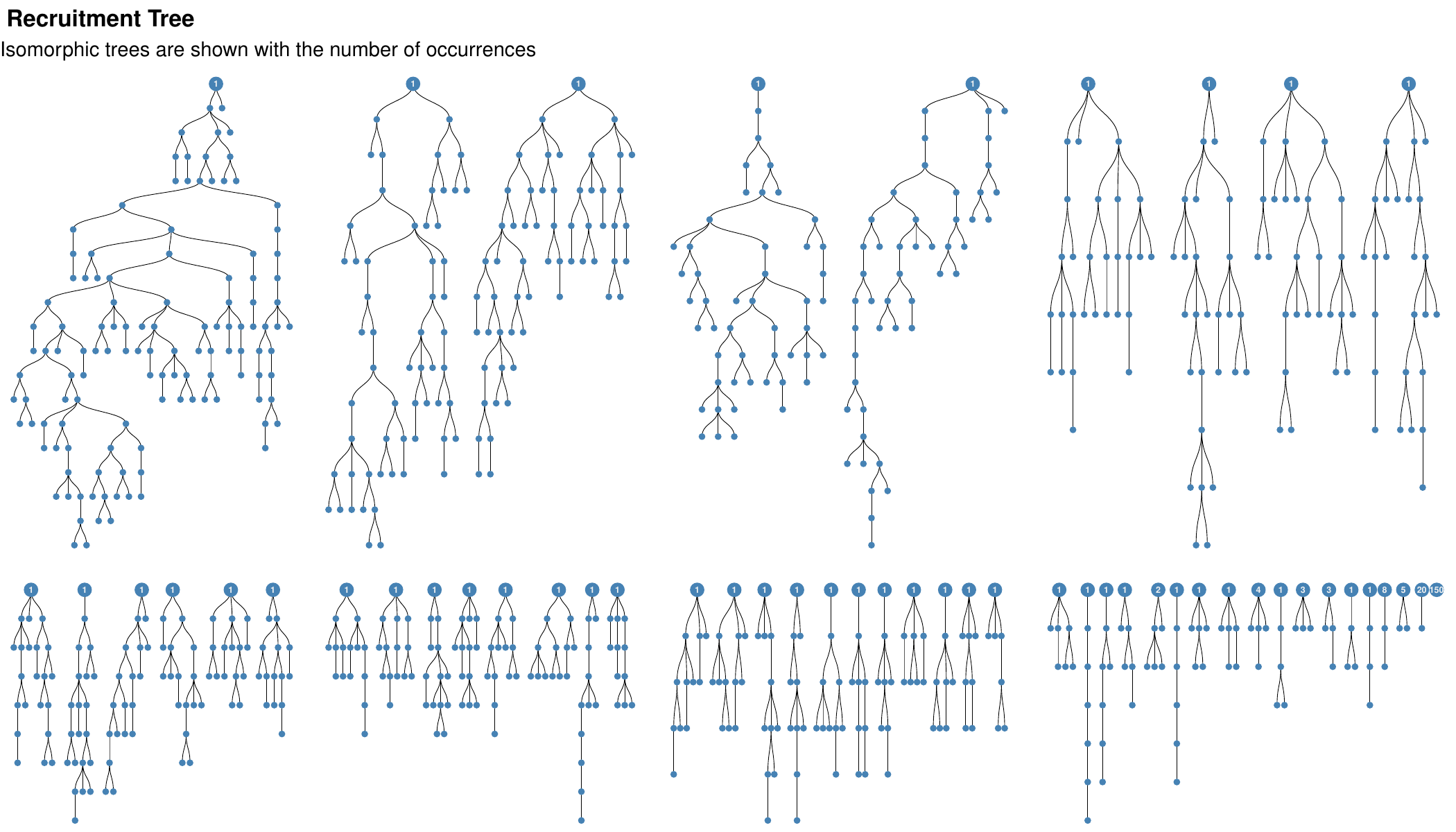}
        \caption{2023 Isomorphism diffusion trees.}
    \end{subfigure}\\
        \begin{subfigure}[t]{1\textwidth}
        \centering
        \includegraphics[width=1\linewidth]{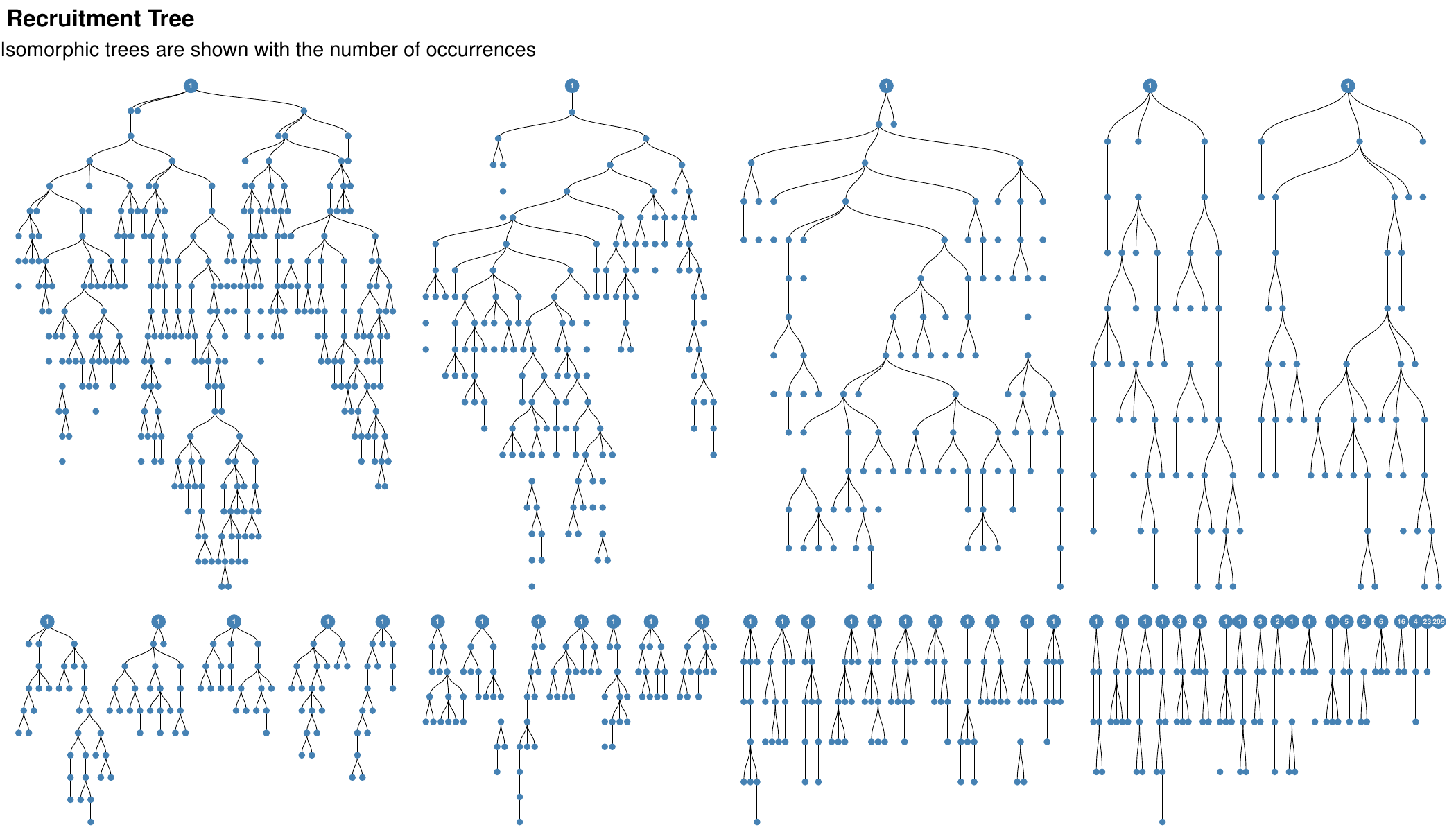}
        \caption{2024 Isomorphism diffusion trees.}
    \end{subfigure}
\end{figure}

\zack{We begin by covering basic descriptive statistics of the RDS sample. Figure~\ref{fig:rdsplots} shows the census of RDS chains observed in the three samples, and Figure~\ref{fig:rdstree} shows isomorphism counts for 2023 and 204 (see Section~\ref{sec:iso} for how the counting is done).} Data collection was vastly improved in 2023 and 2024 after the initial 2022 data collection period.\footnote{Due to technical difficulties with the software used in 2022, the team was forced to shift to paper tracking, and CoC needs included a 90+ minute unstructured qualitative interview beyond the RDS survey.} Here, we also provide some basic statistics of the sample over the three years, including sampled population size, sample size, number of seeds, and max wave recorded (see Table~\ref{tab:summary_stats}). In Table~\ref{tab:mean_recruits} we report the average number of people recruited and the standard error if the person recruited anyone. We observe a steady increase in the number of recruits year over year, but this trend is not statistically significant. For more demographics, see the Table~\ref{tab:rds_combined_all}, where we show the resulting demographics align with what is found in the literature \cite[for a review, see][]{richards2023unsheltered}.

\begin{table}[htp!]
\caption{Summary statistics for each RDS survey by year. }~\label{tab:summary_stats}
\centering
\begin{tabular}{c|cccc|}
\hline
\textbf{Year} & \textbf{Population size} & \textbf{Sample size} & \textbf{\# Seeds} & \textbf{Max wave} \\
\hline
2022 & 7,685 & 616 & 322 & 8 \\
2023 & 8,748 & 1,108 & 239 & 19 \\
2024 & 9,810 & 1,466 & 310 & 20 \\
\hline
\end{tabular}
\end{table}

\begin{table}[htp!]
\caption{Mean degree and standard error of the number of recruits by year if the respondent recruited anyone.}
\label{tab:mean_recruits}
\centering
\begin{tabular}{l|cc}
\toprule
\textbf{Year} & \textbf{Mean degree of recruits} & \textbf{SE of recruits} \\
\midrule
2022 & 1.39 & 0.650 \\
2023 & 1.69 & 0.762 \\
2024 & 1.78 & 0.786 \\
\bottomrule
\end{tabular}
\end{table}

In the following subsections, we will review key network statistics crossed with a standard set of demographics. For details about convergence diagnostics employed, you can go to \citet{almquist2024innovating}, and for a complete list of all demographics and needs assessment questions, see \citet{authorityking22} [2022], \citet{DSSG2024_Understanding_Homelessness} [2023], and \citet{authorityking} [2024].

\subsection{Networks}

\zack{This work includes four egocentric networks \citep{almquist2012random}, where the ego is the respondent and the alters are other people experiencing homelessness (\emph{Acquaintance network}), listed people known to the ego experiencing homelessness (\emph{Close friendship network}), household members [family and other people that ego lives/travels with] (\emph{Kinship network}), and the last network is the peer-referral network itself from the RDS process recruitment process (\emph{Referral network}).}

\subsection{Individual (node) attributes}
\zack{
Over the analysis used in the paper we employ a set of variables measured on three years of data in King County, WA, which are tabulated in Table~\ref{tab:rds_combined_all} and discussed below. We employ the standard Housing and Urban Development demographics, health and homelessness and family/household definitions \citep{HUD_Homelessness_Manual}. These are the same variables typically used in research on homelessness \citep{richards2023unsheltered} and the ones necessary to answer our research questions.

\paragraph{Demographics}
Age (18–24, 25–34, 35–44, 45–54, 55–64, 65+, Missing);
Gender (Female, Male, Other Gender Identity\footnote{Other Gender Identity includes nonbinary, genderfluid, agender, culturally specific gender, transgender, and in the 2024 survey, Two-Spirit due to limited sample size of these groups.}, Missing);
Race (American Indian or Alaska Native or Indigenous (AIANI), Asian/Asian American, Black/African American, White, Multiracial, Native Hawaiian/Pacific Islander (NHOPI)\footnote{Native Hawaiian or Other Pacific Islander}, Other races, Hispanic/Latina/o/x\footnote{Asked only as a race category in 2024.}, Missing/Do not know);
Ethnicity (Hispanic/Latina/o/x, Non-Hispanic/Non-Latina/o/x, Missing);
Veteran (Yes, No, Missing).
See Table~\ref{tab:rds_combined_all} for the surveyed total counts for each option.

\paragraph{Health}
Chronically Homeless (Yes, No, Missing)\footnote{Defined in Section~\ref{study:dem}. This is a composite measure built from the survey based on HUD guidelines.};
Disability (Physical; Yes, No, Missing);
Mental Health (Yes, No, Missing);
Substance Use (Yes, No, Missing).
See Table~\ref{tab:rds_combined_all} for the surveyed total counts for each option.

\paragraph{Homelessness Status}
Sheltered (``Sheltered/Yes''), Unsheltered (“Unsheltered/No”), and Missing.
See Table~\ref{tab:rds_combined_all} for the surveyed total counts for each option.

\paragraph{Alter attributes (household/kinship)}
Each household member not directly surveyed was labeled with gender (Male, Female, Other Gender Identity, Missing) and relationship (parent, partner, sibling, child, other family, missing (NA)).
}

\begin{table}[H]
\centering
\caption{Demographic Characteristics for the 2022 and 2024 King County official RDS PIT Surveys and the UW 2023 Unofficial RDS PIT Survey.}
\label{tab:rds_combined_all}
 \begin{threeparttable}
\scriptsize
\begin{tabular}{llrrrrrrr}
\toprule
\textbf{Category} & \textbf{Response} & \textbf{2022} & \textbf{2022 \%} & \textbf{2023} & \textbf{2023 \%} & \textbf{2024} & \textbf{2024 \%} \\
\midrule
\textbf{Age} & 18–24 & 26 & 3.8 & 26 & 2.3 & 61 & 4.2 \\
             & 25–34 & 115 & 16.8 & 208 & 18.8 & 299 & 20.4 \\
             & 35–44 & 144 & 21.0 & 302 & 27.3 & 407 & 27.8 \\
             & 45–54 & 136 & 19.8 & 240 & 21.7 & 351 & 23.9 \\
             & 55–64 & 86 & 12.6 & 180 & 16.2 & 247 & 16.8 \\
             & 65+ & 20 & 2.9 & 62 & 5.6 & 51 & 3.5 \\
             & Missing & 89 & 13.0 & 90 & 8.1 & 50 & 3.4 \\
\midrule
\textbf{Gender} & Female & 146 & 23.1 & 277 & 25.0 & 394 & 26.9 \\
                & Male & 431 & 68.1 & 717 & 64.7 & 1009 & 68.8 \\
                & Other Gender Identity & 10 & 1.6 & 18 & 1.6 & 12 & 0.8 \\
                & Missing & 29 & 4.6 & 96 & 8.7 & 51 & 3.5 \\
\midrule
\textbf{Race} 
& AIANI\tnote{*} & 53 & 8.60 & 77 & 6.95 & 59 & 4.02 \\
 & Asian/Asian American & 4 & 0.65 & 20 & 1.81 & 25 & 1.71 \\
 & Black/African American & 87 & 14.12 & 163 & 14.71 & 232 & 15.83 \\
 & Hispanic/Latina/o/x\tnote{**}& -- & -- & -- & -- & 219 & 14.94 \\
 & Multiracial & 74 & 12.01 & 129 & 11.64 & 211 & 14.39 \\
 & NHOPI\tnote{***} & 12 & 1.95 & 24 & 2.17 & 30 & 2.05 \\
  & White & 290 & 47.08 & 523 & 47.20 & 576 & 39.29 \\
 & Other races & -- & -- & 14 & 1.26 & 48 & 1.64 \\
  & Missing/Do not know & 96 & 15.58 & 158 & 7.13 & 66 & 4.50 \\
\midrule
\textbf{Ethnicity} & Hispanic/Latina/o/x & 96 & 15.1 & 146 & 13.2 & 317 & 21.6 \\
                   & Non-Hispanic/Non-Latina/o/x & 427 & 67.3 & 820 & 74.0 & 1092 & 74.5 \\
                   & Missing & 93 & 14.7 & 142 & 12.8 & 57 & 3.9 \\
\midrule
\textbf{Shelter} & Unsheltered/No & 391 & 59.6 & 748 & 67.5 & 217 & 14.8 \\
                 & Sheltered/Yes & 115 & 17.5 & 111 & 10.0 & 301 & 20.5 \\
                 & Missing & 110 & 16.8 & 249 & 22.5 & 859 & 58.6 \\
\midrule
\textbf{Veteran} & No & 545 & 85.9 & 972 & 87.7 & 964 & 65.8 \\
                 & Yes & 41 & 6.5 & 111 & 10.0 & 502 & 34.2 \\
                 & Missing & 30 & 4.7 & 25 & 2.3 & 0 & 0.0 \\
\midrule
\textbf{Chronically} & No & 348 & 54.8 & 355 & 32.0 & 835 & 57.0 \\
\textbf{Homeless}   & Yes & 174 & 27.4 & 505 & 45.6 & 591 & 40.3 \\
                          & Missing & 94 & 14.8 & 248 & 22.4 & 40 & 2.7 \\
\midrule
\textbf{Disability} & No & 288 & 45.4 & 379 & 34.2 & 676 & 46.1 \\
\textbf{(Physical)}                    & Yes & 278 & 43.8 & 444 & 40.1 & 726 & 49.5 \\
                    & Missing & 50 & 7.9 & 285 & 25.7 & 64 & 4.4 \\
\midrule
\textbf{Mental} & No & 347 & 54.7 & 508 & 45.8 & 915 & 62.4 \\
 \textbf{Health}                      & Yes & 193 & 30.4 & 312 & 28.2 & 469 & 32.0 \\
                       & Missing & 76 & 12.0 & 288 & 26.0 & 82 & 5.6 \\
\midrule
\textbf{Substance} & No & 300 & 47.3 & 428 & 38.6 & 761 & 51.9 \\
\textbf{Use}                       & Yes & 308 & 48.6 & 407 & 36.7 & 628 & 42.8 \\
                       & Missing & 60 & 9.5 & 273 & 24.6 & 77 & 5.3 \\
\bottomrule
\end{tabular}
 \begin{tablenotes}
 \item[*] American Indian or Alaska Native or Indigenous
\item[**] Asked only as a race in 2024.
\item[***] Native Hawaiian or Other Pacific Islander
\end{tablenotes}
 \end{threeparttable}
\end{table}

\section{Analysis and Results}

\subsection{Descriptive Results}

This section reviews the basic descriptive network statistics that can be obtained from the data, with a focus on personal networks. We review both the sample and population-level estimates of the network descriptive statistics. Due to its robustness and widespread use, we employed the RDS-II estimator \citep{gile2010respondent} with bootstrap confidence intervals (CIs) at the 85\% and/or 95\% levels for all sample- and population-level statistics, as appropriate.

The data, as discussed, contains four core network measures: (i) Acquaintance network, (ii) Close friendship network, (iii) Kinship network, (iv) Peer referral network via coupons. Each of these networks is distinct, with very different profiles. This section will focus primarily on personal networks (i)-(iii).

\subsubsection{Descriptive Statistics}

In this subsection, we will review the population-level estimates of the personal network statistics (including mean degree, density, and degree distribution), gender and racial mixing statistics, family size, and diffusion metrics.

\subsubsection{Personal network statistics: The degree distribution and mean degree}

\paragraph{Sample degree distribution and visibility}

\begin{figure}[htp!]
    \caption{Sample degree distribution over the three core networks.}\label{fig:degdist}
\centering
 \includegraphics[width=.8\linewidth]{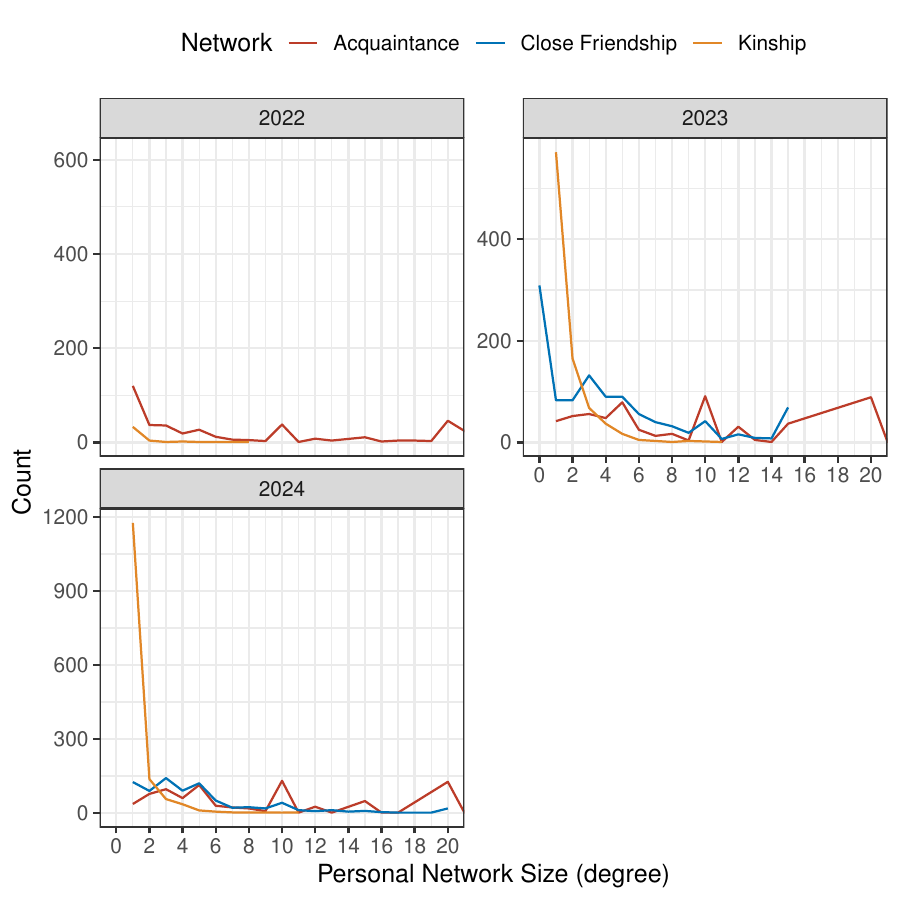}
\end{figure}

We begin by examining the sample degree distribution (or reported degrees of those surveyed over the three years) and the visibility distribution used for our sample weights in the RDS-II estimator. The plot of all three networks in Figure~\ref{fig:degdist} from 2022 to 2023 demonstrates that the Acquaintanceship network (capped at 100) typically peaks around 20, the Close Friendship network (2023 and 2024) peaks more around 4-6 people, and kinship across all three years peaks around 2-3 people. We will see this clearly when we review the mean degree (average personal network size) in the following subsection.

\begin{figure}[htp!]
    \caption{Imputed visibility distribution.}\label{fig:vis}
\centering
 \includegraphics[width=.8\linewidth]{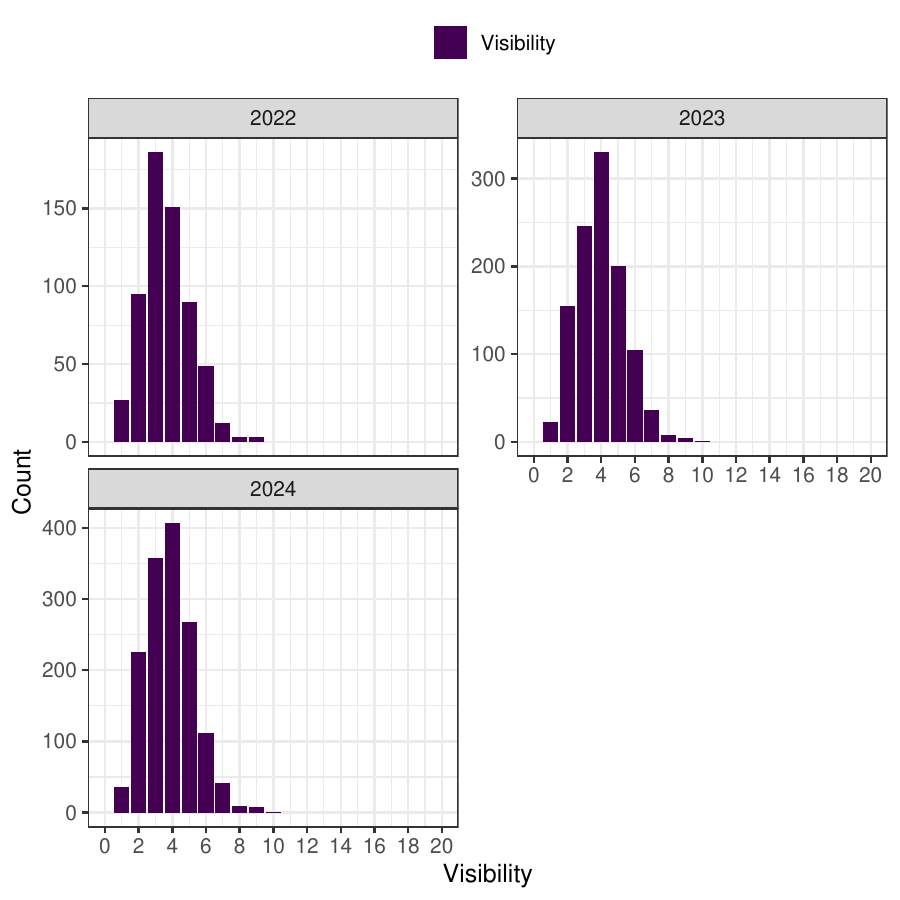}
\end{figure}

Following \citet{mclaughlin2015inference}, we compute the MLE for the visibility measure and use the resulting imputed distribution for our weights in the RDS-II estimator, which can be visualized in Figure~\ref{fig:vis}.

\begin{figure}[htbp!]
    \caption{Estimation of the population level personal networks for Acquaintanceship, Close Friendship, and kinship.}
    \label{fig:three_pop_networks}    
    \begin{subfigure}[b]{0.49\textwidth}
        \centering
        \includegraphics[width=\linewidth]{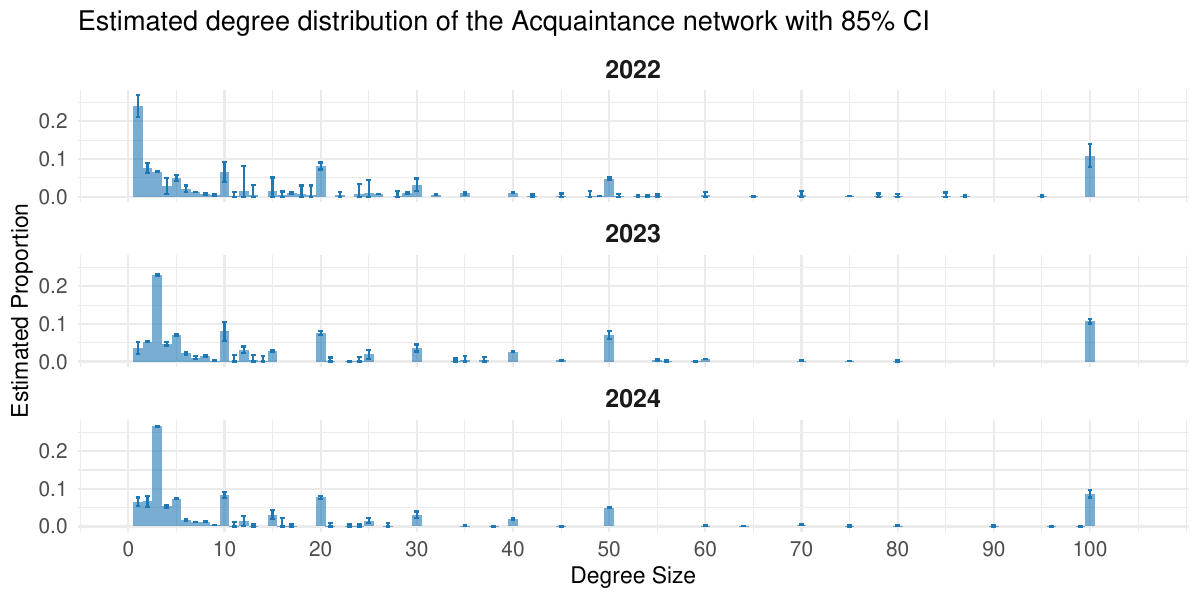}
        \caption{Acquaintanceship}
        \label{fig:acq_pop}
    \end{subfigure}
    \hfill
    \begin{subfigure}[b]{0.49\textwidth}
        \centering
        \includegraphics[width=\linewidth]{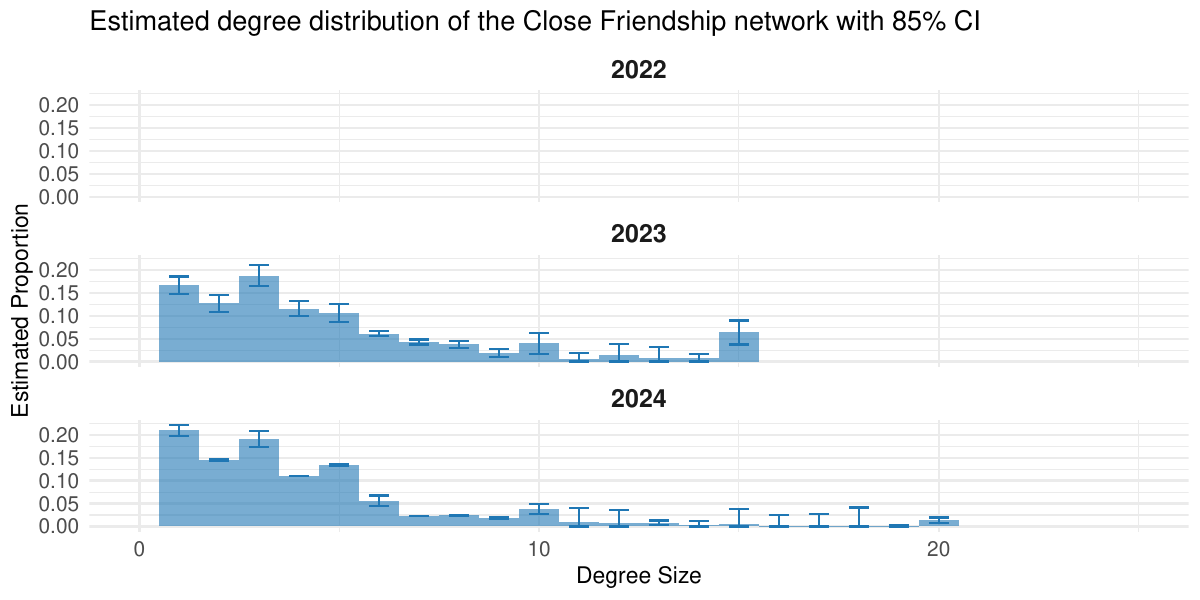}
        \caption{Close Friendship}
        \label{fig:popcf}
    \end{subfigure}
    \hfill
    \begin{subfigure}[b]{0.49\textwidth}
        \centering
        \includegraphics[width=\linewidth]{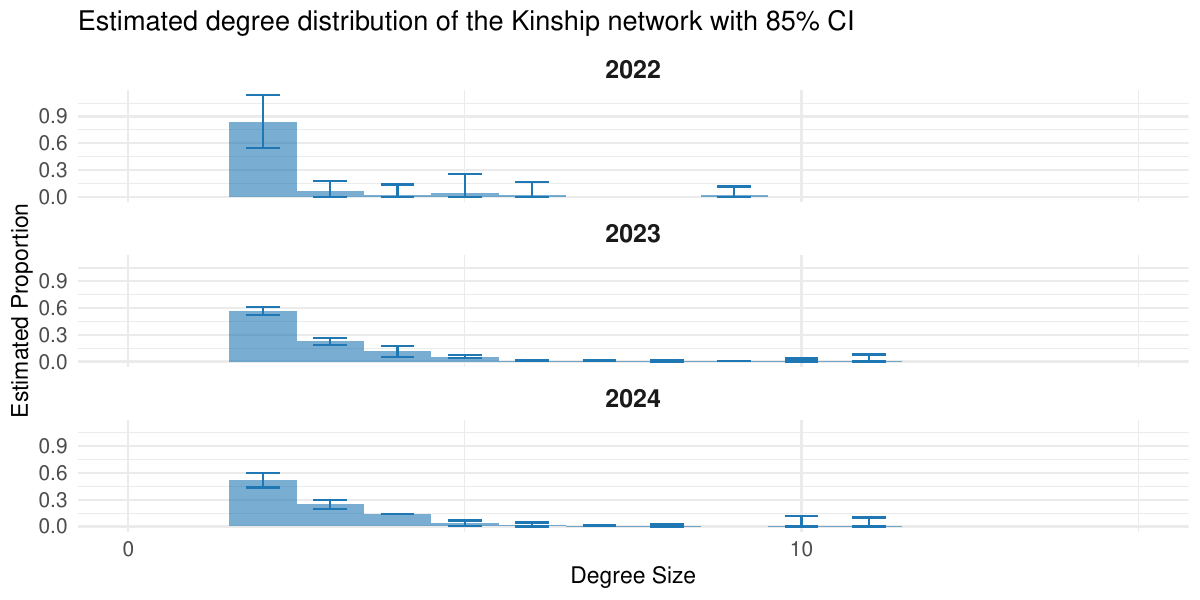}
        \caption{Kinship}
        \label{fig:popkin}
    \end{subfigure}
\end{figure}

\paragraph{Population degree distribution} Using the RDS-II estimator and bootstrap CI at $\alpha=0.15$ level, we plot the resulting estimates of the population-level personal network distribution (see Figure~\ref{fig:three_pop_networks}). The Acquaintanceship network is long-tailed with a large bump at the point where we capped the data (size 100), but all three years look relatively similar; the Close Friendship network in 2023 has a large bump at 15 (where it was capped), but is pretty smooth in 2024 (where it was capped at 20), suggesting 20 is reasonable stopping point for listing out alters. The Kinship network looks similar in all three years.

\paragraph{Mean degree (personal network size)}

We begin our analysis of the personal networks between people experiencing homelessness by estimating one of the most important metrics of personal networks: the mean degree (or size of the personal network). In Table~\ref{tab:combined}, we can see the RDS-II estimate of mean degree for all three networks and their 95\% confidence interval. We can compare the sample personal network size (Acquaintanceship, Close Friendship, Kinship) statistics (26.1, 29.17, 25.21) to the observed population estimates (22.9, 27.5, 24.4), where we see the largest change in the 2022 estimate.

Further, in Table~\ref{tab:combined} we see that there are no statistically significant differences across the three years at the $\alpha=.05$ level, however, Close Friendship is significantly lower in 2024 than 2023 at the  $\alpha=.15$-level with a measurable decrease. We further looked into close friendship by HUD defined shelter usage (e.g., emergency shelter) versus unsheltered (e.g., sleeping outside) from 2023 to 2024, where we found that in 2023 the population mean and 85\% CI for sleeping in shelters was 4.58 (2.90, 6.25), but 3.729 (2.76, 4.70) in 2024; and the population sleep unsheltered in 2023 was  4.934 (4.49, 5.37) and 4.39 (3.81, 4.97) in 2024. In both cases, we saw a decrease of 0.85 (sheltered) and 0.55 (unsheltered), respectively, though neither was statistically significant. Last, we also look at the change in mean degree size by HUD-defined chronically homeless, where in 2023 we find 5.41 (4.88, 5.94) [chronically homeless] and 4.33 (3.55, 5.10) [not chronically homeless]. In 2024, we find 4.90 (4.10, 5.70) [chronically homeless] and 3.71 (3.30, 4.13) [not chronically homeless]. This result largely aligns with prior literature, which finds that those newly experiencing homelessness often have fewer ties to those experiencing homelessness currently \citep{snow1993down}.

\begin{table}[htp!]
\caption{Comparison of mean degree estimates (with 85\% and 95\% confidence intervals) from RDS-II for 2022, 2023, and 2024 surveys of people experiencing homelessness in King County, WA.}
\label{tab:combined}

\centering
\tiny
\begin{tabular}{lrrr|rrr|rrr}
\toprule
\multirow{2}{*}{\bfseries Term} 
& \multicolumn{3}{c|}{\bfseries 2022} 
& \multicolumn{3}{c|}{\bfseries 2023} 
& \multicolumn{3}{c}{\bfseries 2024} \\
\cmidrule(lr){2-4} \cmidrule(lr){5-7} \cmidrule(lr){8-10}
& Mean & 85\% C.I. & 95\% C.I. 
& Mean & 85\% C.I. & 95\% C.I.
& Mean & 85\% C.I. & 95\% C.I. \\
\midrule\addlinespace[2.5pt]

Acquaintance 
& 22.9 & (18.6, 27.1) & (16.8, 28.9) 
& 27.5 & (25.3, 29.6) & (24.8, 30.2) 
& 24.5 & (22.6, 26.3) & (22.0, 27.0) \\ 

\addlinespace[2.5pt]

Close Friendship 
& --- & --- & --- 
& 4.90 & (4.58, 5.22) & (4.45, 5.35) 
& 4.18 & (3.83, 4.54) & (3.69, 4.67) \\ 

\addlinespace[2.5pt]

Kinship 
& 2.40 & (1.21, 3.59) & (0.78, 4.02) 
& 2.84 & (2.61, 3.08) & (2.53, 3.16) 
& 2.95 & (2.59, 3.31) & (2.46, 3.44) \\ 

\addlinespace[2.5pt]
\midrule

Sample Size 
& 616 & & 
& 1,108 & & 
& 1,466 & & \\

\bottomrule
\end{tabular}

\raggedright
\begin{minipage}{\linewidth}
\tiny
Note: Confidence intervals are presented in \((,)\) notation next to estimates. Entries with ''---'' reflect unavailable data.
\end{minipage}
\end{table}

\subsubsection{Gender and race mixing}

From the RDS recruitment procedure, we can also estimate empirical rates of mixing by gender (Tables~\ref{tab:gendermixing2022}, \ref{tab:gendermixing2023}, and \ref{tab:gendermixing2024}), and race (Tables~\ref{tab:racemixing2022}, \ref{tab:racemixing2023} and \ref{tab:mixingrates2024}). In Tables~\ref{tab:gendermixing2022}, \ref{tab:gendermixing2023}, and \ref{tab:gendermixing2024}, we observe high levels of homophily, as expected. Notably, there is a higher proportion of female-male recruitment, given that women comprise approximately 20\% of the population, compared to males. Similarly, in Tables~\ref{tab:racemixing2022}, \ref{tab:racemixing2023}, and \ref{tab:mixingrates2024}, we see high mixing with the largest racial group (White), followed by Black/African American and Multiracial.

\begin{table}[htp!]
\caption{Gender mixing rates estimated from the peer referral network.}

\begin{subtable}{.33\textwidth}\centering
\tiny
    \begin{tabular}{l|ccc}
     & \multicolumn{2}{c}{{\bf Recruitee}}\\
\textbf{Recruiter} & Male & Female \\ 
\midrule
Male & 0.76 & 0.24  \\ 
Female & 0.69 & 0.31 \\
\end{tabular}
\caption{\tiny 2022 Gender mixing between male and female respondents. We have dropped those who responded with other genders (only 10 in the total data set).}\label{tab:gendermixing2022}
\end{subtable}~
\begin{subtable}{.33\textwidth}\centering
\tiny
    \begin{tabular}{l|cc}
     & \multicolumn{2}{c}{{\bf Recruitee}}\\
\textbf{Recruiter} & Male & Female \\ 
\midrule
Male & 0.71 & 0.29 \\
Female & 0.66 & 0.34  \\ 
\end{tabular}
\caption{\tiny The 2023 Gender mixing between male and female respondents. We have dropped those who responded with other genders (only 19 in the total data set).}\label{tab:gendermixing2023}
\end{subtable}~
\begin{subtable}{.33\textwidth}\centering
\tiny
    \begin{tabular}{l|ccc}
     & \multicolumn{2}{c}{{\bf Recruitee}}\\
\textbf{Recruiter} & Male & Female \\ 
\midrule
Male & 0.79 & 0.21  \\ 
Female & 0.60 & 0.40 \\
\end{tabular}
\caption{\tiny 2024 Gender mixing between male and female respondents. We have dropped those who responded with other genders (only 14 in the total data set).}    \label{tab:gendermixing2024}
\end{subtable}
\end{table}

\begin{table}[htp!]
\caption{Racial mixing rates estimated from the peer referral network.}

\begin{subtable}{1\textwidth}\centering
    \tiny
\begin{tabular}{l|cccccccc}
 & \multicolumn{6}{c}{{\bf Recruitee}}\\
\textbf{Recruiter} & AIANI & AAA & BAA & MR & NHPI & W \\ 
\midrule
American Indian, Alaskan Native or Indigenous & 0.14 & 0.00 & 0.07 & 0.36 & 0.00 & 0.43 \\
Asian or Asian American & 0.17 & 0.17 & 0.17 & 0.17 & 0.17 & 0.17 \\
Black, African American & 0.18 & 0.00 & 0.27 & 0.18 & 0.00 & 0.36 \\
Multiracial & 0.00 & 0.00 & 0.00 & 0.15 & 0.15 & 0.69 \\
Native Hawaiian or Pacific Islander & 0.00 & 0.00 & 1.00 & 0.00 & 0.00 & 0.00 \\
White & 0.11 & 0.01 & 0.10 & 0.11 & 0.01 & 0.65 \\
\bottomrule
\end{tabular}
\caption{\tiny The 2022 empirical mixing rates derived from the RDS procedure.} \label{tab:racemixing2022}
\end{subtable}

\begin{subtable}{1\textwidth}\centering
        \tiny
    \begin{tabular}{l|cccccccc}
 & \multicolumn{6}{c}{{\bf Recruitee}}\\
\textbf{Recruiter} & AIANI & AAA & BAA & MR & NHPI & W \\ 
  \hline
American Indian, Alaskan Native or Indigenous & 0.16 & 0.00 & 0.30 & 0.06 & 0.02 & 0.46 \\ 
Asian or Asian American & 0.00 & 0.38 & 0.23 & 0.00 & 0.08 & 0.31 \\ 
Black or African American & 0.04 & 0.01 & 0.24 & 0.15 & 0.06 & 0.50 \\ 
Multiracial & 0.10 & 0.00 & 0.15 & 0.17 & 0.00 & 0.57 \\ 
Native Hawaiian or Pacific Islander & 0.19 & 0.05 & 0.10 & 0.10 & 0.05 & 0.52 \\ 
White & 0.06 & 0.02 & 0.15 & 0.15 & 0.02 & 0.59 \\ 
\end{tabular}
\caption{\tiny The 2023 empirical mixing rates derived from the RDS procedure.} \label{tab:racemixing2023}
\end{subtable}

\begin{subtable}{1\textwidth}\centering
    \centering
        \tiny
    \begin{tabular}{l|cccccccc}
 & \multicolumn{6}{c}{{\bf Recruitee}}\\
\textbf{Recruiter} & AIANI & AAA & BAA & HL & MR & NHPI & W \\ 
  \hline
American Indian, Alaskan Native or Indigenous & 0.08 & 0.03 & 0.20 & 0.20 & 0.18 & 0.00 & 0.31 \\ 
  Asian or Asian American & 0.00 & 0.06 & 0.00 & 0.17 & 0.28 & 0.06 & 0.44 \\ 
  Black or African American & 0.02 & 0.02 & 0.40 & 0.12 & 0.14 & 0.04 & 0.26 \\ 
  Hispanic or Latina/o/x & 0.02 & 0.00 & 0.09 & 0.54 & 0.13 & 0.01 & 0.20 \\ 
  Multiracial & 0.06 & 0.03 & 0.19 & 0.16 & 0.22 & 0.01 & 0.34 \\ 
  Native Hawaiian or Pacific Islander & 0.07 & 0.00 & 0.40 & 0.13 & 0.27 & 0.10 & 0.29 \\ 
  White & 0.04 & 0.02 & 0.12 & 0.07 & 0.15 & 0.01 & 0.61 \\ 
\end{tabular}
\caption{\tiny The 2024 empirical mixing rates derived from the RDS procedure. Note that HUD changed the racial categories in 2024, resulting in different categories (e.g., HUD now treats Hispanic/Latina/o/x as race and ethnicity).} \label{tab:mixingrates2024}
\end{subtable}
\end{table}

\subsubsection{2023 and 2024 Kinship network size and composition}

\begin{figure}[H] 
\caption{2023 Kinship networks counted by their graph (labeled) Isomorphisms.}
\label{fig:family_iso23}
\centering
\includegraphics[width=1\linewidth]{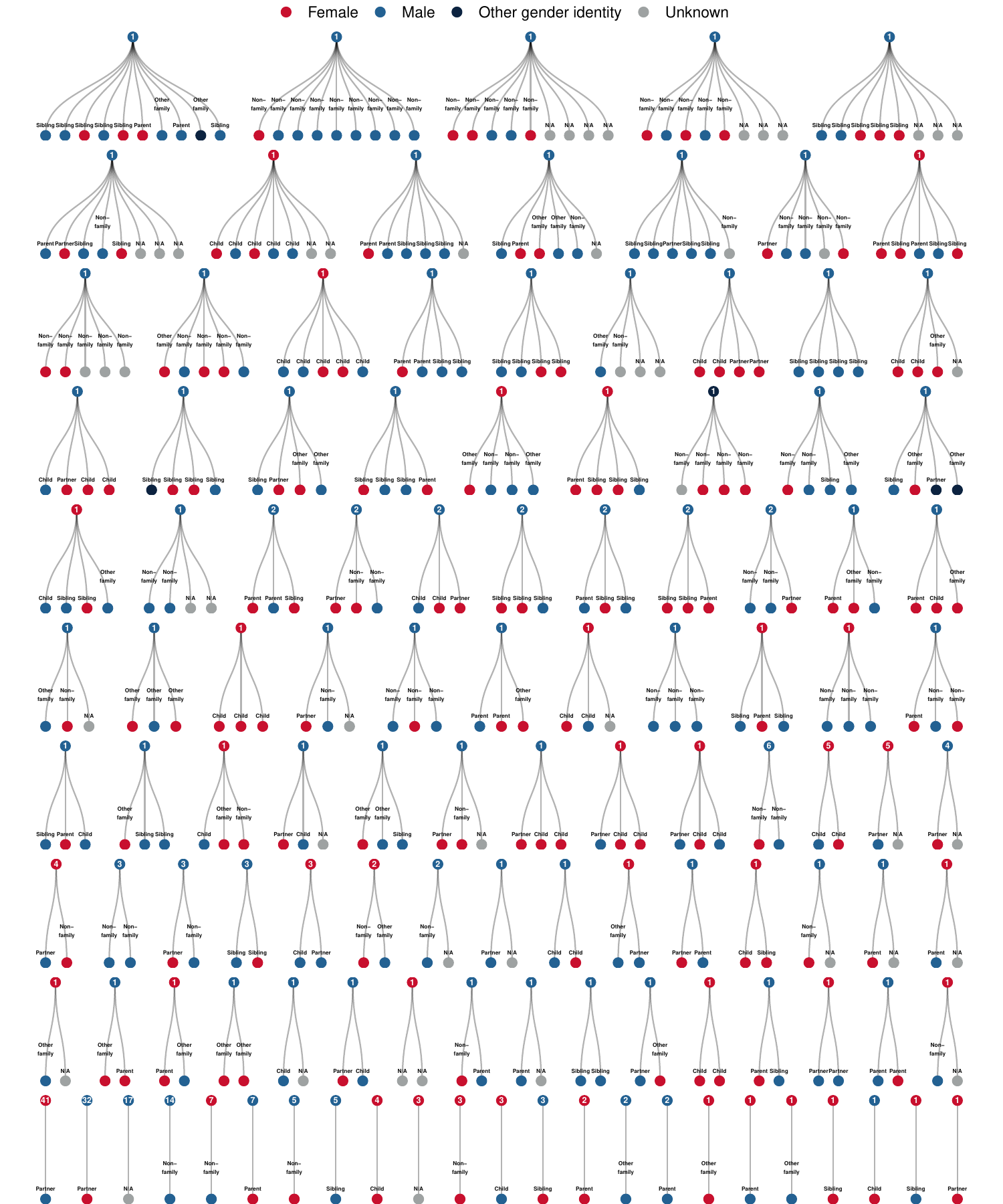}
\end{figure}

\begin{figure}[H]
\caption{2024 Kinship networks counted by their graph (labeled) Isomorphisms.}
\label{fig:family_iso24}
\centering
\includegraphics[width=1\linewidth]{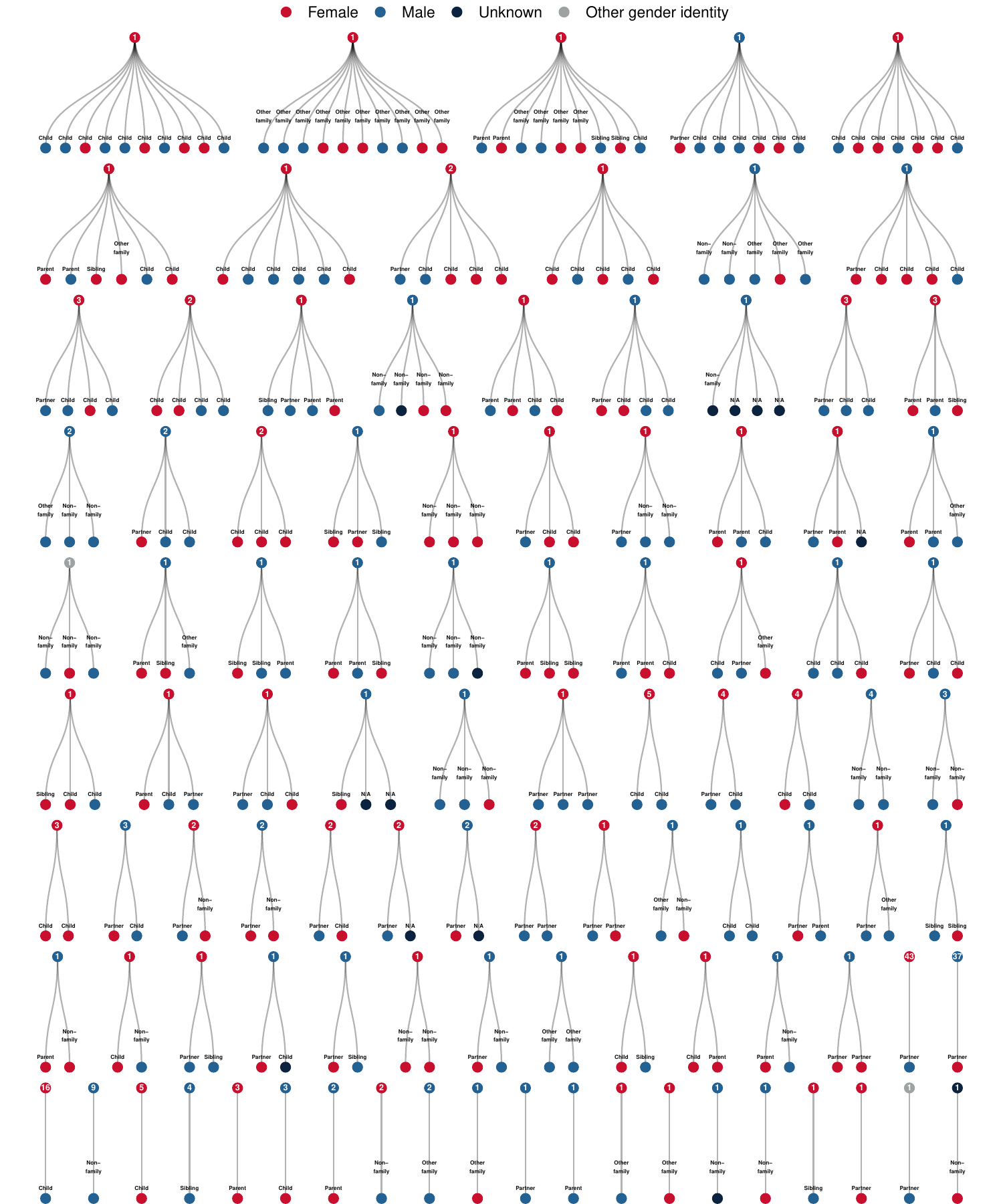}
\end{figure}

 In Figures~\ref{fig:family_iso23} and \ref{fig:family_iso24}, we have mapped the labeled isomorphism by gender and role (e.g., child, spouse, etc.) for all households surveyed in 2023 and 2024 (Kinship network or household network). Details on the visualization are discussed in Section~\ref{sec:iso}. Most relationships involve single parents with a child, but we see several large households, including families with ten children. We see that family sizes for Black or African Americans, Asian or Asian Americans, and Native Hawaiian or Other Pacific Islanders are statistically larger than those of White families, whereas the other groups are statistically distinguishable. Lastly, this model also reveals that chronically homeless individuals tend to have larger family groups. We see an overall increase in household size among people experiencing homelessness from 2022 to 2023 (though not statistically significant). 
 

\subsection{Statistical Modeling Results}

\subsubsection{ZIP Model for degree (personal network size)}

\zack{In this section, we examine zero-inflated negative binomial models, as discussed in the Methods section, to understand the impact of various characteristics on personal network size (e.g., age, gender, race, ethnicity, and HUD-defined mental health, substance use, and disability). We separate out the peer-referral network (coupon-based) and self-identified networks (acquaintanceship, close friends, and kinship) in our analysis. We first perform model fit and diagnostic checks across all four networks, and then review our findings. Because of the limited inferential power in regression analysis, we do not directly discuss the magnitude of significant effects; we discuss only their direction and significance. We leave the magnitude discussion to the discussion section \citep{mize2019best}.}


\subsubsection{Model fit and model diagnostics}

\begin{sidewaysfigure}[htp]
    \centering
    
    \caption{Four diagnostic tests for model fit for the best fitting degree distribution models across networks.}
    \label{fig:model-eval-all-24}
    \begin{subfigure}[b]{0.48\linewidth}
        \centering
        \includegraphics[width=\linewidth]{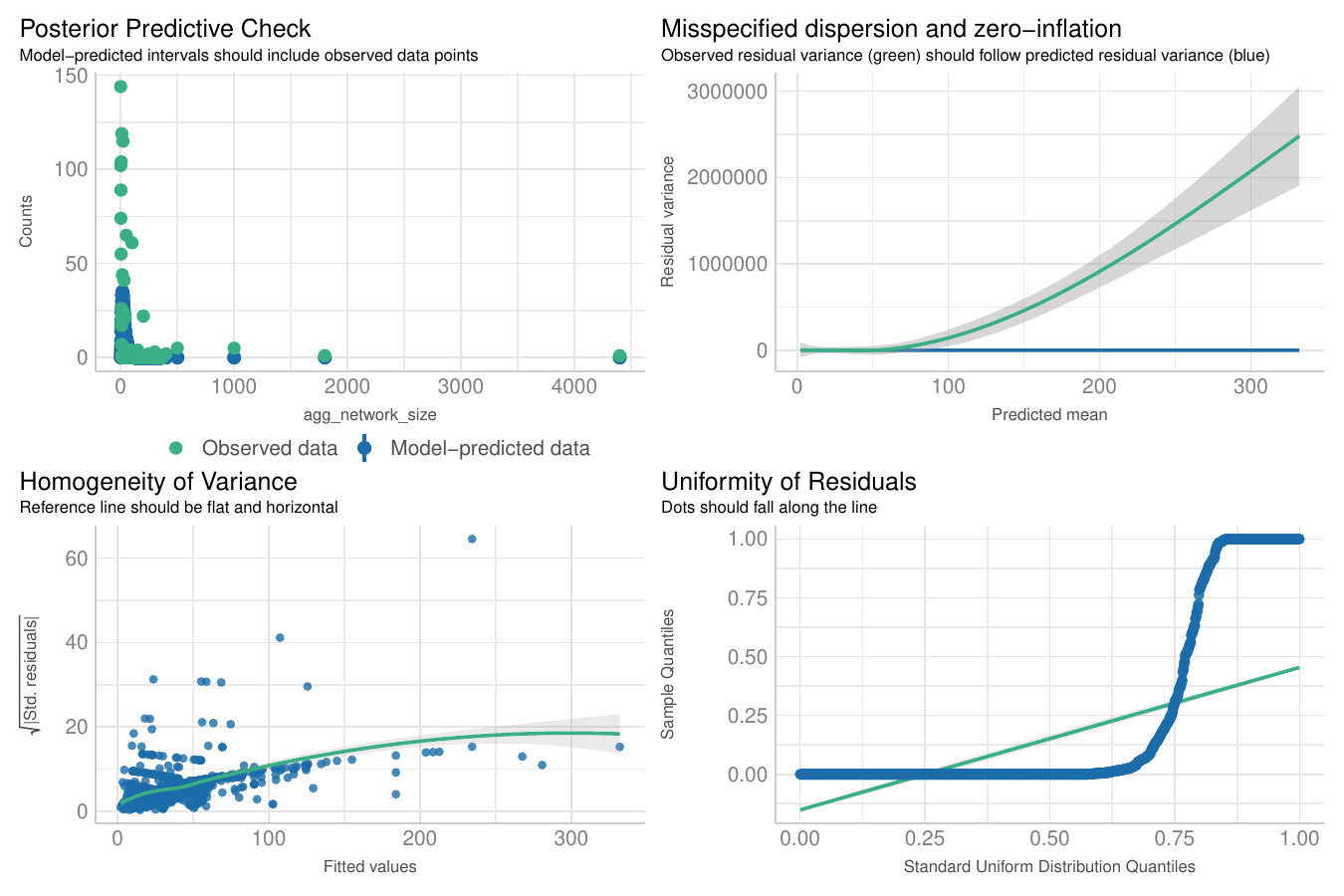}
        \caption{Acquaintance network}
        \label{fig:model-eval-acquaintance-24}
    \end{subfigure}
    \hfill
    \begin{subfigure}[b]{0.48\linewidth}
        \centering
        \includegraphics[width=\linewidth]{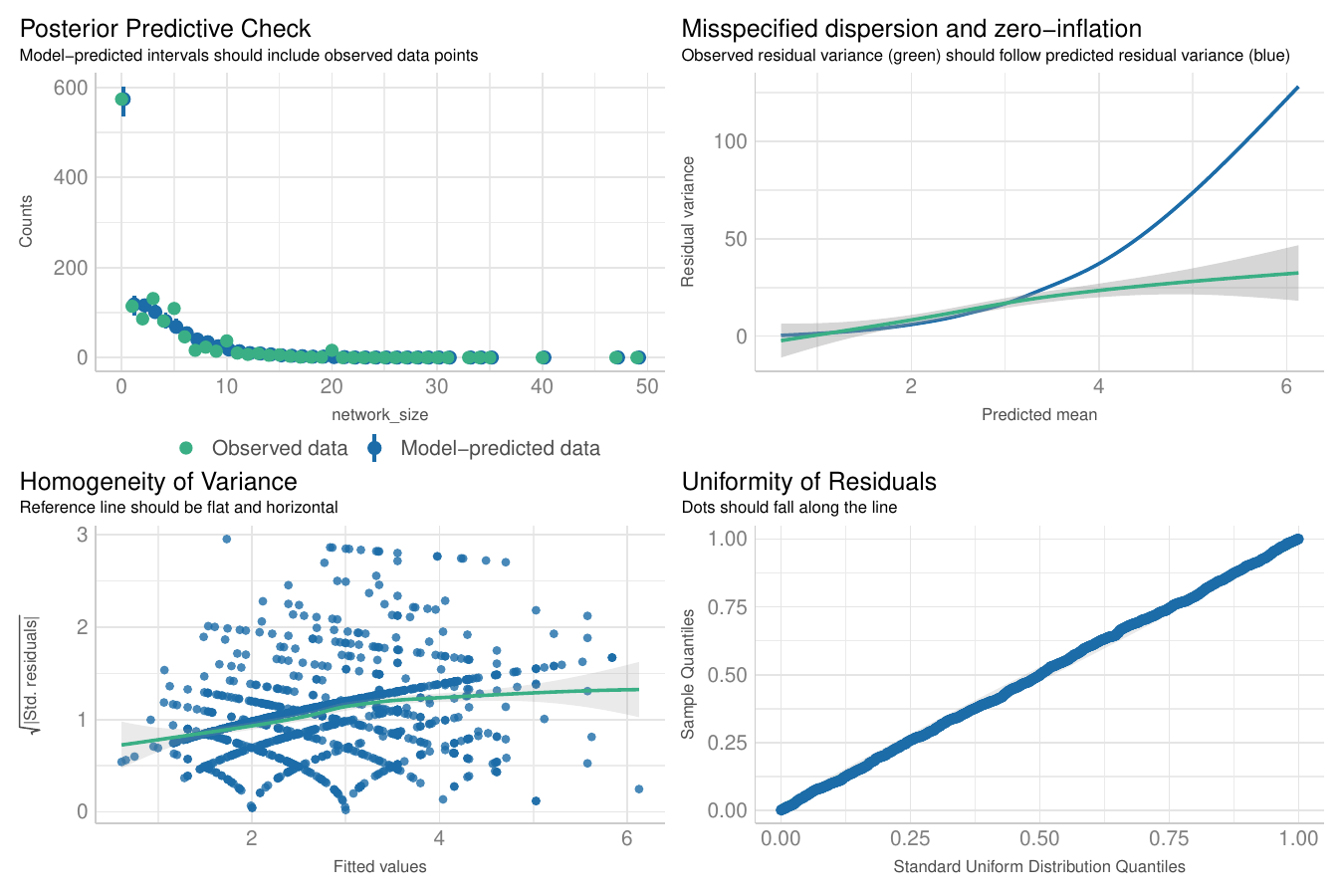}
        \caption{Close friendship network}
        \label{fig:model-eval-friendship-24}
    \end{subfigure}

    \vspace{0.5cm}

    \begin{subfigure}[b]{0.48\linewidth}
        \centering
        \includegraphics[width=\linewidth]{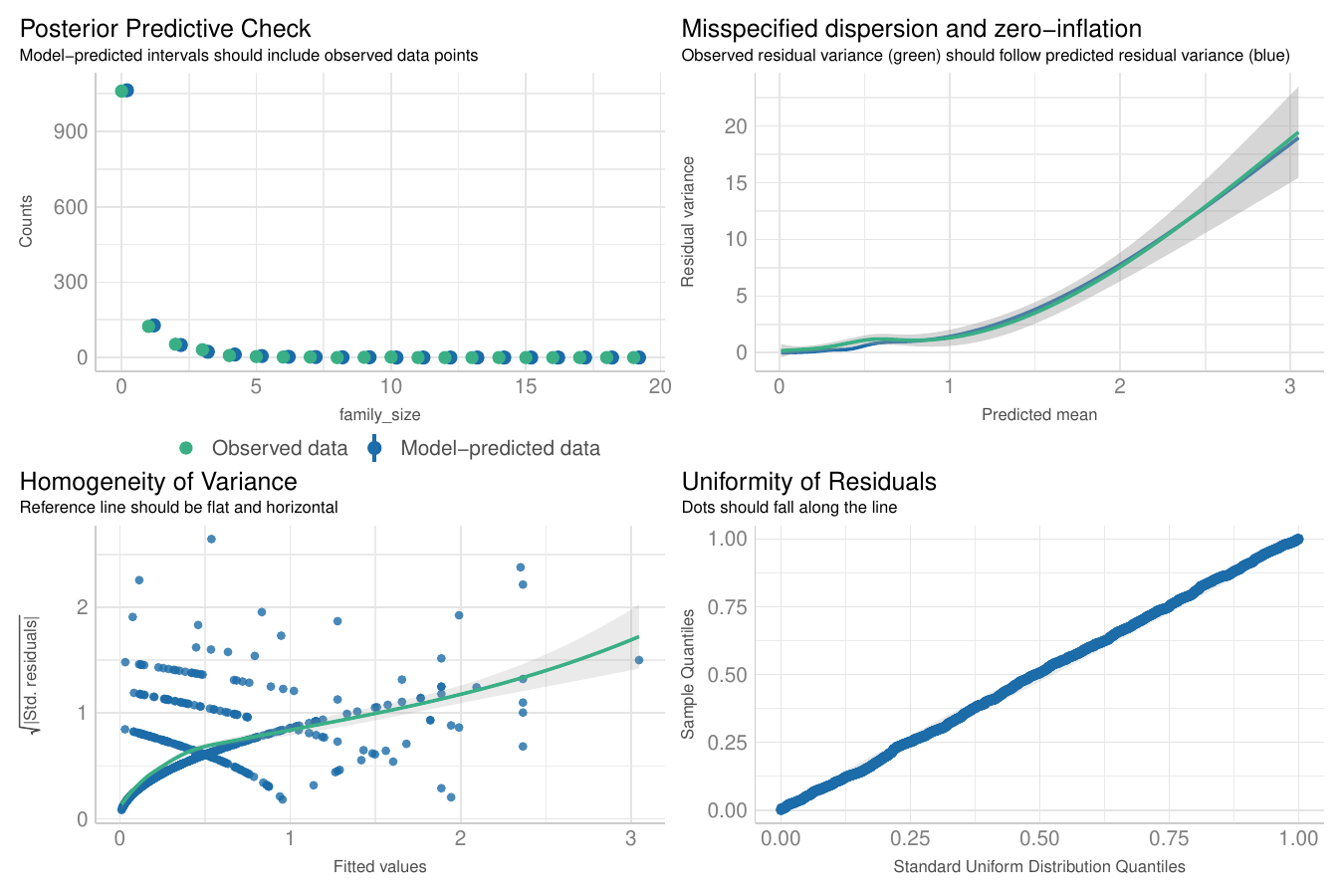}
        \caption{Kinship network}
        \label{fig:model-eval-kinship-24}
    \end{subfigure}
    \hfill
    \begin{subfigure}[b]{0.48\linewidth}
        \centering
        \includegraphics[width=\linewidth]{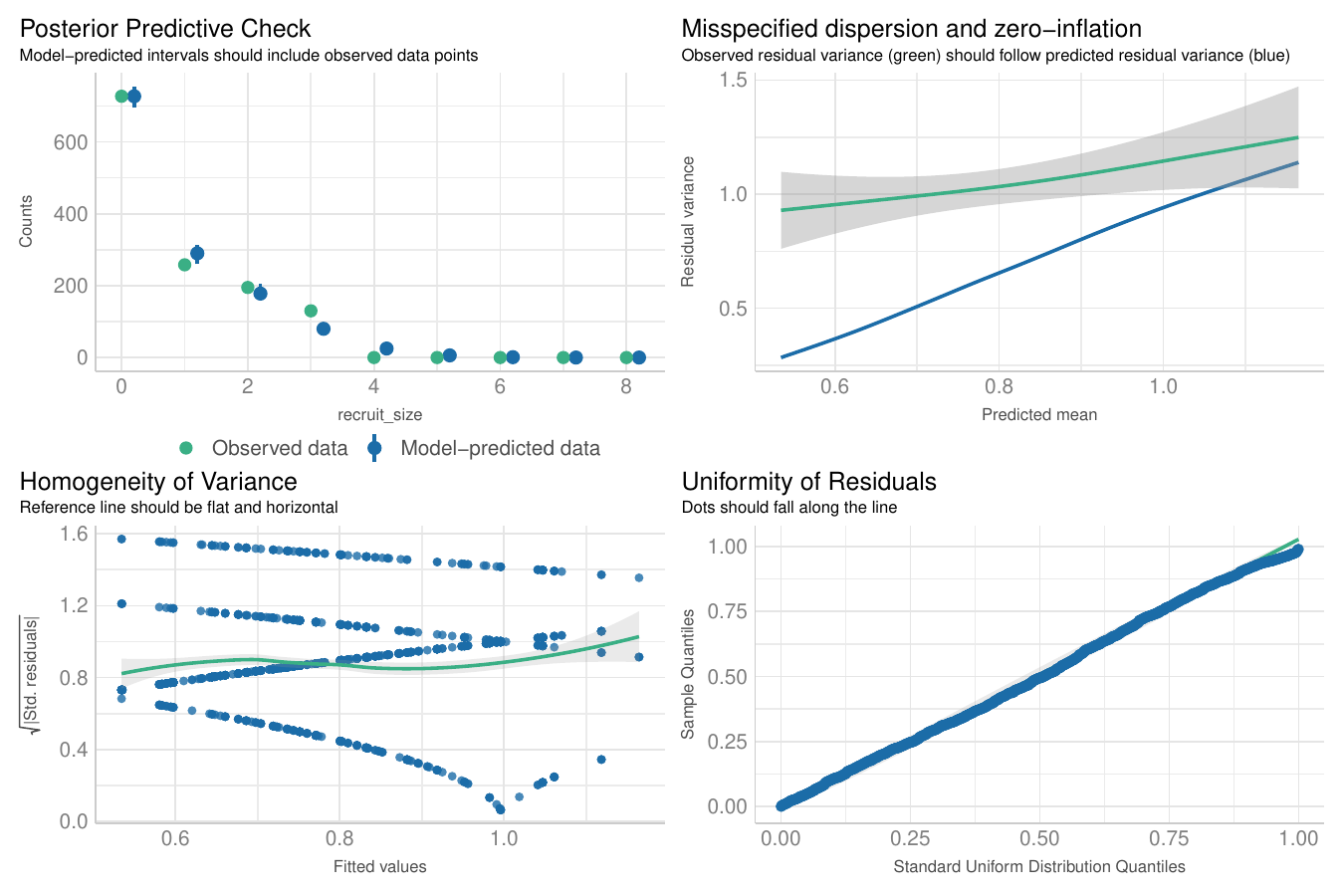}
        \caption{Referral network}
        \label{fig:model-eval-referral-24}
    \end{subfigure}
\end{sidewaysfigure}

\zack{In Tables~\ref{tab:modelfit} and Figures~\ref{fig:model-eval-all-24}, we select the best fitting model based on AICc, AIC, BIC, and RMSE criteria \citep{burnham2004multimodel}. We used likelihood-based methods, including stepwise comparison for variable selection (see Section~\ref{sec:variables} for the full set of variables considered in this analysis) and different categorical models (Poisson, Negative Binomial, and Zero-inflated Negative Binomial and Poisson models). We then do simple model fit diagnostics, which can be seen in Figures~\ref{fig:model-eval-all-24}, which show a decent, but not perfect fit. We only analyze the best-fit models in the next two subsections.}

\begin{table}[H]
\centering
\caption{ZIP Model Goodness-of-Fit Statistics for 2023 and 2024 data.}\label{tab:modelfit}
\tiny
\begin{tabular}{lcccc}
\toprule
\multicolumn{4}{c}{\textbf{2023 Data}} \\
\midrule
& \textbf{Acquaintance} & \textbf{Friendship} & \textbf{Kinship} & \textbf{Referral} \\
\midrule
AICc & 6374.0 & 3629.0 & 1213.0 & 1747.0 \\
AIC  & 6373.0 & 3625.0 & 1212.0 & 1744.0 \\
BIC  & 6455.0 & 3785.0 & 1287.0 & 1899.0 \\
RMSE & 228.0  & 4.39   & 1.22   & 0.976 \\
\midrule
\multicolumn{4}{c}{\textbf{2024 Data}} \\
\midrule
& \textbf{Acquaintance} & \textbf{Friendship} & \textbf{Kinship} & \textbf{Referral} \\
\midrule
AICc & 10050.0 & 5278.0 & 1673.0 & 3141.0 \\
AIC  & 10049.0 & 5274.0 & 1669.0 & 3137.0 \\
BIC  & 10171.0 & 5517.0 & 1912.0 & 3375.0 \\
RMSE & 155.0   & 3.74   & 0.855  & 1.02   \\
\bottomrule
\end{tabular}
\end{table}

\subsubsection{Degree modeling of the acquaintance, close, and kinship networks}

\zack{In 2023, there is no zero-inflated component (this is mirrored in the peer-referral analysis in the following subsection), as there appears to be no significant structure in the RDS seeds (zero degree individuals). In 2024, there is a Zero-inflated component (again, mirrored in the peer-referral analysis in the following subsection), as there appears to be significant structure in the RDS seeds. 

In Table~\ref{tab:ZIC}, for the 2024 King County RDS PIT, we see no zero-inflated component for the acquaintances, but we do for Close Friendship and Kinship (Household) networks. Specifically, for the Close Friends network, we see positive and significant effects in degree size for shelter usage and mental health, but negative and significant effects for Veteran status and (physical) disability. In the Kinship (household) network, we see significant and positive effects for age (55-64 category), race (Asian/Asian American, Black/African American, and Hispanic/Latina/o/x), shelter usage (both categories compared to missing), Veteran status, and Mental Health. We find negative and significant effects for gender (female compared to male) and (physical) disability.

In Table~\ref{tab:ZIC}, we have the conditional component of the zero-inflated negative binomial model for 2023 and 2024. 

\paragraph{Acquaintance network}
In 2023, age (45-65+) and gender (other identifying genders) are both positive and significant for size prediction for the acquaintance network. In 2024, age (all year groups), race (Asian/Asian Americans and multi-racial),  shelter usage, veteran status, mental health, and substance use are all positive and significant for size prediction for the acquaintance network. Only the gender terms are very different in 2023 versus 2024 (race terms are in the same direction, but not significant in 2023). 

In 2023, race (Asian/American, NHPI, and Other race not specified) is negative and significant for predicting the size of the reported acquaintance network. In 2024, gender (other identifying genders), race (Asian/Asian American, Black/African American, Latina/o/x, and NHPI), chronically homeless, and (physical) disability are negative and significant for predicting the size of the reported acquaintance network. No major differences in the negatively predicting variables (either not significant or same direction) between 2023 and 2024. We generally interpret differences as due to geographic and sample size differences between 2023 and 2024, with the gender difference as resulting from the larger sample of other identifying genders in 2024 compared to 2023.

\paragraph{Close friendship network}

In 2023, the only significant effects were race (Asian/Asian American and other races), both negatively associated with close friendship network size.

In 2024, gender (female), chronically homeless, and substance use are positive and significantly associated with close friendship network size; and gender (other identifying genders) and race (American Indian/Indigenous, Black/African American, Hispanic/Latina/o/x, multiracial, and other race) are significantly and negatively associated with close friendship network size.

\paragraph{Kinship (household) network}

In 2023, race (Hispanic/Latina/o/x) and shelter usage are positively associated with kinship (household) size, whereas age (45-54 and 65+) are significant and negatively associated with kinship (household) size.

In 2024, race (Asian/Asian American, Black/African American and NHPI), chronically homeless, and mental health are all significant and positively associated with kinship (household) network size. In 2024, substance use and (physical) disability are significant and negatively associated with kinship (household) size.}

\begin{table}[H]
\centering
\caption{Zero-inflated negative binomial model for degree distribution, 2024 (Zero-Inflation Component), but there is no Zero-inflated component for 2023.} \label{tab:ZIC}
\tiny
\begin{tabular}{p{10.5em}p{10em}p{10em}p{10em}}
\toprule
\textbf{Term} & \textbf{Acquaintance} & \textbf{Close Friendship} & \textbf{Kinship} \\
\midrule
\multicolumn{4}{l}{\textit{Age (ref: 18--24)}} \\
25--34 & --- & & -0.11 (0.70) \\
35--44 & --- & & 0.86 (0.71) \\
45--54 & --- & & 0.92 (0.73) \\
55--64 & --- & & 1.81 (0.77)* \\
65+ & --- & & 1.33 (0.97) \\

\multicolumn{4}{l}{\textit{Gender (ref: Male)}} \\
Female & --- & & -2.82 (0.44)*** \\
Other & --- & & 0.72 (1.44) \\

\multicolumn{4}{l}{\textit{Race (ref: White)}} \\
Am. Indian/Indigenous & --- & & -0.97 (0.97) \\
Asian/Asian Am. & --- & & 3.47 (1.06)** \\
Black/African Am. & --- & & 2.17 (0.57)*** \\
Hispanic/Latina/o/x & --- & & 1.54 (0.90)+ \\
Multiracial & --- & & -0.14 (0.76) \\
Native Hawaiian/PI & --- & & 0.82 (0.91) \\
Other & --- & & -0.74 (1.57) \\

\multicolumn{4}{l}{\textit{Ethnicity (ref: Non-Hispanic)}} \\
Hispanic/Latina/o/x & --- & & 0.42 (0.84) \\

\multicolumn{4}{l}{\textit{Shelter (ref: Housed)}} \\
Sheltered & --- & 0.45 (0.23)+ & 1.95 (0.52)*** \\
Unsheltered & --- & 0.07 (0.25) & 1.57 (0.53)** \\

\multicolumn{4}{l}{\textit{Other Indicators}} \\
Veteran & --- & -0.27 (0.15)+ & \\
Chronic Homeless & --- & 0.23 (0.21) & 1.28 (0.52)* \\
Mental Health & --- & 0.32 (0.15)* & 0.96 (0.46)* \\
Substance Use & --- & -0.28 (0.15)+ & -0.30 (0.42) \\
Disability & --- & & -1.07 (0.43)* \\
\bottomrule
\end{tabular}
\vspace{1mm}
\caption*{\tiny + p < 0.1, * p < 0.05, ** p < 0.01, *** p < 0.001}
\end{table}

\begin{sidewaystable}[htp]
\centering
\caption{Zero-inflated negative binomial model for degree distribution, 2023 and 2024 (Conditional Component)} 
\label{tab:COC_combined_sidebyside}
\tiny
\begin{tabular}{p{8em}p{7em}p{7em}p{7em}p{7em}p{7em}p{7em}}
\toprule
\textbf{Term} & \multicolumn{3}{c}{\textbf{2023}} & \multicolumn{3}{c}{\textbf{2024}} \\
\cmidrule(lr){2-4} \cmidrule(lr){5-7}
 & Acquaintance & Close & Kinship & Acquaintance & Close & Kinship \\
\midrule
\multicolumn{7}{l}{\textit{Age (ref: 18--24)}} \\
25--34 & 0.42 (0.43) & & -0.31 (0.46) & 0.49 (0.04)*** & & \\
35--44 & 0.49 (0.42) & & -0.46 (0.46) & 0.97 (0.04)*** & & \\
45--54 & 1.13 (0.42)** & & -0.86 (0.47)+ & 0.52 (0.04)*** & & \\
55--64 & 0.90 (0.43)* & & -0.47 (0.48) & 0.88 (0.04)*** & & \\
65+ & 1.68 (0.49)*** & & -1.95 (0.68)** & 1.96 (0.04)*** & & \\

\multicolumn{7}{l}{\textit{Gender (ref: Male)}} \\
Female & & & & 0.00 (0.01) & 0.19 (0.07)** & \\
Other & 1.78 (0.55)** & & -18.33 (4283.57) & -1.62 (0.12)*** & -0.92 (0.48)+ & \\

\multicolumn{7}{l}{\textit{Race / Ethnicity (ref: White / Non-Hispanic)}} \\
Am. Indian/Ind. & & & & 0.06 (0.03)* & -0.56 (0.18)** & -0.30 (0.36) \\
Asian/Asian Am. & -1.51 (0.47)** & -0.88 (0.31)** & & -0.55 (0.05)*** & 0.09 (0.24) & 1.52 (0.74)* \\
Black/Afr. Am. & 0.13 (0.26) & -0.19 (0.16) & & -0.09 (0.02)*** & -0.32 (0.10)** & 0.88 (0.29)** \\
Hispanic/Latina/o/x & 0.15 (0.19) & 0.08 (0.12) & 0.40 (0.22)+ & -0.69 (0.02)*** & 0.28 (0.13)* & 0.55 (0.38) \\
Multiracial & 0.16 (0.27) & -0.17 (0.17) & & 0.71 (0.01)*** & -0.23 (0.11)* & -0.20 (0.32) \\
NHPI & -0.88 (0.47)+ & 0.16 (0.28) & & -1.04 (0.06)*** & -0.26 (0.23) & 1.08 (0.40)** \\
Other Race & -1.33 (0.52)* & -1.37 (0.45)** & & 0.05 (0.03)+ & -0.35 (0.18)+ & -0.69 (0.53) \\

\multicolumn{7}{l}{\textit{Shelter (ref: Missing)}} \\
Sheltered & -0.17 (0.19) & & 0.59 (0.21)** & 0.35 (0.02)*** & & \\
Unsheltered & & & & 0.66 (0.02)*** & & \\

\multicolumn{7}{l}{\textit{Other Indicators}} \\
Veteran & 0.14 (0.19) & 0.13 (0.12) & -0.18 (0.28) & 0.53 (0.01)*** & 0.11 (0.07) & \\
Chronic Homeless & & & & -0.14 (0.02)*** & 0.16 (0.07)* & 0.74 (0.22)*** \\
Mental Health & & & & 0.69 (0.01)*** & & 0.41 (0.22)+ \\
Substance Use & & & & 0.39 (0.01)*** & 0.21 (0.07)** & -0.69 (0.20)*** \\
Disability & & & & -0.10 (0.01)*** & 0.07 (0.07) & -0.65 (0.20)*** \\
\bottomrule
\end{tabular}
\vspace{1mm}
\caption*{\tiny + p < 0.1, * p < 0.05, ** p < 0.01, *** p < 0.001}
\end{sidewaystable}

\subsubsection{Peer recruitment}

\zack{In 2023, there is no evidence that key metrics, such as sheltered versus unsheltered, matter for the diffusion chain, see Table~\ref{tab:referral2023_2024} where no effects of the best fitting model are significant. In 2024, we again see no significant effects for the recruitment chains (see Table~\ref{tab:referral2023_2024}), however, we do see some significant effects in the seed (the Zero-inflation component of the model), where we see significant and positive effects in recruitment for females and a negative and significant effect for sheltered use recruitment.  We observe the longest chains within the unsheltered community (see Figure~\ref{fig:rdsplots}), but this is purely a descriptive finding based on the Isomorphism counts and not apparent in the statistical models. }

\begin{table}[H]
\centering
\caption{Zero-inflated Poisson model for referral network for 2023 and 2024.}
\label{tab:referral2023_2024}
\tiny
\begin{tabular}{lcccc}
\toprule
\textbf{Variable / Characteristic} & \multicolumn{2}{c}{\textbf{2023 (Estimate ± SE)}} & \multicolumn{2}{c}{\textbf{2024 Zero-Inflated Poisson}} \\
\cmidrule(lr){2-3} \cmidrule(lr){4-5}
 & Estimate & SE & Beta & SE \\
\midrule
\itshape Age & & & & \\
25-34 & 0.160 & 0.345 & --- & --- \\
35-44 & 0.376 & 0.337 & --- & --- \\
45-54 & 0.544 & 0.340 & --- & --- \\
55-64 & 0.145 & 0.354 & --- & --- \\
65+ & -0.028 & 0.416 & --- & --- \\
\itshape Gender & & & & \\
Male & -0.158 & 0.107 & --- & --- \\
Female / Other identity & -0.225 & 0.420 & 0.44* / 0.35 & 0.179 / 0.848 \\
\itshape Ethnicity / Race & & & & \\
Hispanic/Latina/o/x & -0.033 & 0.546 & --- & --- \\
Asian or Asian American & --- & --- & --- & --- \\
Black, African American & --- & --- & --- & --- \\
Native Hawaiian or Pacific Islander & --- & --- & --- & --- \\
White & --- & --- & --- & --- \\
Multiracial & --- & --- & --- & --- \\
Another Race & --- & --- & --- & --- \\
\itshape Shelter & & & & \\
Shelter Use / Sheltered & 0.244 & 0.370 & -0.80** / 0.15 & 0.300 / 0.229 \\
\itshape Veteran & & & & \\
No & --- & --- & --- & --- \\
Yes & 0.346 & 0.560 & -0.14 / -0.08 & 0.232 / 0.095 \\
\itshape Mental Health & & & & \\
Yes & --- & --- & -0.30 / 0.04 & 0.235 / 0.092 \\
\itshape Substance Use & & & & \\
Yes & --- & --- & -0.46* / -0.16 & 0.224 / 0.092 \\
\itshape Disability & & & & \\
Yes & --- & --- & 0.19 & 0.183 \\
\midrule
No. Obs. & 756 & & 1,310 & \\
AIC & 1815.3 & & 3,131 & \\
BIC & 1880.1 & & 3,198 & \\
RMSE / Sigma & 0.98 & & 1.00 & \\
\bottomrule
\end{tabular}
\end{table}

\subsubsection{ERGM Results}

\begin{figure}[htp!]
\caption{Updated power analysis from \cite{almquist2024innovating} based on \ref{sec:ergm} simulation analysis of the complete network of people experiencing homelessness in King County, WA, 2024.}
    \label{fig:power_ergm}
    \centering
    \includegraphics[width=1\linewidth]{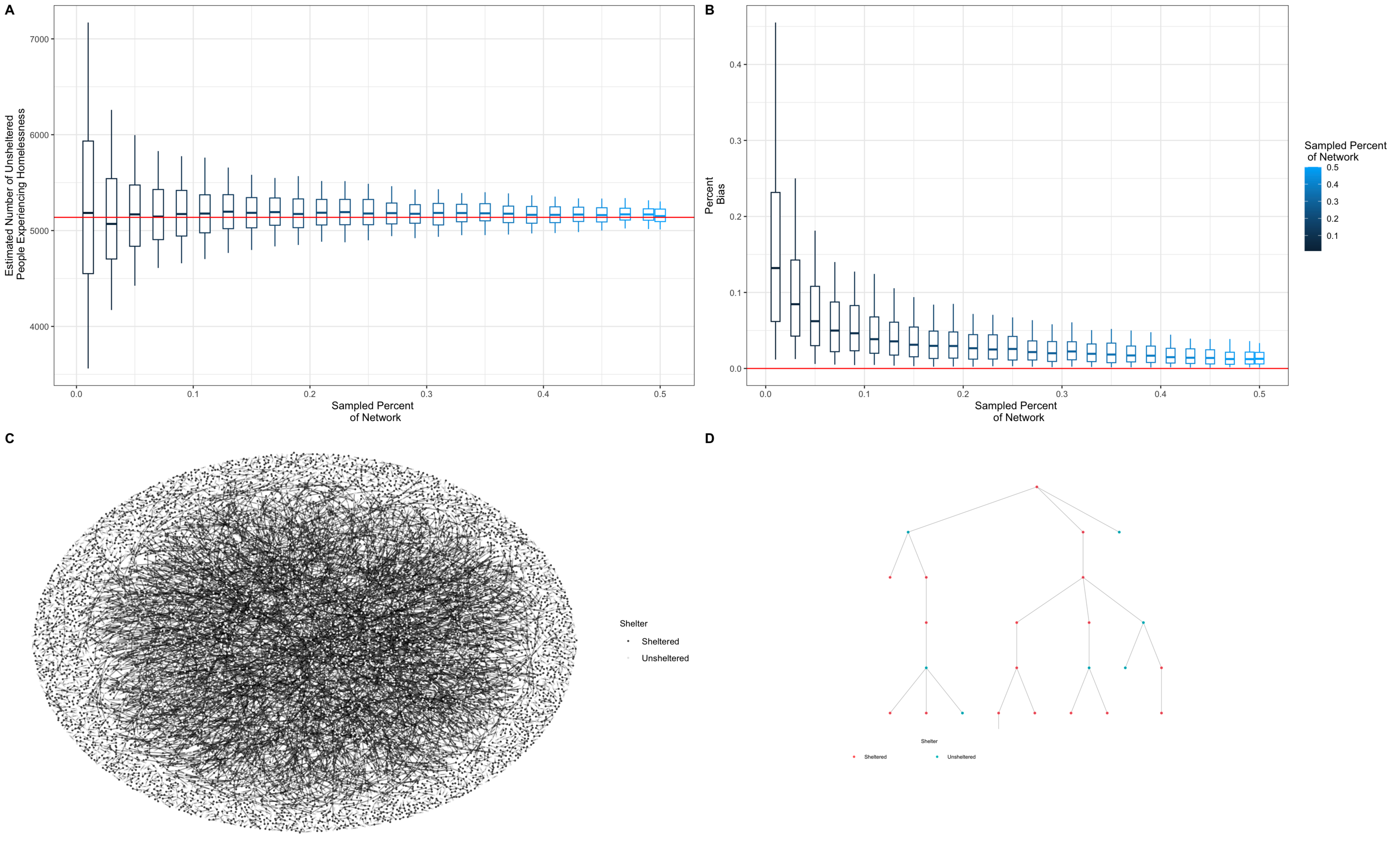}
\end{figure}

\paragraph{Simulated complete networks for power analysis:} Now that we know more about the networks of people experiencing homelessness, we can improve the power analysis introduced in \cite{almquist2024innovating} with updated simulations of this unhoused to unhoused network (see \ref{sec:ergm} for details on the simulation process from Exponential-family Random Graph Models (ERGM) \citep{robins2007introduction}). We can see in Figure~\ref{fig:power_ergm}, what this RDS process might look like, where we have simulated an ERGM based on the observed density and gender mixing.

\section{Discussion}

\zack{In this paper, we review the core network statistics from the King County RDS PIT survey conducted from 2022 to 2024, which may represent the most extensive network study of people experiencing homelessness in the US. We report a decrease in density (stable mean degree for Close Friends) between 2023 and 2024 (significant at the $\alpha=0.15$-level as discussed in the Analysis and Results Section), suggesting that people experiencing homelessness have become less connected (within the community of people experiencing homelessness) over the last year. We extend this argument by noting that HUD-defined shelter usage declined by 0.85 for sheltered individuals and 0.55 for unsheltered individuals in 2024, though neither change was statistically significant, suggesting a direction for future research. We also find some evidence of decreased connectedness, reflected in lower mean degree among those experiencing HUD-defined chronic homelessness compared to those who are not, through our regression analysis. This aligns with \citet{snow1993down}'s thesis on lateral denigration among the unhoused, where many people, when first becoming homeless, distance themselves from the stigmatized population, avoid making friends in this ``community,'' et cetra, as we find that the population of people experiencing chronic homelessness is larger than those not experiencing chronic homelessness. We do not have a way to test whether this is due to composition (i.e., more people entering homelessness), a policy change (e.g., increased displacement events from 2023 to 2024), or both. Lastly, this decrease does not extend to the Acquaintanceship network, where we observe a relatively stable number of Acquaintances over the study period, although this implies a reduction in density as the population size has grown.}

\zack{In our regression analysis, we find that the recruitment network is influenced by gender and shelter usage in 2024 and that key variables influencing the size of reported close friends and acquaintanceship networks are age, race, substance use, and mental and physical disability (self-reported), which aligns with much of the literature around people experiencing homelessness \citep{richards2023unsheltered}. Further, the implication of the diffusion process observed suggests that in the case of getting out emergency information -- e.g., the heat warnings like the heat dome in Seattle, WA in 2021, where the temperatures reached 107 degrees Fahrenheit or 42 degrees Celsius, and information on where cooling stations would be available was utmost importance -- depends on finding a few well-connected people which can result in large and fast cascades of information flow through the unsheltered community. This result backs up the simulation results in \cite{almquist2020large}. }

To conclude our discussion, we review each network, its limitations, and future studies.

\paragraph{Acquaintanceship network}

We interpret the Acquaintanceship network as providing a general sense of how embedded the individual is in the larger homelessness community -- we see this most prominently in the increase in size by age, but also the differing composition in size by race/ethnicity in the ZINB model, where we find that Hispanic or Latina/o/x, American Indian, Alaskan Natives or Indigenous and multiracial groups have larger families on average compared to White, whereas  Asian or Asian American and Native Hawaiian or Other Pacific Islanders have smaller families compared to Whites. Overall, the Acquaintanceship network was found to be stable over time, indicating a slight decrease in density during the study period. This result is striking as the population has grown by 23\% between 2022 and 2024 \citet{almquist2024innovating}. Again, this measure implies the population is becoming less tightly connected over time. This result has large implications for social support, access to resources, information, and safety.

\paragraph{Close friendship network}

We find that the mean degree of the Close Friendship is around 4.9 (2023) and 4.18 (2024), which is about twice the important matters from the General Social Survey (GSS) and political matters from the American National Election Survey (ANES), which range from 2.14 to 2.94 \citep{lee2017important} -- we interpret this as evidence we are capturing a meaningful and close social relationship to that of who people discuss important matters or politics with. These ties are known to be very important for survival \cite{cummings2022social}, access to resources \citep{adams2023nutritional}, and well-being \citep{addo2022social}. Further, we see within these networks from the ZINB model of network size that gender plays an important role (women have larger networks). Last, we note that the stability of the mean degree of Close Friendships means that the density of the social network between people experiencing homelessness must decline as the population grows -- suggesting that the population is becoming more anomic over time within the community of people experiencing homelessness (we cannot comment on the access to the housed population). This finding might be because of increased displacement events or due to the increase in people newly experiencing homelessness -- both of these mechanisms are likely playing out to weaken the social support networks of the unhoused population.

\paragraph{Kinship connections and composition}

One of the most critical networks for survival while living without shelter is one's kinship network (or family/household network). \citet{almquist2024innovating} estimated 17\% of people living unhoused in 2024 lived with their extended kin.  In Figure~\ref{fig:family_iso24}, we have mapped the labeled isomorphism by gender and role (e.g., child, spouse, etc.) for all families surveyed in 2024. Most relationships involve single parents with a child, but we see several large households, including families with ten children. We see that family sizes for Black or African Americans, Asian or Asian Americans, and Native Hawaiian or Other Pacific Islanders are statistically larger than those of White families, whereas the other groups are statistically distinguishable. Lastly, this model also allows us to see that chronically homeless individuals tend to have larger family groups. 

\paragraph{Referral network}

\zack{One of the defining characteristics of an RDS study is that it is a diffusion process, a basic measure of how information or goods flow through the community of people experiencing homelessness. We can categorize the typology of this diffusion process as expressed in the isomorphism \citep{wasserman1994social} of the recruitment trees (See Figure~\ref{fig:rdstree} for the 2023 and 2024 tree isomorphism observed). The method for counting these isomorphisms is in Section~\ref{sec:iso}. Overall, we observe numerous small diffusion chains, with a handful of cases that take off \cite[similar to other diffusion processes studied, such as those over online networks like Twitter or Facebook]{goel2012structure}. 

The referral network exhibits an empirical diffusion process, where we observe some chains taking off, similar to online networks \citep{dow2013anatomy}, with certain sets emerging and many others burning out in short order (see Figure~\ref{fig:rdstree}). In the regression model (Tables~\ref{tab:referral2023_2024}), we see a gender effect in 2024, where women are more likely to recruit women (to the study) than are men; this is important, as recruiting women is typically harder in this setting where they are a smaller percentage of the population, more vulnerable and less likely to be recruited.}

\paragraph{Limitations:} These studies have all the standard limitations of an RDS sample, which include concerns of convergence (see \cite{almquist2024innovating} for details on the convergence statistics in this sample). We worry about known biases in network reporting; see \cite{butts2003network}. However, there is good evidence that this is not a major concern in estimating population-level statistics from the RDS procedure \citep{fellows2022robustness}. Further, we have limited clustering data due to minimal information on alter–alter ties. 
\zack{Lastly, we also note a key limitation in our close friendship and acquaintance networks for providing social support: respondents may list people they ``just'' know rather than true ``friends,'' or potentially friends who engage in risky or dangerous behavior or provide access to illegal substances (\citet{anderson2023norms} shows that when asked to list friends, people experiencing homelessness in Nashville list   ``friends'' rather than people they have a negative association with, but this does not limit out bad influences). Our evidence on this issue is mixed and based on our regression analysis. Our categorical models examining the relationship between self-reported substance use and network size yield mixed results. While the ethnographic literature raises concern that a larger set of ``close Friends'' in this context could represent negative social value, as such ties may provide access to drugs or other socially detrimental resources, and may correlate with chronic homelessness. We find evidence in both directions, where we see that substance use is linked to being more likely to have zero alters, yet conditionally associated with larger networks more generally for both the close friends and acquaintance networks. However, larger kinship (household) networks are consistently associated with lower substance use. This highlights an important direction for future research on the social networks of people experiencing homelessness. Finally, we find that the recruitment network is larger among those who do not have substance use disorder (self-reported).}

\paragraph{Summary:} Here, we have carefully described the personal networks of people experiencing homelessness in King County, WA. Over the three years, we have surveyed more than 3,000 unhoused people, making this the most extensive network study of people experiencing homelessness in the United States. We have found that the population of people experiencing homelessness in King County, WA, appears to be becoming less connected and either more reliant on their local family structures or an increase in households experiencing homelessness, which are growing and heterogeneous in nature (i.e., include many non-kin relationships). This work suggests that policies to improve social connections and build community could benefit this community, such as reducing displacements or providing more places to congregate.

\paragraph{Future studies:} Key future studies should include surveying in closer time intervals with novel methods (i.e., maintaining privacy for the unhoused individual) for estimating change in the population over short periods to better answer questions around whether increased sweeps or more people newly experiencing homelessness, et cetra are the biggest impact on personal network size. Furthermore, we aim to incorporate information on people's alter-alter ties and to explore how individuals transition in and out of homelessness by modifying the instrument to facilitate closer-in-time sampling. We're also interested in the relationship between people experiencing homelessness and the housed community to understand whether having family members who are homeless increases the odds of one becoming homeless themselves. Further, we are interested in exploring the mechanisms for why the network is becoming less connected over time, e.g., this could be because the city has been conducting more frequent encampment displacement operations,  because of the Supreme Court's 2024 \textit{Grants Pass} decision changing the legal landscape of the city and county jurisdictions \cite[It has been shown by][and others the displacement events disrupt people's social networks.]{kaufman2022expulsion} or because of the growth in the population of people experiencing homelessness. 

\section*{Acknowledgments}

\paragraph{Funding} Partial support for this research came from a Eunice Kennedy Shriver National Institute of Child Health and Human Development research infrastructure grant, P2C HD042828, to the Center for Studies in Demography \& Ecology at the University of Washington; NSF CAREER Grant \#SES-2142964; University of Washington Population Health Initiative Tier 2 and 3 Grants. The content is solely the authors' responsibility and does not necessarily represent the official NIH or NSF views. \\

\paragraph{Data-availability} HUD Data exchange provides all official PIT totals from KCRHA (see \href{https://www.hudexchange.info/resource/3031/pit-and-hic-data-since-2007/}{HUD website}). Individual data for 2022 and 2024 is available upon request to KCRHA and the University of Washington through Zack W. Almquist for the 2023 PIT data.\\

\paragraph{Partnership} King County Regional Homelessness Authority partnered with a team of researchers at the University of Washington, led by Zack W. Almquist, to conduct the unsheltered 2022 and 2024 PIT RDS survey design and analysis. The King County Regional Homelessness Authority, as the Community of Care (CoC) lead, conducted and funded the survey project as part of their 2022 and 2024 Point-in-Time count.

\paragraph{Conflict of Interest} Zack Almquist, Ihsan Kahveci, and Amy Hagopian have no conflict of interest to declare. Owen Kajfasz and Janelle Rothfolk, as employees of the King County Regional Homelessness Authority, acknowledge the importance of transparency and accountability in scientific research and peer review. The King County Regional Homelessness Authority employed Kajfasz and Rothfolk during the 2022 and 2024 Point-in-Time count. The King County Regional Homelessness Authority, in its role as the Community of Care (CoC) lead, both conducted and provided funding for the King County 2022 RDS survey, which was carried out as a requirement of the HUD-mandated biannual 2022 Point-in-Time count. \textbf{Financial Interests}: Kajfasz and Rothfolk declare that they have no financial interests, such as stocks, patents, or research grants, that may be perceived as affecting their objectivity in the peer review process. \textbf{Organizational Interests}: The King County Regional Homelessness Authority has a stake in addressing homelessness in King County and may be impacted by the outcomes of this survey project. As the Community of Care (CoC) lead, the King County Regional Homelessness Authority is required by both federal and state statutes to complete a bi-annual unsheltered Point-in-Time count, a mandate for continued funding from the Department of Housing and Urban Development. As such, they have a stake in identifying valid methods for enumerating people experiencing unsheltered homelessness.

\bibliographystyle{elsarticle-harv} 
\bibliography{homelessness}

\begin{thebibliography}{104}
\expandafter\ifx\csname natexlab\endcsname\relax\def\natexlab#1{#1}\fi
\providecommand{\url}[1]{\texttt{#1}}
\providecommand{\href}[2]{#2}
\providecommand{\path}[1]{#1}
\providecommand{\DOIprefix}{doi:}
\providecommand{\ArXivprefix}{arXiv:}
\providecommand{\URLprefix}{URL: }
\providecommand{\Pubmedprefix}{pmid:}
\providecommand{\doi}[1]{\href{http://dx.doi.org/#1}{\path{#1}}}
\providecommand{\Pubmed}[1]{\href{pmid:#1}{\path{#1}}}
\providecommand{\bibinfo}[2]{#2}
\ifx\xfnm\relax \def\xfnm[#1]{\unskip,\space#1}\fi
\bibitem[{Adams et~al.(2023)Adams, Lu, Duan, Chao, Kessler, Miller, Richter,
  Latyshev, Dastoor, Eckburg et~al.}]{adams2023nutritional}
\bibinfo{author}{Adams, E.J.}, \bibinfo{author}{Lu, M.}, \bibinfo{author}{Duan,
  R.}, \bibinfo{author}{Chao, A.K.}, \bibinfo{author}{Kessler, H.C.},
  \bibinfo{author}{Miller, C.D.}, \bibinfo{author}{Richter, A.G.},
  \bibinfo{author}{Latyshev, D.G.}, \bibinfo{author}{Dastoor, J.D.},
  \bibinfo{author}{Eckburg, A.J.}, et~al., \bibinfo{year}{2023}.
\newblock \bibinfo{title}{Nutritional needs, resources, and barriers among
  unhoused adults cared for by a street medicine organization in chicago,
  illinois: a cross-sectional study}.
\newblock \bibinfo{journal}{BMC public health} \bibinfo{volume}{23},
  \bibinfo{pages}{2430}.
\bibitem[{Addo et~al.(2022)Addo, Yuma, Barrera and Layton}]{addo2022social}
\bibinfo{author}{Addo, R.}, \bibinfo{author}{Yuma, P.},
  \bibinfo{author}{Barrera, I.}, \bibinfo{author}{Layton, D.},
  \bibinfo{year}{2022}.
\newblock \bibinfo{title}{Social networks and subjective wellbeing of adults in
  a housing first program}.
\newblock \bibinfo{journal}{Journal of Community Psychology}
  \bibinfo{volume}{50}, \bibinfo{pages}{238--249}.
\bibitem[{{All Home}(2020)}]{home2020seattle}
\bibinfo{author}{{All Home}}, \bibinfo{year}{2020}.
\newblock \bibinfo{title}{Seattle/King County point-in-time count of
  individuals experiencing homelessness}.
\newblock \bibinfo{publisher}{Published by King County Regional Homelessness
  Authority}.
\bibitem[{Almquist et~al.(2024a)Almquist, Yang, Appiah~Kubi, Kaye, Robinson and
  Schachtman}]{DSSG2024_Understanding_Homelessness}
\bibinfo{author}{Almquist, Z.}, \bibinfo{author}{Yang, J.},
  \bibinfo{author}{Appiah~Kubi, F.J.}, \bibinfo{author}{Kaye, B.},
  \bibinfo{author}{Robinson, J.}, \bibinfo{author}{Schachtman, R.},
  \bibinfo{year}{2024}a.
\newblock \bibinfo{title}{Understanding homelessness - data science for social
  good 2024}.
\newblock
  \bibinfo{howpublished}{\url{https://uwescience.github.io/DSSG2024_understanding_homelessness/}}.
\newblock \bibinfo{note}{Accessed: 2024-10-24}.
\bibitem[{Almquist(2012)}]{almquist2012random}
\bibinfo{author}{Almquist, Z.W.}, \bibinfo{year}{2012}.
\newblock \bibinfo{title}{Random errors in egocentric networks}.
\newblock \bibinfo{journal}{Social networks} \bibinfo{volume}{34},
  \bibinfo{pages}{493--505}.
\bibitem[{Almquist(2020)}]{almquist2020large}
\bibinfo{author}{Almquist, Z.W.}, \bibinfo{year}{2020}.
\newblock \bibinfo{title}{Large-scale spatial network models: An application to
  modeling information diffusion through the homeless population of san
  francisco}.
\newblock \bibinfo{journal}{Environment and Planning B: Urban Analytics and
  City Science} \bibinfo{volume}{47}, \bibinfo{pages}{523--540}.
\bibitem[{Almquist et~al.(2025)Almquist, Hebert, Hagopian
  et~al.}]{almquist2025does}
\bibinfo{author}{Almquist, Z.W.}, \bibinfo{author}{Hebert, P.},
  \bibinfo{author}{Hagopian, A.}, et~al., \bibinfo{year}{2025}.
\newblock \bibinfo{title}{Does demography have a role in measuring
  homelessness? insights and approaches in the united states}.
\newblock \bibinfo{journal}{Vienna Yearbook of Population Research}
  \bibinfo{volume}{2025}.
\bibitem[{Almquist et~al.(2020)Almquist, Helwig and
  You}]{almquist2020connecting}
\bibinfo{author}{Almquist, Z.W.}, \bibinfo{author}{Helwig, N.E.},
  \bibinfo{author}{You, Y.}, \bibinfo{year}{2020}.
\newblock \bibinfo{title}{Connecting continuum of care point-in-time homeless
  counts to united states census areal units}.
\newblock \bibinfo{journal}{Mathematical Population Studies}
  \bibinfo{volume}{27}, \bibinfo{pages}{46--58}.
\bibitem[{Almquist et~al.(2024b)Almquist, Kahveci, Hazel, Kajfasz, Rothfolk,
  Guilmette, Anderson, Ozeryansky and Hagopian}]{almquist2024innovating}
\bibinfo{author}{Almquist, Z.W.}, \bibinfo{author}{Kahveci, I.},
  \bibinfo{author}{Hazel, M.A.}, \bibinfo{author}{Kajfasz, O.},
  \bibinfo{author}{Rothfolk, J.}, \bibinfo{author}{Guilmette, C.},
  \bibinfo{author}{Anderson, M.}, \bibinfo{author}{Ozeryansky, L.},
  \bibinfo{author}{Hagopian, A.}, \bibinfo{year}{2024}b.
\newblock \bibinfo{title}{Innovating a community-driven enumeration and needs
  assessment of people experiencing homelessness: A network sampling approach
  for the hud-mandated point-in-time count}.
\newblock \bibinfo{journal}{American Journal of Epidemiology} .
\bibitem[{Anderson et~al.(1999)Anderson, Butts and
  Carley}]{anderson1999interaction}
\bibinfo{author}{Anderson, B.S.}, \bibinfo{author}{Butts, C.},
  \bibinfo{author}{Carley, K.}, \bibinfo{year}{1999}.
\newblock \bibinfo{title}{The interaction of size and density with graph-level
  indices}.
\newblock \bibinfo{journal}{Social networks} \bibinfo{volume}{21},
  \bibinfo{pages}{239--267}.
\bibitem[{Anderson et~al.(1994)Anderson, Burnham and White}]{anderson1994aic}
\bibinfo{author}{Anderson, D.}, \bibinfo{author}{Burnham, K.},
  \bibinfo{author}{White, G.}, \bibinfo{year}{1994}.
\newblock \bibinfo{title}{Aic model selection in overdispersed
  capture-recapture data}.
\newblock \bibinfo{journal}{Ecology} \bibinfo{volume}{75},
  \bibinfo{pages}{1780--1793}.
\bibitem[{Anderson et~al.(2024)Anderson, Hazel, Perkins and
  Almquist}]{anderson2023norms}
\bibinfo{author}{Anderson, M.C.}, \bibinfo{author}{Hazel, A.},
  \bibinfo{author}{Perkins, J.}, \bibinfo{author}{Almquist, Z.},
  \bibinfo{year}{2024}.
\newblock \bibinfo{title}{Norms of fairness and generosity among people
  experiencing homelessness: A dictator game field experiment}.
\newblock \bibinfo{journal}{International Journal on Homelessness} ,
  \bibinfo{pages}{1--13}.
\bibitem[{Anderson et~al.(2021)Anderson, Hazel, Perkins and
  Almquist}]{anderson2021ecology}
\bibinfo{author}{Anderson, M.C.}, \bibinfo{author}{Hazel, A.},
  \bibinfo{author}{Perkins, J.M.}, \bibinfo{author}{Almquist, Z.W.},
  \bibinfo{year}{2021}.
\newblock \bibinfo{title}{The ecology of unsheltered homelessness:
  Environmental and social-network predictors of well-being among an
  unsheltered homeless population}.
\newblock \bibinfo{journal}{International journal of environmental research and
  public health} \bibinfo{volume}{18}, \bibinfo{pages}{7328}.
\bibitem[{Baker(1994)}]{baker1994gender}
\bibinfo{author}{Baker, S.G.}, \bibinfo{year}{1994}.
\newblock \bibinfo{title}{Gender, ethnicity, and homelessness: Accounting for
  demographic diversity on the streets}.
\newblock \bibinfo{journal}{American Behavioral Scientist}
  \bibinfo{volume}{37}, \bibinfo{pages}{476--504}.
\bibitem[{Berkman(2000)}]{berkman2000social}
\bibinfo{author}{Berkman, L.F.}, \bibinfo{year}{2000}.
\newblock \bibinfo{title}{Social support, social networks, social cohesion and
  health}.
\newblock \bibinfo{journal}{Social work in health care} \bibinfo{volume}{31},
  \bibinfo{pages}{3--14}.
\bibitem[{Bernard et~al.(1990)Bernard, Johnsen, Killworth, McCarty, Shelley and
  Robinson}]{bernard1990comparing}
\bibinfo{author}{Bernard, H.R.}, \bibinfo{author}{Johnsen, E.C.},
  \bibinfo{author}{Killworth, P.D.}, \bibinfo{author}{McCarty, C.},
  \bibinfo{author}{Shelley, G.A.}, \bibinfo{author}{Robinson, S.},
  \bibinfo{year}{1990}.
\newblock \bibinfo{title}{Comparing four different methods for measuring
  personal social networks}.
\newblock \bibinfo{journal}{Social networks} \bibinfo{volume}{12},
  \bibinfo{pages}{179--215}.
\bibitem[{Bhandari and Yasunobu(2009)}]{bhandari2009social}
\bibinfo{author}{Bhandari, H.}, \bibinfo{author}{Yasunobu, K.},
  \bibinfo{year}{2009}.
\newblock \bibinfo{title}{What is social capital? a comprehensive review of the
  concept}.
\newblock \bibinfo{journal}{Asian Journal of Social Science}
  \bibinfo{volume}{37}, \bibinfo{pages}{480--510}.
\bibitem[{Brea~Perry and Small(2023)}]{brea2023personal}
\bibinfo{author}{Brea~Perry, A.R.}, \bibinfo{author}{Small, M.},
  \bibinfo{year}{2023}.
\newblock \bibinfo{title}{Personal networks and egocentric analysis}.
\newblock \bibinfo{journal}{The Sage Handbook of Social Network Analysis} ,
  \bibinfo{pages}{439}.
\bibitem[{Breza et~al.(2023)Breza, Chandrasekhar, Lubold, McCormick and
  Pan}]{breza2023consistently}
\bibinfo{author}{Breza, E.}, \bibinfo{author}{Chandrasekhar, A.G.},
  \bibinfo{author}{Lubold, S.}, \bibinfo{author}{McCormick, T.H.},
  \bibinfo{author}{Pan, M.}, \bibinfo{year}{2023}.
\newblock \bibinfo{title}{Consistently estimating network statistics using
  aggregated relational data}.
\newblock \bibinfo{journal}{Proceedings of the National Academy of Sciences}
  \bibinfo{volume}{120}, \bibinfo{pages}{e2207185120}.
\bibitem[{Breza et~al.(2020)Breza, Chandrasekhar, McCormick and
  Pan}]{breza2020using}
\bibinfo{author}{Breza, E.}, \bibinfo{author}{Chandrasekhar, A.G.},
  \bibinfo{author}{McCormick, T.H.}, \bibinfo{author}{Pan, M.},
  \bibinfo{year}{2020}.
\newblock \bibinfo{title}{Using aggregated relational data to feasibly identify
  network structure without network data}.
\newblock \bibinfo{journal}{American Economic Review} \bibinfo{volume}{110},
  \bibinfo{pages}{2454--2484}.
\bibitem[{Browne et~al.(2016)Browne, Varcoe, Lavoie, Smye, Wong, Krause, Tu,
  Godwin, Khan and Fridkin}]{browne2016enhancing}
\bibinfo{author}{Browne, A.J.}, \bibinfo{author}{Varcoe, C.},
  \bibinfo{author}{Lavoie, J.}, \bibinfo{author}{Smye, V.},
  \bibinfo{author}{Wong, S.T.}, \bibinfo{author}{Krause, M.},
  \bibinfo{author}{Tu, D.}, \bibinfo{author}{Godwin, O.},
  \bibinfo{author}{Khan, K.}, \bibinfo{author}{Fridkin, A.},
  \bibinfo{year}{2016}.
\newblock \bibinfo{title}{Enhancing health care equity with indigenous
  populations: evidence-based strategies from an ethnographic study}.
\newblock \bibinfo{journal}{BMC health services research} \bibinfo{volume}{16},
  \bibinfo{pages}{544}.
\bibitem[{Burnham and Anderson(2004)}]{burnham2004multimodel}
\bibinfo{author}{Burnham, K.P.}, \bibinfo{author}{Anderson, D.R.},
  \bibinfo{year}{2004}.
\newblock \bibinfo{title}{Multimodel inference: understanding aic and bic in
  model selection}.
\newblock \bibinfo{journal}{Sociological methods \& research}
  \bibinfo{volume}{33}, \bibinfo{pages}{261--304}.
\bibitem[{Burt(1984)}]{burt1984network}
\bibinfo{author}{Burt, R.S.}, \bibinfo{year}{1984}.
\newblock \bibinfo{title}{Network items and the general social survey}.
\newblock \bibinfo{journal}{Social networks} \bibinfo{volume}{6},
  \bibinfo{pages}{293--339}.
\bibitem[{Burt(2018)}]{burt2018structural}
\bibinfo{author}{Burt, R.S.}, \bibinfo{year}{2018}.
\newblock \bibinfo{title}{Structural holes}, in: \bibinfo{booktitle}{Social
  stratification}. \bibinfo{publisher}{Routledge}, pp.
  \bibinfo{pages}{659--663}.
\bibitem[{Butts(2003)}]{butts2003network}
\bibinfo{author}{Butts, C.T.}, \bibinfo{year}{2003}.
\newblock \bibinfo{title}{Network inference, error, and informant (in)
  accuracy: a bayesian approach}.
\newblock \bibinfo{journal}{social networks} \bibinfo{volume}{25},
  \bibinfo{pages}{103--140}.
\bibitem[{Butts(2008)}]{butts2008social}
\bibinfo{author}{Butts, C.T.}, \bibinfo{year}{2008}.
\newblock \bibinfo{title}{Social network analysis: A methodological
  introduction}.
\newblock \bibinfo{journal}{Asian Journal of Social Psychology}
  \bibinfo{volume}{11}, \bibinfo{pages}{13--41}.
\bibitem[{Corinth and Rossi-de Vries(2018)}]{corinth2018social}
\bibinfo{author}{Corinth, K.}, \bibinfo{author}{Rossi-de Vries, C.},
  \bibinfo{year}{2018}.
\newblock \bibinfo{title}{Social ties and the incidence of homelessness}.
\newblock \bibinfo{journal}{Housing policy debate} \bibinfo{volume}{28},
  \bibinfo{pages}{592--608}.
\bibitem[{Cummings et~al.(2022)Cummings, Lei, Hochberg, Hones and
  Brown}]{cummings2022social}
\bibinfo{author}{Cummings, C.}, \bibinfo{author}{Lei, Q.},
  \bibinfo{author}{Hochberg, L.}, \bibinfo{author}{Hones, V.},
  \bibinfo{author}{Brown, M.}, \bibinfo{year}{2022}.
\newblock \bibinfo{title}{Social support and networks among people experiencing
  chronic homelessness: A systematic review.}
\newblock \bibinfo{journal}{American Journal of Orthopsychiatry}
  \bibinfo{volume}{92}, \bibinfo{pages}{349}.
\bibitem[{Desmond(2012)}]{desmond2012disposable}
\bibinfo{author}{Desmond, M.}, \bibinfo{year}{2012}.
\newblock \bibinfo{title}{Disposable ties and the urban poor}.
\newblock \bibinfo{journal}{American Journal of Sociology}
  \bibinfo{volume}{117}, \bibinfo{pages}{1295--1335}.
\bibitem[{Dow et~al.(2013)Dow, Adamic and Friggeri}]{dow2013anatomy}
\bibinfo{author}{Dow, P.A.}, \bibinfo{author}{Adamic, L.},
  \bibinfo{author}{Friggeri, A.}, \bibinfo{year}{2013}.
\newblock \bibinfo{title}{The anatomy of large facebook cascades}, in:
  \bibinfo{booktitle}{Proceedings of the International AAAI Conference on Web
  and Social Media}, pp. \bibinfo{pages}{145--154}.
\bibitem[{Evans and Forsyth(2004)}]{evans2004risk}
\bibinfo{author}{Evans, R.D.}, \bibinfo{author}{Forsyth, C.J.},
  \bibinfo{year}{2004}.
\newblock \bibinfo{title}{Risk factors, endurance of victimization, and
  survival strategies: The impact of the structural location of men and women
  on their experiences within homeless milieus}.
\newblock \bibinfo{journal}{Sociological Spectrum} \bibinfo{volume}{24},
  \bibinfo{pages}{479--505}.
\bibitem[{Fellows(2022)}]{fellows2022robustness}
\bibinfo{author}{Fellows, I.E.}, \bibinfo{year}{2022}.
\newblock \bibinfo{title}{On the robustness of respondent-driven sampling
  estimators to measurement error}.
\newblock \bibinfo{journal}{Journal of Survey Statistics and Methodology}
  \bibinfo{volume}{10}, \bibinfo{pages}{377--396}.
\bibitem[{Fischer(1982)}]{fischer1982dwell}
\bibinfo{author}{Fischer, C.S.}, \bibinfo{year}{1982}.
\newblock \bibinfo{title}{To dwell among friends: Personal networks in town and
  city}.
\newblock \bibinfo{publisher}{University of chicago Press}.
\bibitem[{Gile(2011)}]{gile2011improved}
\bibinfo{author}{Gile, K.J.}, \bibinfo{year}{2011}.
\newblock \bibinfo{title}{Improved inference for respondent-driven sampling
  data with application to hiv prevalence estimation}.
\newblock \bibinfo{journal}{Journal of the American Statistical Association}
  \bibinfo{volume}{106}, \bibinfo{pages}{135--146}.
\bibitem[{Gile and Handcock(2010)}]{gile2010respondent}
\bibinfo{author}{Gile, K.J.}, \bibinfo{author}{Handcock, M.S.},
  \bibinfo{year}{2010}.
\newblock \bibinfo{title}{Respondent-driven sampling: an assessment of current
  methodology}.
\newblock \bibinfo{journal}{Sociological methodology} \bibinfo{volume}{40},
  \bibinfo{pages}{285--327}.
\bibitem[{Gile et~al.(2015)Gile, Johnston and Salganik}]{gile2015diagnostics}
\bibinfo{author}{Gile, K.J.}, \bibinfo{author}{Johnston, L.G.},
  \bibinfo{author}{Salganik, M.J.}, \bibinfo{year}{2015}.
\newblock \bibinfo{title}{Diagnostics for respondent-driven sampling}.
\newblock \bibinfo{journal}{Journal of the Royal Statistical Society Series A:
  Statistics in Society} \bibinfo{volume}{178}, \bibinfo{pages}{241--269}.
\bibitem[{Goel et~al.(2012)Goel, Watts and Goldstein}]{goel2012structure}
\bibinfo{author}{Goel, S.}, \bibinfo{author}{Watts, D.J.},
  \bibinfo{author}{Goldstein, D.G.}, \bibinfo{year}{2012}.
\newblock \bibinfo{title}{The structure of online diffusion networks}, in:
  \bibinfo{booktitle}{Proceedings of the 13th ACM conference on electronic
  commerce}, pp. \bibinfo{pages}{623--638}.
\bibitem[{Granovetter(1973)}]{granovetter1973strength}
\bibinfo{author}{Granovetter, M.S.}, \bibinfo{year}{1973}.
\newblock \bibinfo{title}{The strength of weak ties}.
\newblock \bibinfo{journal}{American journal of sociology}
  \bibinfo{volume}{78}, \bibinfo{pages}{1360--1380}.
\bibitem[{Green et~al.(2013)Green, Tucker, Golinelli and
  Wenzel}]{green2013social}
\bibinfo{author}{Green, H.D.}, \bibinfo{author}{Tucker, J.S.},
  \bibinfo{author}{Golinelli, D.}, \bibinfo{author}{Wenzel, S.L.},
  \bibinfo{year}{2013}.
\newblock \bibinfo{title}{Social networks, time homeless, and social support: A
  study of men on skid row}.
\newblock \bibinfo{journal}{Network Science} \bibinfo{volume}{1},
  \bibinfo{pages}{305--320}.
\bibitem[{Groton and Radey(2019)}]{groton2019social}
\bibinfo{author}{Groton, D.B.}, \bibinfo{author}{Radey, M.},
  \bibinfo{year}{2019}.
\newblock \bibinfo{title}{Social networks of unaccompanied women experiencing
  homelessness}.
\newblock \bibinfo{journal}{Journal of community psychology}
  \bibinfo{volume}{47}, \bibinfo{pages}{34--48}.
\bibitem[{Handcock et~al.(2024a)Handcock, Fellows and Gile}]{RDS_R}
\bibinfo{author}{Handcock, M.S.}, \bibinfo{author}{Fellows, I.E.},
  \bibinfo{author}{Gile, K.J.}, \bibinfo{year}{2024}a.
\newblock \bibinfo{title}{RDS: Respondent-Driven Sampling}.
\newblock \bibinfo{address}{Los Angeles, CA}.
\newblock \URLprefix \url{https://CRAN.R-project.org/package=RDS}.
  \bibinfo{note}{r package version 0.9-10}.
\bibitem[{Handcock et~al.(2024b)Handcock, Gile, Fellows and
  Neely}]{handcock2024package}
\bibinfo{author}{Handcock, M.S.}, \bibinfo{author}{Gile, K.J.},
  \bibinfo{author}{Fellows, I.E.}, \bibinfo{author}{Neely, W.W.},
  \bibinfo{year}{2024}b.
\newblock \bibinfo{title}{Package ‘rds’}.
\bibitem[{Handcock et~al.(2008)Handcock, Hunter, Butts, Goodreau and
  Morris}]{handcock2008statnet}
\bibinfo{author}{Handcock, M.S.}, \bibinfo{author}{Hunter, D.R.},
  \bibinfo{author}{Butts, C.T.}, \bibinfo{author}{Goodreau, S.M.},
  \bibinfo{author}{Morris, M.}, \bibinfo{year}{2008}.
\newblock \bibinfo{title}{statnet: Software tools for the representation,
  visualization, analysis and simulation of network data}.
\newblock \bibinfo{journal}{Journal of statistical software}
  \bibinfo{volume}{24}, \bibinfo{pages}{1548}.
\bibitem[{Heckathorn(1997)}]{heckathorn1997respondent}
\bibinfo{author}{Heckathorn, D.D.}, \bibinfo{year}{1997}.
\newblock \bibinfo{title}{Respondent-driven sampling: a new approach to the
  study of hidden populations}.
\newblock \bibinfo{journal}{Social problems} \bibinfo{volume}{44},
  \bibinfo{pages}{174--199}.
\bibitem[{Herman et~al.(Jan.-March/2000)Herman, Melancon and
  Marshall}]{herman_graph_2000}
\bibinfo{author}{Herman, I.}, \bibinfo{author}{Melancon, G.},
  \bibinfo{author}{Marshall, M.}, \bibinfo{year}{Jan.-March/2000}.
\newblock \bibinfo{title}{Graph visualization and navigation in information
  visualization: {{A}} survey}.
\newblock \bibinfo{journal}{IEEE Transactions on Visualization and Computer
  Graphics} \bibinfo{volume}{6}, \bibinfo{pages}{24--43}.
\newblock \DOIprefix\doi{10.1109/2945.841119}.
\bibitem[{Hill and Dunbar(2003)}]{hill2003social}
\bibinfo{author}{Hill, R.A.}, \bibinfo{author}{Dunbar, R.I.},
  \bibinfo{year}{2003}.
\newblock \bibinfo{title}{Social network size in humans}.
\newblock \bibinfo{journal}{Human nature} \bibinfo{volume}{14},
  \bibinfo{pages}{53--72}.
\bibitem[{Holten(2006)}]{holten_hierarchical_2006}
\bibinfo{author}{Holten, D.}, \bibinfo{year}{2006}.
\newblock \bibinfo{title}{Hierarchical {{Edge Bundles}}: {{Visualization}} of
  {{Adjacency Relations}} in {{Hierarchical Data}}}.
\newblock \bibinfo{journal}{IEEE Transactions on Visualization and Computer
  Graphics} \bibinfo{volume}{12}, \bibinfo{pages}{741--748}.
\newblock \DOIprefix\doi{10.1109/TVCG.2006.147}.
\bibitem[{Hornik(2012)}]{hornik2012comprehensive}
\bibinfo{author}{Hornik, K.}, \bibinfo{year}{2012}.
\newblock \bibinfo{title}{The comprehensive r archive network}.
\newblock \bibinfo{journal}{Wiley interdisciplinary reviews: Computational
  statistics} \bibinfo{volume}{4}, \bibinfo{pages}{394--398}.
\bibitem[{Hunter et~al.(2012)Hunter, Krivitsky and
  Schweinberger}]{hunter2012computational}
\bibinfo{author}{Hunter, D.R.}, \bibinfo{author}{Krivitsky, P.N.},
  \bibinfo{author}{Schweinberger, M.}, \bibinfo{year}{2012}.
\newblock \bibinfo{title}{Computational statistical methods for social network
  models}.
\newblock \bibinfo{journal}{Journal of Computational and Graphical Statistics}
  \bibinfo{volume}{21}, \bibinfo{pages}{856--882}.
\bibitem[{Jackman et~al.(2015)Jackman, Tahk, Zeileis, Maimone, Fearon, Meers,
  Jackman and Imports}]{jackman2015package}
\bibinfo{author}{Jackman, S.}, \bibinfo{author}{Tahk, A.},
  \bibinfo{author}{Zeileis, A.}, \bibinfo{author}{Maimone, C.},
  \bibinfo{author}{Fearon, J.}, \bibinfo{author}{Meers, Z.},
  \bibinfo{author}{Jackman, M.S.}, \bibinfo{author}{Imports, M.},
  \bibinfo{year}{2015}.
\newblock \bibinfo{title}{Package ‘pscl’}.
\newblock \bibinfo{journal}{Political Science Computational Laboratory}
  \bibinfo{volume}{18}.
\bibitem[{Jenness et~al.(2017)Jenness, Goodreau and
  Morris}]{jenness2017epimodel}
\bibinfo{author}{Jenness, S.M.}, \bibinfo{author}{Goodreau, S.M.},
  \bibinfo{author}{Morris, M.}, \bibinfo{year}{2017}.
\newblock \bibinfo{title}{Epimodel: an r package for mathematical modeling of
  infectious disease over networks}.
\newblock \bibinfo{journal}{bioRxiv} , \bibinfo{pages}{213009}.
\bibitem[{Jenness et~al.(2014)Jenness, Neaigus, Wendel, Gelpi-Acosta and
  Hagan}]{jenness2014spatial}
\bibinfo{author}{Jenness, S.M.}, \bibinfo{author}{Neaigus, A.},
  \bibinfo{author}{Wendel, T.}, \bibinfo{author}{Gelpi-Acosta, C.},
  \bibinfo{author}{Hagan, H.}, \bibinfo{year}{2014}.
\newblock \bibinfo{title}{Spatial recruitment bias in respondent-driven
  sampling: implications for hiv prevalence estimation in urban heterosexuals}.
\newblock \bibinfo{journal}{AIDS and Behavior} \bibinfo{volume}{18},
  \bibinfo{pages}{2366--2373}.
\bibitem[{Joly et~al.(2014)Joly, Cornes and Manthorpe}]{joly2014supporting}
\bibinfo{author}{Joly, L.}, \bibinfo{author}{Cornes, M.},
  \bibinfo{author}{Manthorpe, J.}, \bibinfo{year}{2014}.
\newblock \bibinfo{title}{Supporting the social networks of homeless people}.
\newblock \bibinfo{journal}{Housing, Care and Support} \bibinfo{volume}{17},
  \bibinfo{pages}{198--207}.
\bibitem[{Kaufman(2022)}]{kaufman2022expulsion}
\bibinfo{author}{Kaufman, D.}, \bibinfo{year}{2022}.
\newblock \bibinfo{title}{Expulsion: A type of forced mobility experienced by
  homeless people in canada}.
\newblock \bibinfo{journal}{Urban Geography} \bibinfo{volume}{43},
  \bibinfo{pages}{321--343}.
\bibitem[{Kennedy et~al.(2022)Kennedy, Osilla, Hunter, Golinelli,
  Maksabedian~Hernandez and Tucker}]{kennedy2022restructuring}
\bibinfo{author}{Kennedy, D.P.}, \bibinfo{author}{Osilla, K.C.},
  \bibinfo{author}{Hunter, S.B.}, \bibinfo{author}{Golinelli, D.},
  \bibinfo{author}{Maksabedian~Hernandez, E.}, \bibinfo{author}{Tucker, J.S.},
  \bibinfo{year}{2022}.
\newblock \bibinfo{title}{Restructuring personal networks with a motivational
  interviewing social network intervention to assist the transition out of
  homelessness: A randomized control pilot study}.
\newblock \bibinfo{journal}{Plos one} \bibinfo{volume}{17},
  \bibinfo{pages}{e0262210}.
\bibitem[{Killworth et~al.(1998)Killworth, McCarty, Bernard, Shelley and
  Johnsen}]{killworth1998estimation}
\bibinfo{author}{Killworth, P.D.}, \bibinfo{author}{McCarty, C.},
  \bibinfo{author}{Bernard, H.R.}, \bibinfo{author}{Shelley, G.A.},
  \bibinfo{author}{Johnsen, E.C.}, \bibinfo{year}{1998}.
\newblock \bibinfo{title}{Estimation of seroprevalence, rape, and homelessness
  in the united states using a social network approach}.
\newblock \bibinfo{journal}{Evaluation review} \bibinfo{volume}{22},
  \bibinfo{pages}{289--308}.
\bibitem[{{King County Regional Homelessness Authority}()}]{authorityking22}
\bibinfo{author}{{King County Regional Homelessness Authority}}, .
\newblock \bibinfo{title}{King county 2022 point-in-time count}.
\newblock \URLprefix
  \url{https://kcrha.org/wp-content/uploads/2022/06/PIT-2022-Infograph-v7.pdf}.
\bibitem[{{King County Regional Homelessness Authority}(2025)}]{authorityking}
\bibinfo{author}{{King County Regional Homelessness Authority}},
  \bibinfo{year}{2025}.
\newblock \bibinfo{title}{King county 2024 point-in-time count}.
\newblock \URLprefix
  \url{https://kcrha.org/wp-content/uploads/2025/05/Point-in-Time-Count-2024_King-County_final.pdf}.
\bibitem[{Krivitsky and Morris(2017)}]{krivitsky2017inference}
\bibinfo{author}{Krivitsky, P.N.}, \bibinfo{author}{Morris, M.},
  \bibinfo{year}{2017}.
\newblock \bibinfo{title}{Inference for social network models from
  egocentrically sampled data, with application to understanding persistent
  racial disparities in hiv prevalence in the us}.
\newblock \bibinfo{journal}{The annals of applied statistics}
  \bibinfo{volume}{11}, \bibinfo{pages}{427}.
\bibitem[{Kuha(2004)}]{kuha2004aic}
\bibinfo{author}{Kuha, J.}, \bibinfo{year}{2004}.
\newblock \bibinfo{title}{Aic and bic: Comparisons of assumptions and
  performance}.
\newblock \bibinfo{journal}{Sociological methods \& research}
  \bibinfo{volume}{33}, \bibinfo{pages}{188--229}.
\bibitem[{Kuhn et~al.(2023)Kuhn, Henwood and Chien}]{kuhn2023encampment}
\bibinfo{author}{Kuhn, R.}, \bibinfo{author}{Henwood, B.},
  \bibinfo{author}{Chien, J.}, \bibinfo{year}{2023}.
\newblock \bibinfo{title}{Periodic Assessment of Trajectories of Housing,
  Homelessness and Health (PATHS): Fall 2023 Update: Encampment Sweeps and
  Housing Trajectories}.
\newblock \bibinfo{type}{Technical Report}. UCLA Campuswide Homelessness
  Initiative / UC eScholarship.
\newblock \URLprefix \url{https://escholarship.org/uc/item/46n649n0}.
\bibitem[{Kuhn et~al.(2020)Kuhn, Richards and Roth}]{kuhn2020homelessness}
\bibinfo{author}{Kuhn, R.}, \bibinfo{author}{Richards, J.},
  \bibinfo{author}{Roth, S.}, \bibinfo{year}{2020}.
\newblock \bibinfo{title}{Homelessness and Public Health in Los Angeles}.
\newblock \bibinfo{type}{Technical Report}. eScholarship, University of
  California.
\newblock \bibinfo{note}{Available at
  \url{https://escholarship.org/uc/item/2gn3x56s}}.
\bibitem[{Lachaud et~al.(2024)Lachaud, Yusuf, Maelzer, Perri, Gogosis, Ziegler,
  Mejia-Lancheros and Hwang}]{lachaud2024social}
\bibinfo{author}{Lachaud, J.}, \bibinfo{author}{Yusuf, A.A.},
  \bibinfo{author}{Maelzer, F.}, \bibinfo{author}{Perri, M.},
  \bibinfo{author}{Gogosis, E.}, \bibinfo{author}{Ziegler, C.},
  \bibinfo{author}{Mejia-Lancheros, C.}, \bibinfo{author}{Hwang, S.W.},
  \bibinfo{year}{2024}.
\newblock \bibinfo{title}{Social isolation and loneliness among people living
  with experience of homelessness: a scoping review}.
\newblock \bibinfo{journal}{BMC Public Health} \bibinfo{volume}{24},
  \bibinfo{pages}{2515}.
\bibitem[{Lal et~al.(2021)Lal, Halicki-Asakawa and Fauvelle}]{lal2021scoping}
\bibinfo{author}{Lal, S.}, \bibinfo{author}{Halicki-Asakawa, A.},
  \bibinfo{author}{Fauvelle, A.}, \bibinfo{year}{2021}.
\newblock \bibinfo{title}{A scoping review on access and use of technology in
  youth experiencing homelessness: implications for healthcare}.
\newblock \bibinfo{journal}{Frontiers in digital health} \bibinfo{volume}{3},
  \bibinfo{pages}{782145}.
\bibitem[{Lee and Bearman(2017)}]{lee2017important}
\bibinfo{author}{Lee, B.}, \bibinfo{author}{Bearman, P.}, \bibinfo{year}{2017}.
\newblock \bibinfo{title}{Important matters in political context}.
\newblock \bibinfo{journal}{Sociological Science} \bibinfo{volume}{4},
  \bibinfo{pages}{1--30}.
\bibitem[{Lin(2017)}]{lin2017building}
\bibinfo{author}{Lin, N.}, \bibinfo{year}{2017}.
\newblock \bibinfo{title}{Building a network theory of social capital}.
\newblock \bibinfo{journal}{Social capital} , \bibinfo{pages}{3--28}.
\bibitem[{Marsden(1990)}]{marsden1990network}
\bibinfo{author}{Marsden, P.V.}, \bibinfo{year}{1990}.
\newblock \bibinfo{title}{Network data and measurement}.
\newblock \bibinfo{journal}{Annual review of sociology} \bibinfo{volume}{16},
  \bibinfo{pages}{435--463}.
\bibitem[{Mazerolle(2023)}]{AICcmodavg-package}
\bibinfo{author}{Mazerolle, M.J.}, \bibinfo{year}{2023}.
\newblock \bibinfo{title}{AICcmodavg: Model selection and multimodel inference
  based on (Q)AIC(c)}.
\newblock \URLprefix \url{https://cran.r-project.org/package=AICcmodavg}.
  \bibinfo{note}{r package version 2.3.3}.
\bibitem[{McCallister and Fischer(1978)}]{mccallister1978procedure}
\bibinfo{author}{McCallister, L.}, \bibinfo{author}{Fischer, C.S.},
  \bibinfo{year}{1978}.
\newblock \bibinfo{title}{A procedure for surveying personal networks}.
\newblock \bibinfo{journal}{Sociological Methods \& Research}
  \bibinfo{volume}{7}, \bibinfo{pages}{131--148}.
\bibitem[{McGrath et~al.(2023)McGrath, Crossley, Lhussier and
  Forster}]{mcgrath2023social}
\bibinfo{author}{McGrath, J.}, \bibinfo{author}{Crossley, S.},
  \bibinfo{author}{Lhussier, M.}, \bibinfo{author}{Forster, N.},
  \bibinfo{year}{2023}.
\newblock \bibinfo{title}{Social capital and women’s narratives of
  homelessness and multiple exclusion in northern england}.
\newblock \bibinfo{journal}{International Journal for Equity in Health}
  \bibinfo{volume}{22}, \bibinfo{pages}{41}.
\bibitem[{McLaughlin et~al.(2015)McLaughlin, Handcock, Johnston, Japuki,
  Gexha-Bunjaku, Deva et~al.}]{mclaughlin2015inference}
\bibinfo{author}{McLaughlin, K.R.}, \bibinfo{author}{Handcock, M.S.},
  \bibinfo{author}{Johnston, L.G.}, \bibinfo{author}{Japuki, X.},
  \bibinfo{author}{Gexha-Bunjaku, D.}, \bibinfo{author}{Deva, E.}, et~al.,
  \bibinfo{year}{2015}.
\newblock \bibinfo{title}{Inference for the visibility distribution for
  respondent-driven sampling}.
\newblock \bibinfo{journal}{JSM Proceedings. Alexandria, VA: American
  Statistical Association} , \bibinfo{pages}{2259--2267}.
\bibitem[{McLaughlin et~al.(2024)McLaughlin, Johnston, Jakupi, Gexha-Bunjaku,
  Deva and Handcock}]{mclaughlin2024modeling}
\bibinfo{author}{McLaughlin, K.R.}, \bibinfo{author}{Johnston, L.G.},
  \bibinfo{author}{Jakupi, X.}, \bibinfo{author}{Gexha-Bunjaku, D.},
  \bibinfo{author}{Deva, E.}, \bibinfo{author}{Handcock, M.S.},
  \bibinfo{year}{2024}.
\newblock \bibinfo{title}{Modeling the visibility distribution for
  respondent-driven sampling with application to population size estimation}.
\newblock \bibinfo{journal}{The Annals of Applied Statistics}
  \bibinfo{volume}{18}, \bibinfo{pages}{683--703}.
\bibitem[{Meanwell(2012)}]{meanwell2012experiencing}
\bibinfo{author}{Meanwell, E.}, \bibinfo{year}{2012}.
\newblock \bibinfo{title}{Experiencing homelessness: A review of recent
  literature}.
\newblock \bibinfo{journal}{Sociology Compass} \bibinfo{volume}{6},
  \bibinfo{pages}{72--85}.
\bibitem[{Mize(2019)}]{mize2019best}
\bibinfo{author}{Mize, T.D.}, \bibinfo{year}{2019}.
\newblock \bibinfo{title}{Best practices for estimating, interpreting, and
  presenting nonlinear interaction effects}.
\newblock \bibinfo{journal}{Sociological Science} \bibinfo{volume}{6},
  \bibinfo{pages}{81--117}.
\bibitem[{Moore(1990)}]{moore1990structural}
\bibinfo{author}{Moore, G.}, \bibinfo{year}{1990}.
\newblock \bibinfo{title}{Structural determinants of men's and women's personal
  networks}.
\newblock \bibinfo{journal}{American sociological review} ,
  \bibinfo{pages}{726--735}.
\bibitem[{Nelder and Wedderburn(1972)}]{nelder1972generalized}
\bibinfo{author}{Nelder, J.A.}, \bibinfo{author}{Wedderburn, R.W.},
  \bibinfo{year}{1972}.
\newblock \bibinfo{title}{Generalized linear models}.
\newblock \bibinfo{journal}{Journal of the Royal Statistical Society Series A:
  Statistics in Society} \bibinfo{volume}{135}, \bibinfo{pages}{370--384}.
\bibitem[{Olusegun et~al.(2015)Olusegun, Dikko and
  Gulumbe}]{olusegun2015identifying}
\bibinfo{author}{Olusegun, A.M.}, \bibinfo{author}{Dikko, H.G.},
  \bibinfo{author}{Gulumbe, S.U.}, \bibinfo{year}{2015}.
\newblock \bibinfo{title}{Identifying the limitation of stepwise selection for
  variable selection in regression analysis}.
\newblock \bibinfo{journal}{American Journal of Theoretical and Applied
  Statistics} \bibinfo{volume}{4}, \bibinfo{pages}{414--419}.
\bibitem[{Padgett et~al.(2016)Padgett, Henwood and
  Tsemberis}]{padgett2016housing}
\bibinfo{author}{Padgett, D.}, \bibinfo{author}{Henwood, B.F.},
  \bibinfo{author}{Tsemberis, S.J.}, \bibinfo{year}{2016}.
\newblock \bibinfo{title}{Housing First: Ending homelessness, transforming
  systems, and changing lives}.
\newblock \bibinfo{publisher}{Oxford University Press}.
\bibitem[{Pattison and Wasserman(1999)}]{pattison1999logit}
\bibinfo{author}{Pattison, P.}, \bibinfo{author}{Wasserman, S.},
  \bibinfo{year}{1999}.
\newblock \bibinfo{title}{Logit models and logistic regressions for social
  networks: Ii. multivariate relations}.
\newblock \bibinfo{journal}{British journal of mathematical and statistical
  psychology} \bibinfo{volume}{52}, \bibinfo{pages}{169--193}.
\bibitem[{{R Core Team}(2024)}]{rprog}
\bibinfo{author}{{R Core Team}}, \bibinfo{year}{2024}.
\newblock \bibinfo{title}{R: A Language and Environment for Statistical
  Computing}.
\newblock \bibinfo{organization}{R Foundation for Statistical Computing}.
  \bibinfo{address}{Vienna, Austria}.
\newblock \URLprefix \url{https://www.R-project.org/}.
\bibitem[{Rapier et~al.(2019)Rapier, McKernan and Stauffer}]{rapier2019inverse}
\bibinfo{author}{Rapier, R.}, \bibinfo{author}{McKernan, S.},
  \bibinfo{author}{Stauffer, C.S.}, \bibinfo{year}{2019}.
\newblock \bibinfo{title}{An inverse relationship between perceived social
  support and substance use frequency in socially stigmatized populations}.
\newblock \bibinfo{journal}{Addictive behaviors reports} \bibinfo{volume}{10},
  \bibinfo{pages}{100188}.
\bibitem[{Rice et~al.(2025)Rice, Casiraghi, Gildee, Almquist, Hagopian, Martin
  and de~la Iglesia}]{rice2025sleep}
\bibinfo{author}{Rice, A.}, \bibinfo{author}{Casiraghi, L.P.},
  \bibinfo{author}{Gildee, C.}, \bibinfo{author}{Almquist, Z.W.},
  \bibinfo{author}{Hagopian, A.}, \bibinfo{author}{Martin, M.A.},
  \bibinfo{author}{de~la Iglesia, H.O.}, \bibinfo{year}{2025}.
\newblock \bibinfo{title}{Sleep in people experiencing homelessness under
  different conditions and seasons}.
\newblock \bibinfo{journal}{bioRxiv} .
\bibitem[{Richards and Kuhn(2023)}]{richards2023unsheltered}
\bibinfo{author}{Richards, J.}, \bibinfo{author}{Kuhn, R.},
  \bibinfo{year}{2023}.
\newblock \bibinfo{title}{Unsheltered homelessness and health: a literature
  review}.
\newblock \bibinfo{journal}{AJPM focus} \bibinfo{volume}{2},
  \bibinfo{pages}{100043}.
\bibitem[{Roberts and Dunbar(2011)}]{roberts2011communication}
\bibinfo{author}{Roberts, S.G.}, \bibinfo{author}{Dunbar, R.I.},
  \bibinfo{year}{2011}.
\newblock \bibinfo{title}{Communication in social networks: Effects of kinship,
  network size, and emotional closeness}.
\newblock \bibinfo{journal}{Personal Relationships} \bibinfo{volume}{18},
  \bibinfo{pages}{439--452}.
\bibitem[{Robins et~al.(2007)Robins, Pattison, Kalish and
  Lusher}]{robins2007introduction}
\bibinfo{author}{Robins, G.}, \bibinfo{author}{Pattison, P.},
  \bibinfo{author}{Kalish, Y.}, \bibinfo{author}{Lusher, D.},
  \bibinfo{year}{2007}.
\newblock \bibinfo{title}{An introduction to exponential random graph (p*)
  models for social networks}.
\newblock \bibinfo{journal}{Social networks} \bibinfo{volume}{29},
  \bibinfo{pages}{173--191}.
\bibitem[{Salganik and Heckathorn(2004)}]{salganik20045}
\bibinfo{author}{Salganik, M.J.}, \bibinfo{author}{Heckathorn, D.D.},
  \bibinfo{year}{2004}.
\newblock \bibinfo{title}{5. sampling and estimation in hidden populations
  using respondent-driven sampling}.
\newblock \bibinfo{journal}{Sociological methodology} \bibinfo{volume}{34},
  \bibinfo{pages}{193--240}.
\bibitem[{{\v{S}}imon et~al.(2019){\v{S}}imon, Va{\v{s}}{\'a}t,
  Pol{\'a}kov{\'a}, Gibas and Da{\v{n}}kov{\'a}}]{vsimon2019activity}
\bibinfo{author}{{\v{S}}imon, M.}, \bibinfo{author}{Va{\v{s}}{\'a}t, P.},
  \bibinfo{author}{Pol{\'a}kov{\'a}, M.}, \bibinfo{author}{Gibas, P.},
  \bibinfo{author}{Da{\v{n}}kov{\'a}, H.}, \bibinfo{year}{2019}.
\newblock \bibinfo{title}{Activity spaces of homeless men and women measured by
  gps tracking data: A comparative analysis of prague and pilsen}.
\newblock \bibinfo{journal}{Cities} \bibinfo{volume}{86},
  \bibinfo{pages}{145--153}.
\bibitem[{Small(2013)}]{small2013weak}
\bibinfo{author}{Small, M.L.}, \bibinfo{year}{2013}.
\newblock \bibinfo{title}{Weak ties and the core discussion network: Why people
  regularly discuss important matters with unimportant alters}.
\newblock \bibinfo{journal}{Social networks} \bibinfo{volume}{35},
  \bibinfo{pages}{470--483}.
\bibitem[{Small(2017)}]{small2017someone}
\bibinfo{author}{Small, M.L.}, \bibinfo{year}{2017}.
\newblock \bibinfo{title}{Someone to talk to}.
\newblock \bibinfo{publisher}{Oxford University Press}.
\bibitem[{Small et~al.(2021)Small, Perry, Pescosolido and
  Smith}]{small2021personal}
\bibinfo{author}{Small, M.L.}, \bibinfo{author}{Perry, B.L.},
  \bibinfo{author}{Pescosolido, B.}, \bibinfo{author}{Smith, E.B.},
  \bibinfo{year}{2021}.
\newblock \bibinfo{title}{Personal networks: Classic readings and new
  directions in egocentric analysis}.
\newblock \bibinfo{publisher}{Cambridge University Press}.
\bibitem[{Smith(2008)}]{smith2008searching}
\bibinfo{author}{Smith, H.}, \bibinfo{year}{2008}.
\newblock \bibinfo{title}{Searching for kinship: The creation of street
  families among homeless youth}.
\newblock \bibinfo{journal}{American behavioral scientist}
  \bibinfo{volume}{51}, \bibinfo{pages}{756--771}.
\bibitem[{Snow and Anderson(1987)}]{snow1987identity}
\bibinfo{author}{Snow, D.A.}, \bibinfo{author}{Anderson, L.},
  \bibinfo{year}{1987}.
\newblock \bibinfo{title}{Identity work among the homeless: The verbal
  construction and avowal of personal identities}.
\newblock \bibinfo{journal}{American journal of sociology}
  \bibinfo{volume}{92}, \bibinfo{pages}{1336--1371}.
\bibitem[{Snow and Anderson(1993)}]{snow1993down}
\bibinfo{author}{Snow, D.A.}, \bibinfo{author}{Anderson, L.},
  \bibinfo{year}{1993}.
\newblock \bibinfo{title}{Down on their luck: A study of homeless street
  people}.
\newblock \bibinfo{publisher}{University of California Press},
  \bibinfo{address}{Oakland, CA}.
\bibitem[{Solarz and Bogat(1990)}]{solarz1990social}
\bibinfo{author}{Solarz, A.}, \bibinfo{author}{Bogat, G.A.},
  \bibinfo{year}{1990}.
\newblock \bibinfo{title}{When social support fails: The homeless}.
\newblock \bibinfo{journal}{Journal of Community Psychology}
  \bibinfo{volume}{18}, \bibinfo{pages}{79--96}.
\bibitem[{Stablein(2011)}]{stablein2011helping}
\bibinfo{author}{Stablein, T.}, \bibinfo{year}{2011}.
\newblock \bibinfo{title}{Helping friends and the homeless milieu: Social
  capital and the utility of street peers}.
\newblock \bibinfo{journal}{Journal of Contemporary Ethnography}
  \bibinfo{volume}{40}, \bibinfo{pages}{290--317}.
\bibitem[{Stack(1997)}]{stack1997all}
\bibinfo{author}{Stack, C.B.}, \bibinfo{year}{1997}.
\newblock \bibinfo{title}{All our kin: Strategies for survival in a black
  community}.
\newblock \bibinfo{publisher}{Basic books}.
\bibitem[{Toledo et~al.(2011)Toledo, Code{\c{c}}o, Bertoni, Albuquerque, Malta,
  Bastos, on~Drug~Misuse et~al.}]{toledo2011putting}
\bibinfo{author}{Toledo, L.}, \bibinfo{author}{Code{\c{c}}o, C.T.},
  \bibinfo{author}{Bertoni, N.}, \bibinfo{author}{Albuquerque, E.},
  \bibinfo{author}{Malta, M.}, \bibinfo{author}{Bastos, F.I.},
  \bibinfo{author}{on~Drug~Misuse, B.M.S.G.}, et~al., \bibinfo{year}{2011}.
\newblock \bibinfo{title}{Putting respondent-driven sampling on the map:
  insights from rio de janeiro, brazil}.
\newblock \bibinfo{journal}{JAIDS Journal of Acquired Immune Deficiency
  Syndromes} \bibinfo{volume}{57}, \bibinfo{pages}{S136--S143}.
\bibitem[{Tsai and Alarc{\'o}n(2022)}]{tsai2022annual}
\bibinfo{author}{Tsai, J.}, \bibinfo{author}{Alarc{\'o}n, J.},
  \bibinfo{year}{2022}.
\newblock \bibinfo{title}{The annual homeless point-in-time count: Limitations
  and two different solutions}.
\newblock \bibinfo{journal}{American Journal of Public Health}
  \bibinfo{volume}{112}, \bibinfo{pages}{633--637}.
\bibitem[{{U.S. Department of Housing and Urban
  Development}(2012)}]{HUD_Homelessness_Manual}
\bibinfo{author}{{U.S. Department of Housing and Urban Development}},
  \bibinfo{year}{2012}.
\newblock \bibinfo{title}{Homeless Emergency Assistance and Rapid Transition to
  Housing (HEARTH): Defining "Homeless" Final Rule}.
\newblock \bibinfo{organization}{U.S. Department of Housing and Urban
  Development}.
\newblock
  \bibinfo{note}{\url{https://www.hudexchange.info/resource/1928/hearth-defining-homeless-final-rule/}}.
\bibitem[{Volz and Heckathorn(2008)}]{volz2008probability}
\bibinfo{author}{Volz, E.}, \bibinfo{author}{Heckathorn, D.D.},
  \bibinfo{year}{2008}.
\newblock \bibinfo{title}{Probability based estimation theory for respondent
  driven sampling}.
\newblock \bibinfo{journal}{Journal of official statistics}
  \bibinfo{volume}{24}, \bibinfo{pages}{79}.
\bibitem[{Wasserman(1994)}]{wasserman1994social}
\bibinfo{author}{Wasserman, S.}, \bibinfo{year}{1994}.
\newblock \bibinfo{title}{Social network analysis: methods and applications}.
\newblock \bibinfo{journal}{Cambridge University Press google schola}
  \bibinfo{volume}{2}, \bibinfo{pages}{1--22}.
\bibitem[{Wellman and Wortley(1990)}]{wellman1990different}
\bibinfo{author}{Wellman, B.}, \bibinfo{author}{Wortley, S.},
  \bibinfo{year}{1990}.
\newblock \bibinfo{title}{Different strokes from different folks: Community
  ties and social support}.
\newblock \bibinfo{journal}{American journal of Sociology}
  \bibinfo{volume}{96}, \bibinfo{pages}{558--588}.
\bibitem[{Wesson et~al.(2025)Wesson, Graham-Squire, Perry, Assaf and
  Kushel}]{wesson2025novel}
\bibinfo{author}{Wesson, P.}, \bibinfo{author}{Graham-Squire, D.},
  \bibinfo{author}{Perry, E.}, \bibinfo{author}{Assaf, R.D.},
  \bibinfo{author}{Kushel, M.}, \bibinfo{year}{2025}.
\newblock \bibinfo{title}{Novel methods to construct a representative sample
  for surveying california’s unhoused population: the california statewide
  study of people experiencing homelessness}.
\newblock \bibinfo{journal}{American journal of epidemiology}
  \bibinfo{volume}{194}, \bibinfo{pages}{1238--1248}.
\bibitem[{Zheng et~al.(2006)Zheng, Salganik and Gelman}]{zheng2006many}
\bibinfo{author}{Zheng, T.}, \bibinfo{author}{Salganik, M.J.},
  \bibinfo{author}{Gelman, A.}, \bibinfo{year}{2006}.
\newblock \bibinfo{title}{How many people do you know in prison? using
  overdispersion in count data to estimate social structure in networks}.
\newblock \bibinfo{journal}{Journal of the American Statistical Association}
  \bibinfo{volume}{101}, \bibinfo{pages}{409--423}.

\end{thebibliography}





\clearpage

\section*{Online Appendix To: Understanding the Personal Networks of People Experiencing Homelessness in King County, WA with Aggregate Relational Data}
\appendix
\counterwithin{figure}{section}

\startcontents[sections]
\printcontents[sections]{l}{1}{\setcounter{tocdepth}{2}}

\section{Introduction to the appendix}

This appendix provides the necessary extra details for the paper, including a discussion of statistical models, additional tables, and basic information on the survey and sample statistics.

\section{Network questions from the surveys}
\label{sec:network_questions}

\subsubsection{2022 Network and Family Questions}

\begin{figure}[htp!]
 \centering
    \begin{subfigure}[t]{1\textwidth}
        \centering
        \includegraphics[width=.6\linewidth]{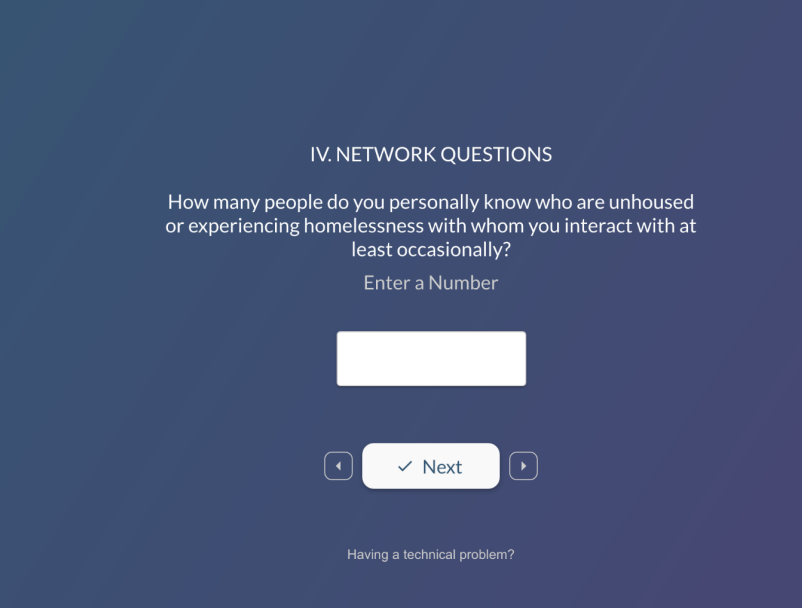}
        \caption{2022 Aggregate network question.}
    \end{subfigure}\\
    \begin{subfigure}[t]{1\textwidth}
        \centering
        \includegraphics[width=.6\linewidth]{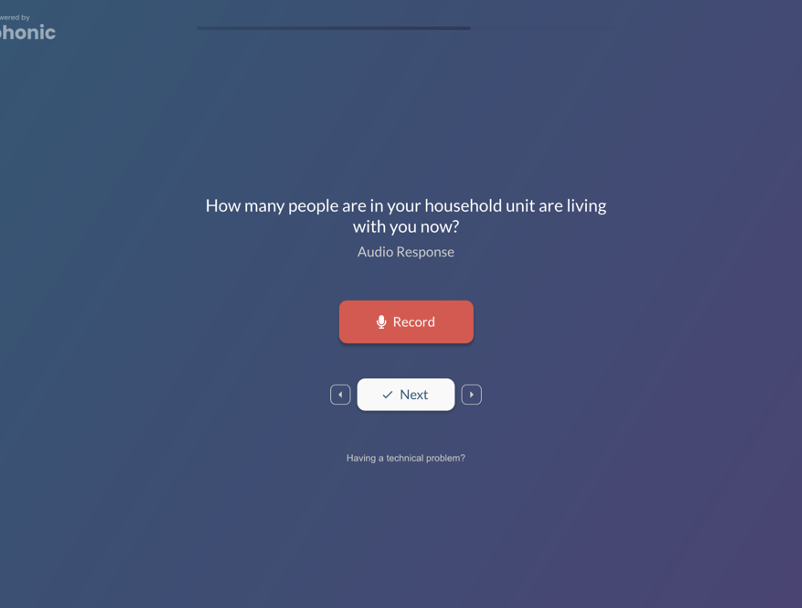}
        \caption{Family list-based question.}
    \end{subfigure}
    \caption{The 2022 network questions asked for the official RDS PIT study.}
    \label{fig:2022nq}
\end{figure}

In 2022, we asked only the aggregate network question of how many people you know experiencing homelessness, followed by the aggregate question of how many are in tents, RVs, etc. (See Figure~\ref{fig:2022nq}), due to privacy concerns. As well as basic family information (also in aggregate). 

\subsubsection{2023 Network and Family Questions}

\begin{figure}[htp!]
 \centering
    \begin{subfigure}[t]{1\textwidth}
        \centering
        \includegraphics[width=.6\linewidth]{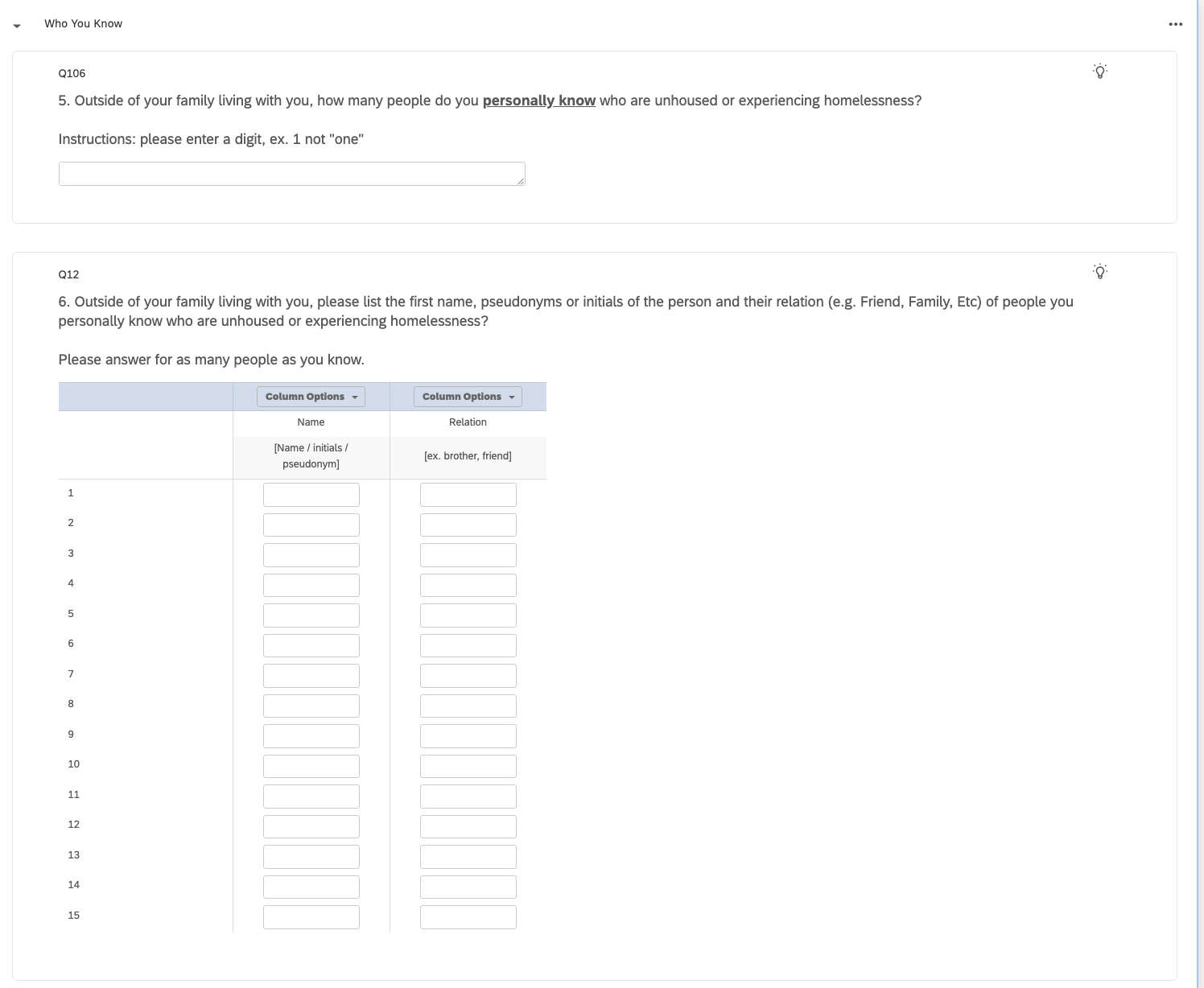}
        \caption{2023 Aggregate and list-based network question.}
    \end{subfigure}\\
    \begin{subfigure}[t]{1\textwidth}
        \centering
        \includegraphics[width=.6\linewidth]{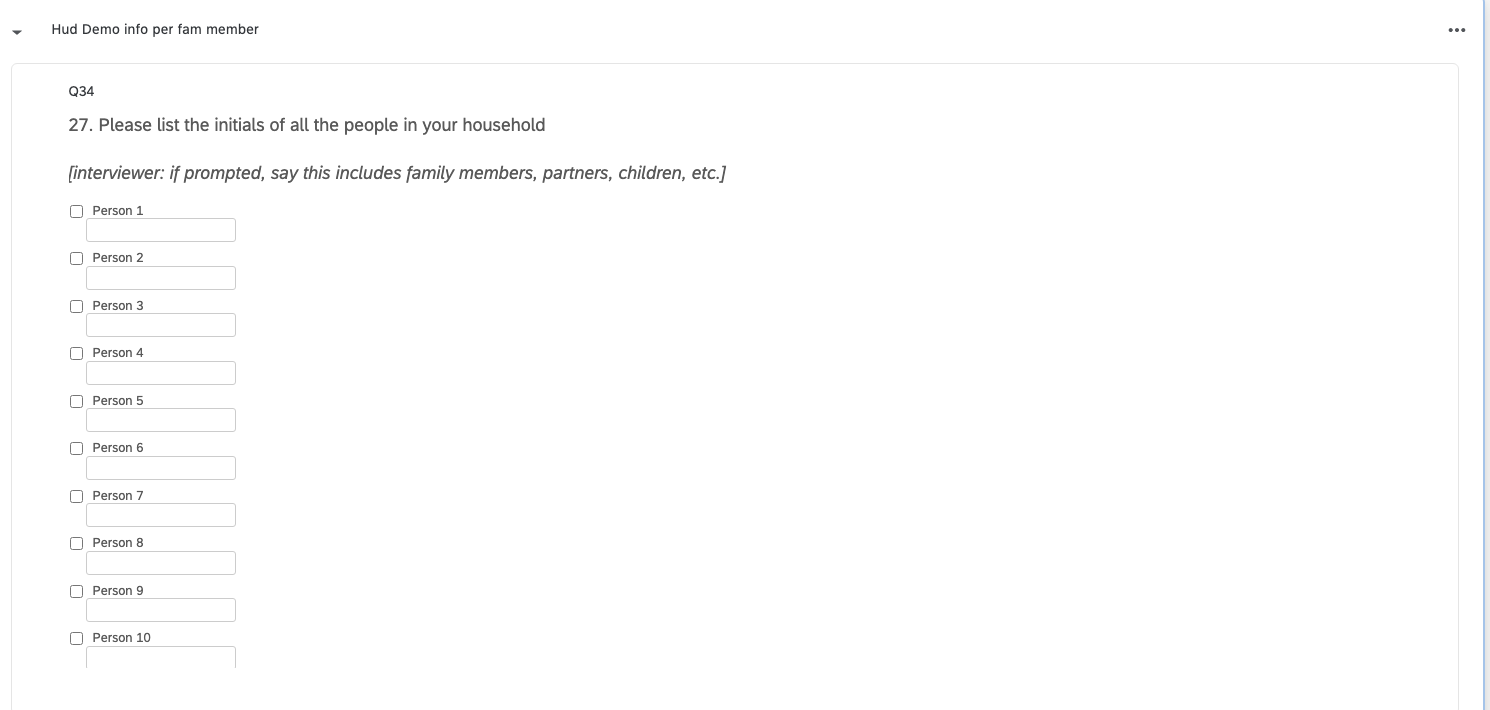}
        \caption{Family list-based question.}
    \end{subfigure}
    \caption{The 2023 network questions asked for the UW non-official RDS PIT study.}
    \label{fig:2023qn}
\end{figure}

In 2023, we expanded our method to include aggregate-level and list-based measures for individuals who know they are experiencing homelessness and their family members. See Figure~\ref{fig:2023qn}.

\subsubsection{2024 Network and Family Questions}

\begin{figure}[htp!]
 \centering
    \begin{subfigure}[t]{1\textwidth}
        \centering
        \includegraphics[width=.6\linewidth]{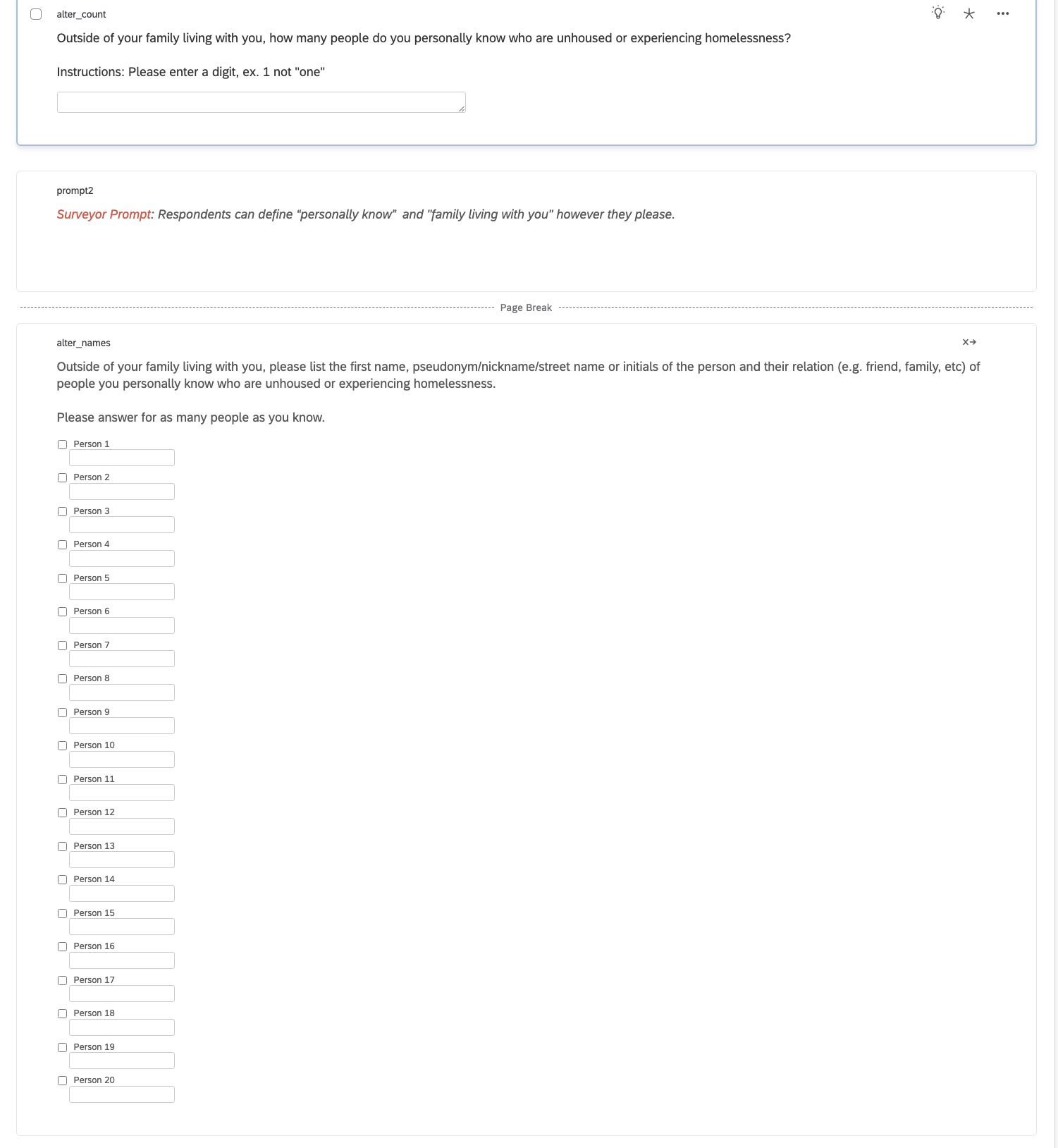}
        \caption{2024 Aggregate and list-based network question.}
    \end{subfigure}\\
    \begin{subfigure}[t]{1\textwidth}
        \centering
        \includegraphics[width=.6\linewidth]{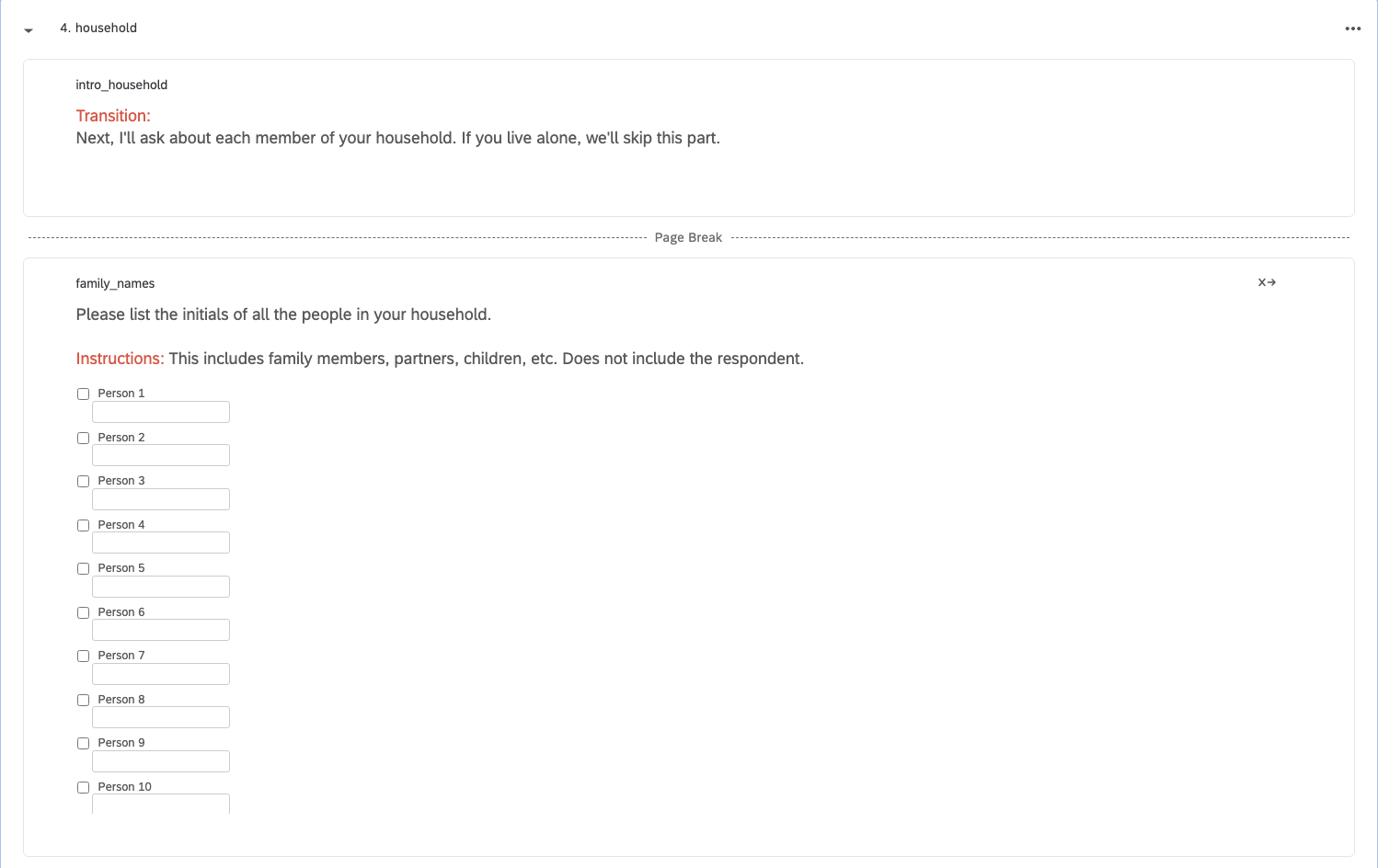}
        \caption{Family list-based question.}
    \end{subfigure}
    \caption{The 2024 network questions asked for the official RDS PIT study.}
    \label{fig:2024qn}
\end{figure}

In 2024, we further improved our family and network questions; see Figure~\ref{fig:2024qn}.

\end{document}